\documentclass[runningheads]{llncs}
\usepackage[T1]{fontenc}
\usepackage{amssymb}
\usepackage[english]{babel}
\usepackage{setspace}
\usepackage{lineno}
\usepackage{color}
\usepackage[T1]{fontenc}
\usepackage[section]{placeins}
\usepackage[pdftex]{graphicx}
\usepackage{floatflt}
\usepackage{float}
\usepackage{amsmath}
\usepackage{orcidlink}
\usepackage{mathtools}

\usepackage{array}
\newcolumntype{M}[1]{>{\centering\arraybackslash}m{#1}}
\newcolumntype{N}{@{}m{0pt}@{}}

\begin{document}

\title{Foundations of Future Communication Systems: Innovations in Communication --
A Report}
\titlerunning{Foundations of Future Communication Systems}

\author{Christian Deppe\inst{1,2}\orcidlink{0000-0002-2265-4887} \and
Eduard Jorswieck\inst{1}\orcidlink{0000-0001-7893-8435} \and \\ Pin-Hsun Lin \inst{1}\orcidlink{}\and Vida Gholamian\inst{1,2}\orcidlink{0000-0002-0513-7378}\and Marcel A.  Mross\inst{1}\orcidlink{0000-0003-1747-6876}}

\authorrunning{C. Deppe, E. Jorswieck, P-H. Lin, V. Gholamian, M. A. Mross}
%
\institute{
Technical University of Braunschweig, Institute for Communications Technology, Braunschweig, Germany \and 6G-life$^2$, 6G research hub, Germany\\
\email{christian.deppe@tu-bs.de, eduard.jorswieck@tu-bs.de, pin-hsun.lin@tu-bs.de, v.gholamian-sefiddarboni@tu-bs.de, m.mross@tu-bs.de}}

\maketitle

\begin{abstract}
The Foundations of Future Communication Systems (FFCS) conference brought together leading researchers from information theory, quantum communication, molecular communication, semantic communication, and secure network design to explore the fundamental principles shaping next-generation communication systems. The event serves as a platform for interdisciplinary exchange, bridging classical Shannon theory, post-Shannon paradigms, quantum information science, and emerging physically grounded communication models. This report compiles the abstracts of all invited talks, contributed presentations, and poster contributions presented at FFCS. The collected works reflect the breadth of contemporary research directions, including identification-based communication, entanglement-assisted networks, semantic and goal-oriented communication, coding for molecular and nano\-scale systems, secure authentication mechanisms, and information-theo\-retic limits of novel phys\-ical-layer architectures.
A central theme of the conference was the re-examination of foundational limits under realistic physical, architectural, and security constraints. Many contributions move beyond traditional rate-centric perspectives and instead investigate reliability, identification, semantics, resource efficiency, and trust in complex and heterogeneous networks. The inclusion of poster abstracts further highlights emerging ideas, early-stage research results, and innovative cross-disciplinary approaches that contribute to shaping future communication paradigms.
By documenting the intellectual landscape presented at FFCS, this report aims to provide a structured overview of current research frontiers and to stimulate continued collaboration across theoretical and experimental domains.
\end{abstract}
\sloppy
\section{Introduction}
The Foundations of Future Communication Systems (FFCS 2026) conference was held from 23rd to 27th February 2026 at Technische Universität Braunschweig (TUBS), Germany. The event brought together international experts from information theory, quantum communication, molecular communication, semantic and goal-oriented communication, and secure network design to discuss the theoretical and technological foundations of next-generation communication systems.

FFCS 2026 was organized by Christian Deppe, Boulat Bash, Holger Boche, Eduard Jorswieck, Uzi Pereg, and Shun Watanabe, representing a broad spectrum of expertise across classical and quantum information theory, network science, and emerging interdisciplinary communication paradigms. The conference was designed as a focused, high-level scientific meeting fostering deep technical exchange, open discussion, and cross-community interaction.

The central aim of FFCS is to revisit and extend the foundations of communication theory in light of new physical resources, architectural constraints, and security requirements. Topics ranged from transmission- and identification-based communication to entanglement-assisted networks, molecular and nanoscale communication systems, semantic and context-aware communication, as well as novel cryptographic and authentication primitives grounded in information theory.

Over the course of five days, the program featured invited talks by leading researchers, panel discussions addressing open challenges in future communication systems, and poster sessions highlighting emerging ideas and ongoing research projects. The interdisciplinary nature of the conference encouraged dialogue between theorists and experimentalists, as well as between researchers working in classical, quantum, and biologically inspired communication frameworks.

This report documents the abstracts of all talks and poster presentations delivered during FFCS 2026 and captures the breadth of current research directions as well as the intellectual momentum generated during the conference week in Braunschweig. It is organized as a collection of individual contributions, each presented on a single page and providing a concise, self-contained summary of the respective presentation, including its motivation, main ideas, and key results. The contributions cover a wide range of topics related to future communication systems, including information-theoretic foundations, identification-based communication, integrated sensing and communication, quantum communication, and emerging application domains. While each contribution can be read independently, the report as a whole offers a comprehensive overview of current research directions and key challenges in the design of next-generation communication systems.

\section{Poster Sessions}

The poster sessions at FFCS 2026 provided a vibrant platform for presenting emerging ideas, ongoing research, and interdisciplinary approaches across a wide spectrum of modern communication theory. The contributions covered areas including classical and quantum information theory, combinatorics, communication and network theory, optimization, and computational methods. These works collectively illustrate the strong influence of foundational research on the development of future communication systems, with applications ranging from 6G and quantum communication to molecular communication, network optimization, resilience, low-latency systems, post-Shannon paradigms, and secure as well as privacy-preserving communication.

A distinctive feature of the poster sessions was the close interplay between deep theoretical insights and practically relevant open problems. This combination fostered intensive discussions during the sessions and breaks, leading not only to refined research directions but also to the initiation of new collaborations, publications, and even patent ideas. In this sense, the poster sessions played a central role in shaping the scientific exchange and intellectual momentum of the conference.

To recognize outstanding contributions, a \textbf{Best Poster Award sponsored by \emph{Entropy}} was presented during the conference. The awards were given as follows:
\begin{itemize}
    \item \textbf{First Prize:} ``Security of Super Dense Coding under Pauli Noise'' by Ghislaine Coulter de Wit
    \item \textbf{Second Prize:} ``Second-Order Rates for Identification with Feedback'' by Marcel Mross
\end{itemize}

\begin{figure}[t]
    \centering
    \includegraphics[width=0.45\textwidth]{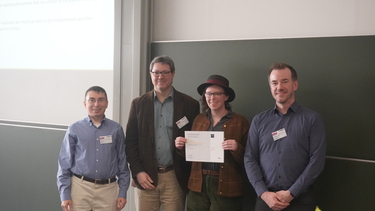}
    \caption{Presentation of the poster award to Ghislaine.}
    \label{fig:poster_award1}
\end{figure}

\begin{figure}[t]
    \centering
    \includegraphics[width=0.45\textwidth]{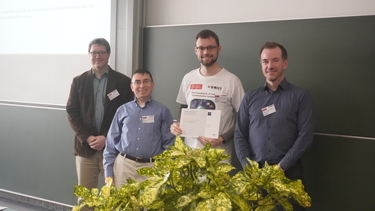}
    \caption{Presentation of the poster award to Marcel.}
    \label{fig:poster_award2}
\end{figure}

The award ceremony is illustrated in Fig.~\ref{fig:poster_award1} and Fig.~\ref{fig:poster_award2}, highlighting the high quality and impact of the presented work and underscoring the importance of early-stage and exploratory research within the FFCS community.

In the following, we present the poster contributions in detail.

\newpage
\subsection{Hadi Aghaee: "Network Oblivious Transfer via Noisy Channels: A Quantum Leap"}
\begin{floatingfigure}[r]{6cm}
\mbox{\includegraphics[width=5.5cm, height=9.78cm]{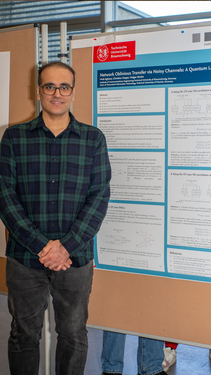}}
\caption{Hadi Aghaee}\label{Hadi}
\end{floatingfigure}
Hadi Aghaee (Fig.~\ref{Hadi}) presented the contribution entitled “On (Im)possibility of Network Oblivious Transfer via Noisy Channels and Non-Signaling Correlations,” ~\cite{aghaee2025networkoblivioustransfernoisy} which studies the fundamental limits of information-theoretic oblivious transfer (OT) in network settings when the communicating parties are assisted by non-signaling (NS) correlations. The paper considers two central multi-user models: discrete memoryless multiple-access channels (DM-MACs) and discrete memoryless broadcast channels (DM-BCs). Its main goal is to understand whether general tripartite NS resources—covering classical, quantum, and even super-quantum correlations—can enhance or enable OT over shared noisy channels.

In the proposed framework, the shared auxiliary resource is modeled as an arbitrary tripartite NS-box, which gives a unified description of channel-induced and externally supplied correlations. Building on this model, the paper formulates correctness and secrecy requirements for network OT in both MAC and BC settings under honest-but-curious users, and distinguishes carefully between sender security, receiver security, and the causal role of the receivers’ private choices. This yields a common language for comparing classical noisy-channel OT with NS-assisted network OT.

The core result of the paper is negative: perfect OT is impossible whenever all relevant parties share a tripartite NS correlation. For the DM-MAC, the paper shows that if the NS resource is nontrivial, Bob’s observation inevitably becomes statistically dependent on both senders’ inputs, which violates the secrecy of the unselected messages. Moreover, the paper argues that even trivial tripartite NS assistance does not rescue OT in the considered deterministic-encoder framework, because the non-signaling conditional-independence constraints conflict with the choice-dependent causal structure that OT requires. An analogous impossibility result is then established for the DM-BC model.

\vspace*{0.5cm}

\newpage
\subsection{Houman Asgari: "Remote Monitoring of two-state Markov sources via random access Channels"}
\begin{floatingfigure}[r]{6cm}
\mbox{\includegraphics[width=5.5cm, height=9.78cm]{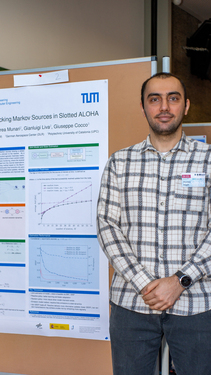}}
\caption{Houman Asgari}\label{Houman}
\end{floatingfigure}
Houman Asgari (Fig.~\ref{Houman}) presented the poster contribution entitled ``Learning and Tracking Markov Sources in Slotted ALOHA,'' which studies the problem of monitoring multiple stochastic sources that share a random-access wireless channel.
This work studies the problem of remote monitoring of multiple two-state Markov sources that transmit status updates to a common receiver over a slotted random access channel. In such systems, the receiver aims to accurately estimate the current source states based on sporadic and possibly collided transmissions. Classical metrics such as Age of Information (AoI) capture the freshness of received updates but do not fully characterize the receiver’s uncertainty about the monitored processes when the sources exhibit temporal correlation. Recent work has therefore proposed state-centric metrics such as state estimation error probability (SEEP), which explicitly account for the dynamics of the underlying Markov sources.
Building on this framework, prior work models the monitoring problem through a hidden Markov structure and derives optimal state estimators at the receiver. In particular, the maximum a posteriori (MAP) detector can be implemented through a recursive update of a log-likelihood ratio, where the posterior belief about the source state is updated sequentially based on channel observations and the Markov transition probabilities. This recursive formulation enables tractable inference. Such analyses reveal a fundamental trade-off between different performance objectives: reactive transmission policies, where nodes transmit upon detecting a state change, can significantly reduce SEEP, while random transmission policies may yield lower AoI.
In many practical scenarios, however, the transition statistics of the monitored sources are not known a priori at the receiver. This paper addresses this limitation by proposing a joint model and state estimation framework that learns the unknown source dynamics directly from channel observations. The findings show that integrating statistical learning with recursive Bayesian state estimation enables effective monitoring even when the source model is initially unknown.

\vspace*{0.5cm}

\newpage
\subsection{Igor Bernard: "One-Shot Distributed Instrument Simulation"}
\begin{floatingfigure}[r]{6cm}
\mbox{\includegraphics[width=5.5cm, height=9.78cm]{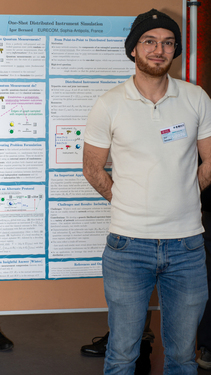}}
\caption{Igor Bernard}\label{IgorBernard}
\end{floatingfigure}
Igor Bernard (Fig.~\ref{IgorBernard}) presented the poster contribution ``One-Shot Distributed Instrument Simulation,'' which studies the distributed simulation of quantum instruments in a one-shot setting. Quantum measurements generate both classical outcomes and post-measurement quantum states, thereby establishing correlations between measurement results and the resulting quantum system.

Quantum instruments generalize POVMs by producing a classical outcome together with the associated post-measurement quantum state. In the point-to-point case, Winter’s measurement compression theorem characterizes the classical communication and shared randomness required to simulate an instrument while preserving the induced correlations. A natural extension is to consider network settings where subsystems of an entangled state are distributed among different parties.

In this work, one-shot distributed instrument simulation is studied for a bipartite quantum state shared by two encoders and an inaccessible reference, where a receiver must recover the measurement outcome. Focusing on separable instruments, the work analyzes the communication rates and common randomness required to simulate the instrument up to small trace-norm error without relying on asymptotic tools.

A likelihood-operator-based simulation framework is developed using proxy states and a compatible-operator-sliding (COS) technique to control the simulation error in the one-shot regime. This yields an achievable one-shot rate region expressed in terms of hypothesis-testing mutual information, smoothed Rényi-2 quantities, and other one-shot divergence measures.

As an application, the framework provides a one-shot inner bound for three-party purity distillation, where two senders communicate classical information to a receiver while all parties distill local purity from a shared tripartite state, highlighting distributed instrument simulation as a useful primitive for quantum Shannon theory beyond point-to-point scenarios.

\vspace*{0.5cm}

\newpage
\subsection{Holger Boche: "Integrated Sensing and Communication with Distributed Rate-Limited Helpers"}
\begin{floatingfigure}[r]{6cm}
\mbox{\includegraphics[width=5.5cm, height=9.78cm]{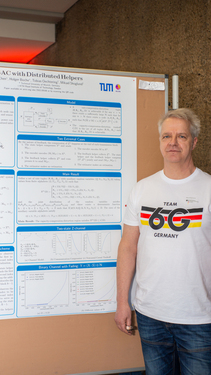}}
\caption{Holger Boche}\label{Holger1}
\end{floatingfigure}
Holger Boche (Fig.~\ref{Holger1}) presented integrated sensing and communication (ISAC) systems with two rate-limited helpers who observe the channel state sequence and the feedback sequence, respectively. The state helper observes the state sequence noncausally and can therefore use the state information either causally or noncausally. In the noncausal case, it sends a description of the state to the encoder at the beginning of the transmission block to facilitate communication at the cost of higher compression. In the causal case, it waits until the end of the block and compresses the state jointly with the feedback helper, which does not improve the communication but can reduce the compression rate. They propose a coding scheme combining both approaches by splitting the message of the state helper into two parts: the first part is decoded by the sender and used as channel state information, while the second part is decoded only after the sender receives the feedback at the end of the block; the state estimation is performed at the end of each block. The scheme provides an inner bound on the capacity--compression--distortion tradeoff region. When the state helper is aware of the decoding result, the scheme is revised to a block Markov coding strategy so that the helper can incorporate the decoding outcome into its compression step, and two special cases of the capacity--compression--distortion region are fully characterized. Two numerical examples illustrate the impact of causal and noncausal state usage on the communication and compression rates, and the tradeoff between communication rate and compression rate under different distortion constraints, highlighting the necessity of the state helper when the distortion constraint is small. Part of this work was presented at the 2025 International Communication Conference (ICC)~\cite{chen2025integrated}. This work was partly supported by the German Ministry of Education and Research (BMBF) within the program ``Souver\"an. Digital. Vernetzt'' in the joint project 6G-life under Grant~16KISK002 and by the Bavarian State Ministry of Science and the Arts within the project Next Generation AI Computing (GAIn).

\vspace*{0.5cm}

\newpage
\subsection{Holger Boche: "Complete Characterization of the Arithmetic Complexity of the Capacity of Gaussian Fading Channels"}
\begin{floatingfigure}[r]{6cm}
\mbox{\includegraphics[width=5.5cm, height=9.78cm]{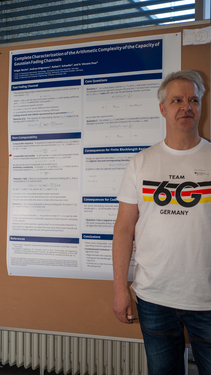}}
\caption{Holger Boche}\label{Holger2}
\end{floatingfigure}
Holger Boche (Fig.~\ref{Holger2}) presented this poster contribution, which studies the algorithmic computability of the secrecy capacity of fast-fading Gaussian wiretap channels from a foundational perspective~\cite{boche2025arithmeticcomplexitysecrecy}. The reliable design of wireless communication systems, particularly for emerging 6G applications, requires a precise understanding of channel capacity under realistic fading conditions. The work investigates the algorithmic and arithmetic complexity of the capacity of fast-fading Gaussian channels with input--output relation $y_i=\alpha_i x_i+z_i$, where $\alpha_i$ denotes the fading coefficient and $z_i$ is additive white Gaussian noise of variance $\sigma^2$. The fading process is described by a density $f_\alpha$, and the capacity under an average power constraint $P$ is given by $C(P,f_\alpha,\sigma^2)=\int_{-\infty}^{+\infty}\log\!\left(1+\frac{P|\alpha|^2}{\sigma^2}\right)f_\alpha(\alpha)\,d\alpha$. The authors characterize the computability and arithmetic complexity of this capacity and show that there exist computable continuous fading distributions $f^*$ for which the capacity $C(P,f_\alpha,\sigma^2)$ is not a computable real number for all computable $P>0$ and $\sigma^2>0$, implying that no universal algorithm can approximate the capacity to arbitrary precision. Moreover, the capacity is shown to belong to the class $\Sigma_1$ of the arithmetical hierarchy, meaning it can be represented as the supremum of a computable sequence of rational lower bounds, while still being non-computable for certain channels, which reveals fundamental algorithmic limitations in determining the capacity of such fading channels.
\vspace*{0.5cm}

\newpage
\subsection{Holger Boche: "Convexification in Non-Convex Optimization is not Turing Computable"}
\begin{floatingfigure}[r]{6cm}
\mbox{\includegraphics[width=5.5cm, height=9.78cm]{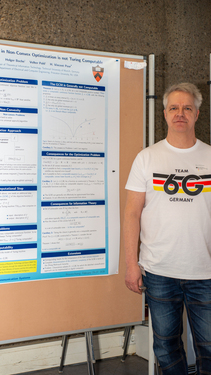}}
\caption{Holger Boche}\label{Holger3}
\end{floatingfigure}
Holger Boche (Fig.~\ref{Holger3}) presented this poster contribution, which investigates the algorithmic limits of convexification procedures in continuous optimization problems. Optimization techniques play a central role in communications, signal processing, information theory, and control, where convex optimization is particularly attractive due to its computational tractability and the availability of efficient numerical algorithms. However, many practically relevant problems are non-convex, for which no universal efficient solution methods exist, and convexification is often used to exploit convex optimization techniques. In this context, the convexification operator maps a continuous real-valued function to its greatest convex minorant. The poster shows that this operator cannot be implemented on a Turing machine even when restricted to computable continuous functions~\cite{BPP_CDC25}. In particular, there exist piecewise linear continuous functions $f$ with a unique global minimum whose greatest convex minorant is not Turing-computable. Moreover, for such functions there is no monotone increasing computable sequence of convex computable continuous functions that converges to the greatest convex minorant, implying that the standard maximization-based characterization does not yield a universal numerical algorithm. The same conclusions hold analogously for the least concave majorant. The poster also discusses implications for information theory, where computing convex hulls of sets frequently arises; it is shown that even for simple sets this task can be infeasible, and the work further examines whether the sets of achievable and non-achievable rates are semi-decidable.
\vspace*{0.5cm}

\newpage
\subsection{Minglai Cai: "Quantum Byzantine Multiple Access Channels"}
\begin{floatingfigure}[r]{6cm}
\mbox{\includegraphics[width=5.5cm, height=9.78cm]{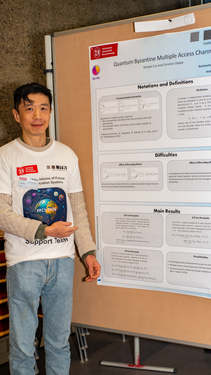}}
\caption{Minglai Cai}\label{Minglai}
\end{floatingfigure}
Minglai Cai (Fig.~\ref{Minglai}) presented the poster contribution ``Quantum Byzantine Multiple Access Channels,'' which studies classical--quantum multiple-access channels in the presence of an adversarial transmitter. In such Byzantine communication scenarios, one of the legitimate senders may behave maliciously and attempt to disrupt the communication of honest users. The model extends the Byzantine multiple-access channel introduced in~\cite{Sa/Ba/De/Pr,Sa/Ba/De/Pr2} to the quantum setting. In this framework, a classical--quantum multiple-access channel is considered where one sender may inject malicious inputs as a jamming attack, while the receiver does not know which sender is adversarial. Nevertheless, the receiver is required to reliably decode the messages of the honest users. Extending this model to the quantum domain introduces additional challenges: unlike classical arbitrarily varying channels, whose capacity can be achieved using deterministic codes~\cite{Ahl/Ca}, quantum arbitrarily varying channels currently rely on random coding techniques~\cite{Da/Wa/Ha}. Moreover, quantum measurements can disturb the quantum state, so that decoding earlier messages may affect the decodability of subsequent ones. The poster introduces the formal model of a $k$-user Byzantine classical--quantum multiple-access channel and derives the corresponding random coding capacity region when at most one sender is adversarial and the receiver decodes the messages successively.
```

\vspace*{0.5cm}

\newpage
\subsection{Ghislaine Coulter-de Wit: "Security of Super Dense Coding under Pauli Noise"}
\begin{floatingfigure}[r]{6cm}
\mbox{\includegraphics[width=5.5cm, height=9.78cm]{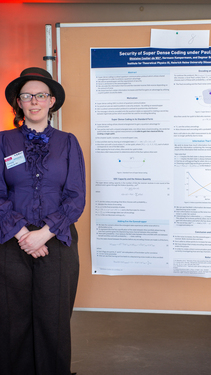}}
\caption{Ghislaine Coulter-de Wit}\label{Ghislaine}
\end{floatingfigure}
Ghislaine Coulter-de Wit's poster (Fig.~\ref{Ghislaine}) examined the security of super dense coding in the presence of Pauli noise, analyzing how noise impacts confidentiality and decoding performance, and deriving conditions under which security guarantees can be preserved. Super dense coding is a form of quantum communication utilizing shared entanglement such that—in the simplest formulation—Bob receives a message two bits long from one qubit sent by Alice. To accomplish this, Alice and Bob share a maximally entangled Bell state, Alice applies one of four unitary encodings to her qubit, and then sends the qubit to Bob. Bob is then able to perform a global Bell measurement and recover which unitary Alice chose, thereby decoding the message.

The real world contains noise and untrustworthy parties (eavesdroppers). Building on the work of Zarah Shadman et al.~\cite{Shadman} on noisy super dense coding, the security of the transmitted classical data is investigated. An eavesdropper Eve is introduced, who distributes a purification of the initial state. This purification is such that Alice and Bob experience a depolarizing noise channel, representable with Pauli matrices. There are two cases: with probability $\lambda$, Eve distributes a qubit between herself and Alice, thereby gaining full information. Bob then receives a qubit between himself and Eve to disguise the attack but never receives any legitimate message. With probability $(1-\lambda)$, Eve distributes the qubit correctly between Alice and Bob and gains no information about the message. Alice then chooses her encodings from the same group of four unitary operators, and Bob and Eve perform their measurements once Alice has sent her qubit.

Eve utilizes qutrit flags to determine which case she is in, thereby gaining more information than expected, and more than Bob loses for the same amount of noise. Further work aims to extend this approach by utilizing wiretap codes as an alternative encoding procedure.

\vspace*{0.5cm}

\newpage
\subsection{Christian Eckrich: "Distributed Radar Sensor Networks Under Communication Constraints"}
\begin{floatingfigure}[r]{6cm}
\mbox{\includegraphics[width=5.5cm, height=9.78cm]{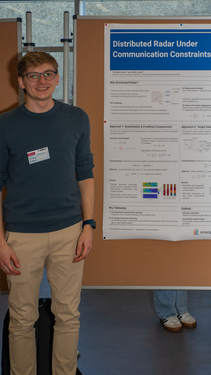}}
\caption{Christian Eckrich}\label{Eckrich}
\end{floatingfigure}
Christian Eckrich's poster (Fig.~\ref{Eckrich}) studied distributed radar sensor networks under communication constraints, focusing on trade-offs between sensing accuracy, network topology, and limited-rate links between sensors and fusion centers~\cite{Eckrich2024}. Distributed radar systems can localize and track targets more accurately than a single radar because spatially separated sensors provide diverse viewpoints and redundancy. However, exploiting these advantages requires transmitting sensing data to a fusion center over capacity-constrained fronthaul or backhaul links, which limits the information available for estimation. The work links communication constraints to estimation performance by expressing localization uncertainty via the Cram\'er--Rao lower bound (CRLB) derived from the Fisher information available at the fusion center. This leads to a resource allocation problem that minimizes localization uncertainty subject to limited communication resources. Two strategies are investigated. The first uses quantization to compress sensor observations before transmission, where the resulting distortion can be interpreted as a degradation of the information reaching the fusion center; sensors that contribute less information are allocated fewer bits~\cite{Eckrich2024}. The second approach uses discrete resource allocation through sensor selection: radar nodes perform spectral windowing to separate targets locally and decide which targets to transmit. The selection of radar--target pairs is optimized such that the CRLB across all targets is minimized while satisfying the communication constraints. Both approaches prioritize transmissions according to marginal information gain and can preserve much of the sensing benefit even under severe communication limitations.

\vspace*{0.5cm}

\newpage
\subsection{Pol Juli\`a Farr\'e: "An Alternative to Conventional Quantum Key Distribution for Secret-Key Growing"}
\begin{floatingfigure}[r]{6cm}
\mbox{\includegraphics[width=5.5cm, height=9.78cm]{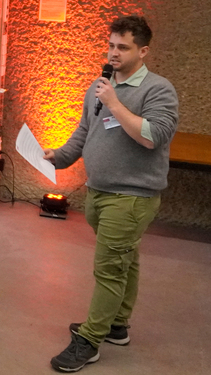}}
\caption{Pol Juli\`a Farr\'e}\label{Pol}
\end{floatingfigure}
Pol Juli\`a Farr\'e (Fig.~\ref{Pol}) presented a secret key growing protocol that incorporates several novelties with respect to the state of the art in the field. The poster is the result of a project that lasted around six months and concluded in autumn 2025, when it was presented at the European Wireless Conference in Nice, France~\cite{farré2025securehybridkeygrowing}. While in the initial stages of the project the proposed ideas appeared promising and incorporated real-world considerations, some of them turned out to be impractical.

The scheme allows, in theory, the secure growth of an initial shared key between two parties by exploiting two different degrees of freedom within light pulses: the photon number and the time bin. The latter, together with a classical assumption on the physical layer, is actively exploited to prevent those adversaries that accumulate information over several protocol rounds, while the former motivates the use of quantum coherence witnessing for eavesdropping detection. To the best of current knowledge, this approach to key growing, falling into the subcategory of Hybrid Key Growing, has not been reported in the literature prior to this contribution. However, after careful analysis, it is observed that the present state of quantum technologies does not allow for a practical realization of the scheme due to the delicateness of photon-number-state manipulation. Nevertheless, the work in \cite{doi:10.1073/pnas.2219208120} provides some indication that such methods may become implementable in the future.

As a final positive aspect of the research presented in the poster, the method ensures entity authentication via the quantum channel instead of, as is typically done, the classical one. This idea has already been proposed by other authors, e.g., \cite{Park2023}, but the authentication mechanism, based on a pre-shared secrecy update after each round, is again novel and appears promising. In this direction, an ongoing project benefits from the foundational ideas presented in this poster.

\newpage
\subsection{Rami Ezzine: "Common Randomness Generation from Sources with Infinite Polish Alphabet Aided by Unidirectional Communication"}
\begin{floatingfigure}[r]{6cm}
\mbox{\includegraphics[width=5.5cm, height=9cm]{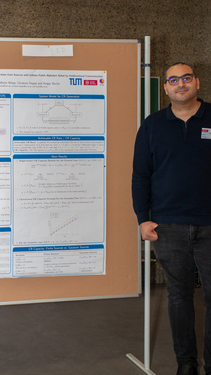}}
\caption{Rami Ezzine}\label{Rami}
\end{floatingfigure}
Rami Ezzine (Fig.~\ref{Rami}) presented the poster entitled \emph{“Common Randomness Generation from Sources with Infinite Polish Alphabet Aided by Unidirectional Communication.”} The work studies the generation of common randomness (CR) between communicating parties observing correlated sources while using limited communication. CR is a fundamental resource in several information-theoretic applications, including message identification, secret key generation, and modular coding schemes. Classical CR generation problems were introduced by Ahlswede and Csiszár \cite{AhlswedeCsiszar1998} and mainly studied for finite-alphabet sources.

The poster extends this framework to sources defined on infinite alphabets. In particular, the sources take values in Polish spaces, i.e., complete separable metric spaces that naturally model continuous random variables. The considered model involves two correlated sources and unidirectional communication over a noisy memoryless channel. The objective is to characterize the CR capacity, defined as the maximum achievable rate at which the parties can agree on common randomness.

For general infinite-alphabet sources, single-letter lower and upper bounds on the CR capacity are derived. As an important example, the Gaussian source model is analyzed, for which a closed-form expression for the CR capacity is obtained as a function of the source parameters and the channel capacity. Interestingly, the CR capacity becomes unbounded when the Gaussian sources are perfectly correlated. These results reveal a fundamental difference between finite-alphabet and continuous source models and highlight how moving to infinite alphabets leads to qualitatively different behavior in common randomness generation.

\vspace*{0.5cm}

\newpage
\subsection{Vida Gholamian: "Joint Secure Randomized Identification and Sensing over State-Dependent Wiretap Channels with Feedback"}
\begin{floatingfigure}[r]{6cm}
\mbox{\includegraphics[width=5.5cm, height=9.78cm]{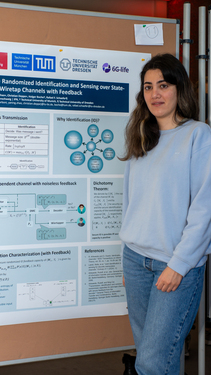}}
\caption{Vida Gholamian}\label{Vida}
\end{floatingfigure}
Vida Gholamian (Fig.~\ref{Vida}) presented the poster contribution entitled ``Secure Randomized Identification with Feedback and Sensing over State-Dependent Wiretap Channels,'' which studies identification via channels in a setting that jointly incorporates feedback, sensing, and secrecy constraints. In contrast to classical transmission, identification only requires the receiver to decide whether a specific message was sent, leading to a fundamentally different double-exponential scaling of the message set.

The considered model is a state-dependent discrete memoryless wiretap channel with noiseless feedback, where the channel state is unknown to both legitimate terminals and a sensing distortion constraint is imposed. Besides reliable identification at the legitimate receiver, the system must ensure strong secrecy with respect to an external wiretapper observing the channel output through a second channel. This unifies identification, joint sensing and communication, and information-theoretic security within a single framework.
A central result of the work is a complete characterization of the secure randomized identification capacity–distortion function. The analysis reveals a clear dichotomy: if the secrecy capacity of the underlying wiretap channel is strictly positive, then the secure identification capacity with feedback coincides with the non-secure one; if it is zero, secure identification is impossible. In the positive case, the fundamental limit is given by the entropy of the induced output distribution at the legitimate receiver, maximized over all input distributions satisfying the sensing distortion constraint.
The achievability combines feedback-generated common randomness, typicality arguments, and channel resolvability techniques to guarantee both reliability and strong secrecy. Overall, the work demonstrates that, under mild conditions, secrecy does not reduce the identification performance even in the presence of joint sensing requirements, thereby clarifying the fundamental limits of secure identification in state-dependent environments.

\vspace*{0.5cm}

\newpage
\subsection{Fatma Gouiaa: "Achievable Rate Regions for Classical Quantum Broadcast and Interference channels using Coset Codes"}
\begin{floatingfigure}[r]{6cm}
\mbox{\includegraphics[width=5.5cm, height=9.78cm]{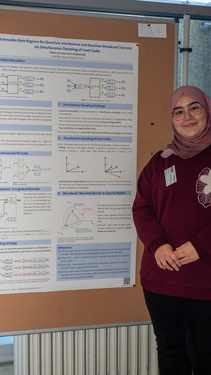}}
\caption{Fatma Gouiaa}\label{Fatma}
\end{floatingfigure}
Fatma Gouiaa (Fig.~\ref{Fatma}) presented achievable rate regions for classical–quantum broadcast and interference channels using coset codes, emphasizing structured coding techniques that exploit algebraic properties to manage interference and attain improved trade-offs. Fatma undertakes a Shannon-theoretic study of communicating classical bit streams over multi-terminal quantum channels, specifically the three-user quantum interference channel (3-CQIC) and the three-user quantum broadcast channel (3-CQBC).
In this work, Fatma characterizes a new achievable rate region for the classical–quantum capacity regions of the 3-CQIC and the 3-CQBC. Our approach is motivated by insights from classical network information theory, where structured codes such as coset codes have been shown to outperform unstructured IID codes in certain multi-user scenarios. Inspired by these results, Fatma designs a coding strategy that exploits the algebraic structure of coset codes to manage and decode the bivariate interference more efficiently.
A key technical challenge in our work is the analysis of simultaneous decoding in the quantum setting. Our main contribution is to extend the powerful technique of tilting, smoothing, and augmentation, introduced by Sen recently \cite{202103SAD_Sen}, to the analysis of coset codes. This extension enables the construction of suitable simultaneous decoders for our coding strategy. Fatma also proposes a likelihood encoder  that facilitates analysis. Among others, one contribution of our work is the integration of several new elements - coset codes, simultaneous decoding via tilting smoothing and augmentation, and a likelihood encoder - in a manner that permits analysis.
Analyzing the information-theoretic performance of the proposed coset code based coding strategy, Fatma characterizes new inner bounds that strictly subsumes all currently known inner bounds for the 3-CQIC and 3-CQBC. Moreover, for certain identified channel instances, Fatma shows that our inner bound is strictly larger than previously known inner bounds.

\vspace*{0.5cm}

\newpage
\subsection{Rodrigo Kloster Albarrac\'in: "Continuous-Variable Quantum Key Distribution (CV-QKD) for Inter-Satellite Links (ISL)"}
\begin{floatingfigure}[r]{6cm}
\mbox{\includegraphics[width=5.5cm, height=9.78cm]{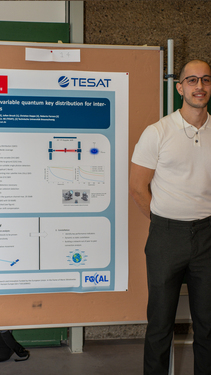}}
\caption{Rodrigo Kloster Albarrac\'in}\label{Rodrigo}
\end{floatingfigure}
Rodrigo Kloster Albarracín (Fig.~\ref{Rodrigo}) presented the poster contribution ``Continuous Variable Quantum Key Distribution for Inter-Satellite Links,'' which investigates the feasibility of continuous-variable quantum key distribution (CV-QKD) for secure communication in satellite networks. Satellite-based QKD is considered a key enabling technology for global-scale quantum communication, as satellites can provide long-distance links beyond the reach of terrestrial fiber networks.

This poster introduces a new doctoral research project focused on exploring the feasibility of continuous-variable quantum key distribution (CV-QKD) for inter-satellite links (ISLs). Current discrete-variable (DV) QKD systems are limited to satellite-to-ground links due to their reliance on specialized single-photon detectors. In contrast, CV-QKD leverages space-proven coherent detection techniques, making it a promising alternative for ISLs \cite{usenko_continuous-variable_2025}.
The research aims to analyze peer-to-peer connections within satellite constellations, evaluate different CV-QKD protocols, and address challenges such as Doppler shifts and detector noise \cite{schlake_pulse_2025}. The goal is to develop secure and implementable protocols tailored to ISLs, incorporating rigorous security analyses and performance simulations.
Funded by the European Union under the Marie Skłodowska-Curie Actions (MSCA), this project seeks to advance Europe’s capabilities in quantum-secure space infrastructures, paving the way for globally scalable quantum networks.

\vspace*{0.5cm}

\newpage
\subsection{Pin-Hsun Lin: "Oblivious transfer through AWGN channel by polar codes"}
\begin{floatingfigure}[r]{6cm}
\mbox{\includegraphics[width=5.5cm, height=9.78cm]{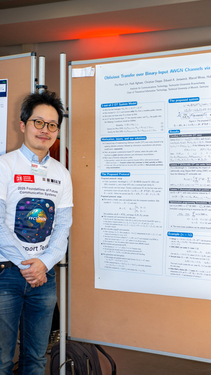}}
\caption{Pin-Hsun Lin}\label{PHL}
\end{floatingfigure}
  Pin-Hsun Lin (Fig.~\ref{PHL}) introduced a polar-code-based OT framework~\cite{PHL_OT} that views the underlying binary-input AWGN channel through a \emph{virtual} BEC: indices in the good set $\mathcal{G}$ play the role of reliable, non-erased positions for Bob's chosen message, while indices in the bad set $\mathcal{B}$ are used to convey essentially no information about the unchosen message. This BEC emulation is tailored specifically to the analysis of OT and is fundamentally different from the alphabet extension/GEC (labeling induced by linear constraints from the polar transform) as in~\cite{SudaWatanabe25}.

A special set of permutations, namely \textit{automorphisms} of the polar transform $\mathbf{T}$, is introduced to generate different ``views'' of $\mathcal{G}$ and $\mathcal{B}$ at Alice and Bob. A controlled amount of reliability is traded for a more symmetric virtual BEC by letting a carefully selected small subset of BBCs carry independent random bits unknown to Bob. From Bob's viewpoint, these bits behave as virtual erasures for the undesired message. The construction achieves the desired BEC-like structure using only polarization and permutations from the automorphism group. An information-theoretic security analysis of the resulting protocol is provided, establishing both secrecy for Alice and secrecy for Bob. A relaxed reliability constraint is also introduced that reflects the nonstandard two-view use of polar codes.

To make the permutation step in the OT construction explicit, efficiently implementable, and analytically tractable, a complete characterization of the automorphism group of the polar transform is provided. Concretely, every permutation from the automorphism is induced by a unique permutation of the $m$ bit positions, and the size of the automorphism set is $m!$. This characterization yields two practical benefits: (i) it provides a complete search space of permissible permutations for hiding the GBC and BBC structure without breaking the polar transform, and (ii) it enables uniform sampling and enumeration of automorphisms rather than relying on ad hoc permutations.

\vspace*{0.5cm}

\newpage

\subsection{Karol Lukanowski: "Security of Optical Key Distribution Against quantum-enhanced Coherent Eavesdropping"}
\begin{floatingfigure}[r]{6cm}
\mbox{\includegraphics[width=5.5cm, height=9.78cm]{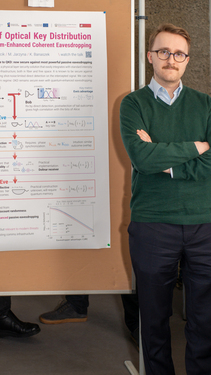}}
\caption{Karol Lukanowski}\label{Karol}
\end{floatingfigure}
Karol Łukanowski (Fig.~\ref{Karol}) presented the poster contribution ``Security of Optical Key Distribution Against Quantum-Enhanced Coherent Eavesdropping,'' which analyzes the security of optical key distribution (OKD) schemes under increasingly powerful eavesdropping models. Optical key distribution is a physical-layer security technique that can be integrated into standard intensity-modulation and direct-detection optical communication systems and provides an alternative to quantum key distribution (QKD) with high practical compatibility to existing infrastructure.

The poster investigates the robustness of OKD against different types of passive eavesdroppers with increasing capabilities. In the considered protocol, Alice encodes information using intensity modulation of coherent optical states, while Bob performs direct detection and extracts correlated key bits from the received signal. The analysis studies several possible measurement strategies available to an eavesdropper, including direct detection, coherent detection via homodyne measurement, the Helstrom measurement minimizing the quantum detection error probability, and the optimal collective measurement described by the Gordon–Holevo bound.

By comparing the achievable key rates under these increasingly powerful attacks, the work evaluates the security margin of OKD systems in the presence of quantum-enhanced eavesdroppers. The results indicate that the protocol remains secure even when the eavesdropper employs advanced quantum measurement strategies, while still maintaining favorable key rates and compatibility with existing optical communication technologies. The study highlights optical key distribution as a promising low-hanging-fruit approach for practical physical-layer security that can complement or extend current QKD solutions.

\vspace*{0.5cm}

\newpage
\subsection{Marcel Mross: "Second-Order Rates for Identification with Feedback"}
\begin{floatingfigure}[r]{6cm}
\mbox{\includegraphics[width=5.5cm, height=9.78cm]{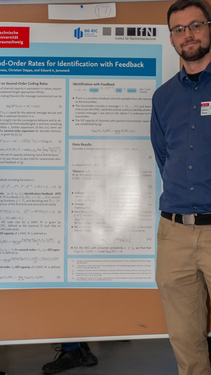}}
\caption{Marcel Mross}\label{Marcel}
\end{floatingfigure}
Marcel Mross' poster (Fig.~\ref{Marcel}) addressed second-order coding rates for identification with feedback, refining asymptotic identification results by quantifying dispersion-type terms and the role of feedback. Second-order coding rates show how quickly the rates converge towards capacity as the blocklength $n$ approaches infinity. In this case, these rates were given for identification via channels, where the receiver asks whether a specific message $i'$ was sent or not, in the presence of noiseless feedback. It is known that feedback increases the identification capacity from the mutual information to the the output entropy of the channel. 

The second-order coding rate for randomized identification with feedback was fully characterized in the work presented in this poster, while the converse for deterministic identification with feedback is valid for the class of simple-dispersion channels. It was shown that for channels with certain symmetry properties, the second-order rate of deterministic identification with feedback is equal to that of transmission, even though the (first-order) capacity differs. For the same class of channels, the second-order coding rate of randomized identification with feedback is shown to be zero, meaning that the rates converge quickly towards capacity. The details of this work can be found in \cite{mross.second-order}.

\vspace*{0.5cm}

\newpage
\subsection{Husein Natur: "Quantum Secret Sharing Rates"}
\begin{floatingfigure}[r]{6cm}
\mbox{\includegraphics[width=5.5cm, height=9.78cm]{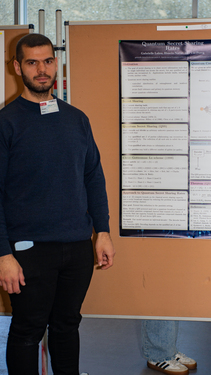}}
\caption{Husein Natur}\label{Husein}
\end{floatingfigure}
Husein Natić (Fig.~\ref{Husein}) presented the poster contribution ``Quantum Secret Sharing Rates,'' which investigates achievable rates for quantum secret sharing (QSS) over noisy quantum channels. Secret sharing is a fundamental cryptographic primitive in which a secret is distributed among multiple parties such that only qualified subsets of participants can reconstruct the secret, while non-qualified subsets obtain no information. Quantum secret sharing extends this concept to the quantum domain, where both classical and quantum information may be protected using quantum states and entanglement.

The poster studies QSS in the presence of channel uncertainty modeled as a quantum compound channel. In this setting, the sender distributes quantum shares through a family of possible quantum channels, while the exact channel realization is unknown but assumed to belong to a known set. The problem is to determine the maximum rate at which quantum secrets can be reliably shared among multiple receivers while satisfying the required reconstruction and secrecy constraints.

Building on classical results for secret sharing over noisy broadcast channels, the work develops a quantum information-theoretic framework for analyzing QSS rates under such compound channel models. The protocol uses an encoder that distributes quantum states to multiple receivers and a decoder that allows qualified subsets of users to reconstruct the secret quantum state. The achievable rate is characterized using coherent information expressions associated with the corresponding quantum channels.

These results contribute to a deeper understanding of the fundamental limits of secure multi-party quantum communication and provide insights into how quantum secret sharing protocols can be implemented reliably in realistic quantum network environments.

\vspace*{0.5cm}

\newpage
\subsection{Kumar Nilesh: "Quantum Biometrics as a Critical Component in Future Secure Quantum Networks"}
\begin{floatingfigure}[r]{6cm}
\mbox{\includegraphics[width=5.5cm, height=9.78cm]{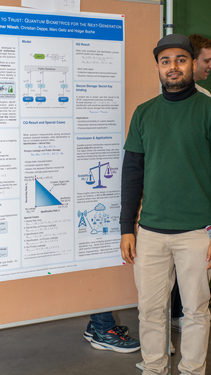}}
\caption{Kumar Nilesh}\label{Nilesh}
\end{floatingfigure}
Kumar Nilesh (Fig.~\ref{Nilesh}) presented the poster “From Hardware to Trust: Quantum Biometrics for the Next Generation,” focusing on quantum-enhanced biometric systems for secure identification and authentication. Classical biometric systems, while widely used, rely on heuristic security assumptions and are vulnerable to spoofing and privacy leakage. In contrast, quantum biometrics leverage intrinsic physical randomness and quantum measurement processes to provide stronger, information-theoretic security guarantees.
Driven by the needs of large-scale digital infrastructures such as IoT and future 6G networks, the work addresses the limitations of classical hardware security primitives like PUFs, which are increasingly susceptible to machine-learning attacks. Instead, it explores Quantum Physical Unclonable Functions (QPUFs) and quantum biometric modalities, whose security is grounded in fundamental quantum principles such as the no-cloning theorem and measurement disturbance.
Building on prior work on QPUF authentication and quantum-secure storage \cite{ref1,ref2}, a unified information-theoretic framework for enrollment-based identification systems is developed \cite{ref3}. This framework jointly supports identification, authentication, and secret-key establishment, modeling biometric or device fingerprints as mappings from classical or quantum challenges to quantum states.
Within this model, the fundamental trade-off between identification rate, secret-key rate, privacy leakage, and storage is characterized. The complete single-letter capacity region is derived for the classical-quantum setting, along with achievable results for the fully quantum case, demonstrating gains from coherent quantum processing. Furthermore, the equivalence between secret-key generation and binding is established \cite{ref4}, providing flexibility for practical implementations.
Overall, the work establishes rigorous information-theoretic limits and design principles for scalable, quantum-secure identification and authentication systems.
\vspace*{0.5cm}

\newpage
\subsection{Nikhitha Nunavath: "Communicating Properties of Quantum States over a Classical Wireless Channel"}
\begin{floatingfigure}[r]{6cm}
\mbox{\includegraphics[width=5.5cm, height=9.78cm]{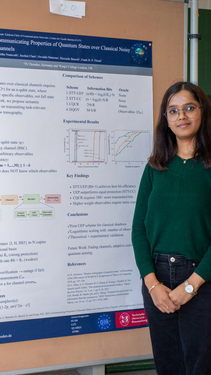}}
\caption{Nikhitha Nunavath}\label{Nikhitha}
\end{floatingfigure}
Nikhitha Nunavath (Fig.~\ref{Nikhitha}) presented the poster contribution ``Communicating Properties of Quantum States over Classical Noisy Channels,'' which investigates how task-relevant properties of quantum states can be communicated efficiently using classical communication. In many applications, transmitting a full quantum state description requires exponential resources in the number of qubits, while practical tasks often require only specific observables or properties of the state. The work therefore focuses on semantic communication of quantum information, where only the information relevant for a given task is transmitted.

The considered setting involves an encoder that receives multiple copies of an $n$-qubit quantum state and communicates classical information through a noisy classical channel, modeled for instance as a binary symmetric channel. The decoder aims to estimate a set of observables associated with the quantum state while satisfying prescribed accuracy and reliability constraints. Importantly, the encoder does not know in advance which observables will be requested by the decoder, which introduces additional challenges for the communication protocol.
To address this problem, the poster proposes communication schemes based on classical shadow tomography combined with unequal error protection (UEP) coding strategies. The approach separates the transmission of measurement bases and measurement outcomes, allowing different levels of protection depending on their importance for the reconstruction task. Experimental and theoretical evaluations compare several encoding schemes and demonstrate that the proposed STT-UEP protocol achieves improved bit efficiency compared to alternative methods.
The results show that the number of transmitted bits can scale logarithmically with the number of observables to be estimated while maintaining reliable reconstruction. This work highlights how semantic communication principles can significantly reduce the communication overhead for distributed quantum information processing and quantum sensing applications.

\vspace*{0.5cm}

\newpage
\subsection{Kl\"are Wienecke: "Find Pinsker-type Bounds for any Quantum Divergence using our simple 2-Step Method" }
\begin{floatingfigure}[r]{6cm}
\mbox{\includegraphics[width=5.5cm, height=9.78cm]{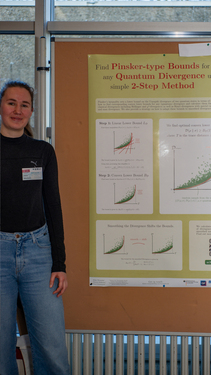}}
\caption{Kl\"are Wienecke}\label{Wienecke}
\end{floatingfigure}

Kläre Wienecke (Fig.~\ref{Wienecke}) presented new Pinsker-type lower bounds for a broad class of quantum divergences based on a simple and systematic two-step approach. The problem of quantifying the distance between two quantum states is inherently task-dependent, since the appropriate measure of distinguishability depends on the operational scenario under consideration. Consequently, quantum information theory provides a variety of divergences that quantify how much a state $\sigma$ deviates from a reference state $\rho$. Although there is no single universally accepted formal definition of a quantum divergence, it is commonly required that a proper divergence $\mathcal{D}(\rho \,\Vert\, \sigma)$ is non-negative and satisfies the data-processing inequality; in contrast, symmetry or the triangle inequality are generally not imposed.

Among the different distinguishability measures, the trace distance 
\[
T(\rho,\sigma)=\tfrac{1}{2}\|\rho-\sigma\|_1
\]
is often regarded as the canonical metric distance between quantum states. This naturally raises the question of how a given divergence $\mathcal{D}(\rho \,\Vert\, \sigma)$ relates quantitatively to the trace distance. A prominent example is Pinsker’s inequality, which lower bounds the quantum relative entropy in terms of $T$. However, this classical bound is tight only for small trace distances and becomes increasingly loose as $T$ approaches $1$.
The presentation demonstrated that optimal Pinsker-type lower bounds can be derived in a unified and systematic manner for a wide range of quantum divergences, including smoothed variants. The proposed two-step method therefore provides a general framework for relating operationally meaningful divergences to the trace distance and for obtaining sharp lower bounds beyond the classical Pinsker regime. Further details can be found in~\cite{wienecke2026collection}.
\vspace*{0.5cm}

\newpage
\subsection{Moritz Wiese: "Improved upper and lower bounds for the security performance of wiretap channels"}
\begin{floatingfigure}[r]{6cm}
\mbox{\includegraphics[width=5.5cm, height=9.78cm]{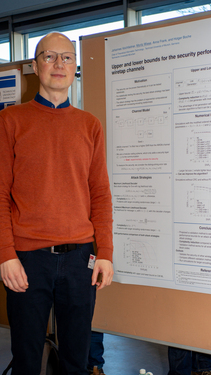}}
\caption{Moritz Wiese}\label{Moritz}
\end{floatingfigure}
Moritz Wiese (Fig.~\ref{Moritz}) presented improved upper and lower bounds for the security performance of wiretap channels. When assessing the security of a wiretap code, one usually has to rely on theoretical upper bounds and on the assumption that the underlying channel model is accurate. An overview of recent work in \cite{Voichtleitner2023a}, \cite{Voichtleitner2023}, \cite{Voichtleitner2026} is given, where such assumptions are unnecessary. This approach is based on the ``distinguishing security'' criterion, which is equivalent to semantic security and has an operational meaning in terms of certain eavesdropper attacks. In a nutshell, distinguishing security holds with respect to a given security threshold if the eavesdropper's success probability (i.e., the probability that the eavesdropper correctly distinguishes any two possible confidential messages known to it solely based on its channel observations) is below the given threshold.

While the optimal eavesdropper attack is an exponential-complexity maximum-likelihood attack, a lower-complexity method is introduced that provides upper and lower bounds on the eavesdropper's success probability. This method uses any list-generating method as an ingredient, e.g., the successive cancellation list decoder of polar codes or a modification of the universal ordered statistics decoder (OSD) of Fossorier and Lin~\cite{Fossorier1995}. Given such a list, the eavesdropper's decision as to which of the two possible messages $m_1$ or $m_2$ has been sent is a maximum-likelihood decision based only on the codewords contained in the list and pertaining to $m_1$ and $m_2$.

An advantage of such a statistical evaluation is that it can be applied to actual channel outputs without requiring a channel model. While the distinguishing security criterion does not require arbitrary eavesdropper attacks in the case where the eavesdropper knows that more than two messages are transmitted, it provides a means of testing the security against wiretap channels about which only limited or no knowledge is available. Experimental results from simulations and hardware implementations show that the presented method enables the assessment of the security of wiretap codes in realistic scenarios.

\vspace*{0.5cm}

\newpage
\subsection{Jyun-Sian Wu: "Asynchronous UMAC with and without latency constraint"}
\begin{floatingfigure}[r]{6cm}
\mbox{\includegraphics[width=5.5cm, height=9.78cm]{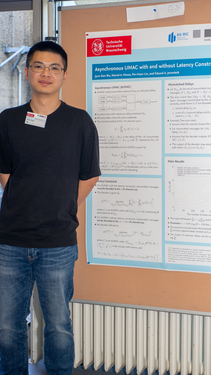}}
\caption{Jyun-Sian Wu}\label{Wu}
\end{floatingfigure}
Jyun-Sian Wu (Fig.~\ref{Wu}) investigated in this poster asynchronous unsourced multiple access (AUMAC) with and without latency constraints. In AUMAC, multiple active users transmit codewords selected from a common codebook, each subject to an individual delay bounded by a maximum delay constraint $\textsf{D}_m$~\cite{ISIT_worst_case}, \cite{ISIT2025_arXiv}. Unlike the synchronous UMAC model~\cite{Polyanskiy_perspective_on_UMAC}, asynchronicity introduces additional combinatorial ambiguity due to unknown relative shifts between codewords, which fundamentally impacts decoding performance and energy efficiency.

Two decoding regimes are considered. In the latency-constrained setting, all messages must be decoded at the $(n)$-th channel use, restricting the decoder to a truncated observation window. In contrast, without a latency constraint, decoding may be postponed until $(n + \textsf{D}_m)$, enabling circular-shift modeling and wrap decoding. The per-user probability of error (PUPE) is formalized, and error events arising from both message collisions and mismatched delay assignments are characterized. Particular attention is given to the impact of unknown delays, including cases where incorrectly decoded messages are paired with correct or permuted delay vectors.

Numerical results demonstrate the trade-off between energy-per-bit-to-noise ratio ($E_b/N_0$) and the number of active users. Asynchronicity reduces the user threshold below which ($E_b/N_0$) remains approximately constant. However, the performance degradation due to unknown delays can be largely mitigated with only a modest increase in ($E_b/N_0$). These findings clarify the fundamental limits of AUMAC under realistic timing uncertainty and provide insight into the cost of latency constraints in asynchronous unsourced access systems.
\vspace*{0.5cm}

\newpage
\subsection{Lorenzo Zaniboni: "Radar Parameter Estimation in Near-Field Integrated Sensing and Communication System"}
\begin{floatingfigure}[r]{6cm}
\mbox{\includegraphics[width=5.5cm, height=9.78cm]{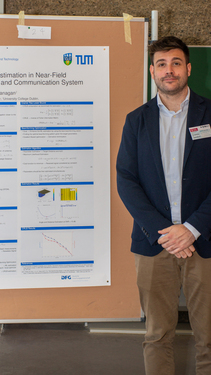}}
\caption{Lorenzo Zaniboni}\label{Lorenzo}
\end{floatingfigure}
Lorenzo Zaniboni (Fig.~\ref{Lorenzo}) presented the poster contribution ``Radar Parameter Estimation in Near-Field Integrated Sensing and Communication Systems,'' which investigates parameter estimation techniques for radar sensing in integrated sensing and communication (ISAC) systems operating in the near-field regime. 

From 1G to 5G wireless networks, the near-field (NF) was generally limited to a few meters, even centimeters, because of low-dimensional antenna arrays and low frequencies~\cite{liu2023nearfield}. With the advent of millimeter wave (mmWave) technologies, the NF region can extend to the order of hundreds of meters. As a result, NF has become increasingly relevant for future 6G communication systems and opens new research directions. At the same time, integrated sensing and communication (ISAC) is emerging as a promising technology for 6G systems, motivating the search for new solutions that incorporate both functionalities.

The work considers an alternative model design based on a continuous-time channel approach, which differs from the discrete models widely used in the literature~\cite{giovannetti2025asymptotic}, \cite{zhang2025fair}. This enables a more realistic channel representation by accounting for important parameters such as delay and Doppler shifts, which are often approximated or neglected in discrete models. Uniform Circular Arrays (UCA) are considered as a suitable array response model to address NF nonlinearities. In particular, such array structures are more compatible with the spherical wavefront behavior in NF systems than Uniform Linear Arrays (ULA).

The goal of the research is to provide improved insight into UCA geometry in NF systems and to develop an algorithm that jointly optimizes beamforming gain for communication and enables estimation of the angle and distance of a target for sensing. The results presented in the poster include the estimation accuracy of the proposed model as well as the Cramér–Rao lower bound (CRLB) used as a benchmark for estimation performance.
\vspace*{0.5cm}

\newpage
\subsection{Yaning Zhao: "Joint Identification and Sensing with Noisy Feedback"}
\begin{floatingfigure}[r]{6cm}
\mbox{\includegraphics[width=5.5cm, height=9.78cm]{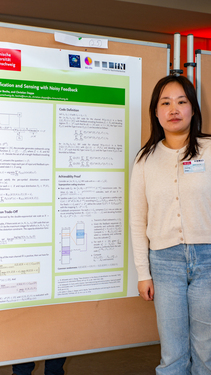}}
\caption{Yaning Zhao}\label{Yaning1}
\end{floatingfigure}
Yaning Zhao’s poster (Fig.~\ref{Yaning1}) addressed joint identification and sensing with noisy feedback and positioned the work within the broader paradigm of task-oriented communication, a key enabler of emerging 6G systems where the objective is to support decisions and actions rather than full message reconstruction. From an information-theoretic perspective, identification (ID) codes naturally capture this paradigm by allowing receivers to identify task-relevant messages without decoding entire codewords. Motivated both by the strong impact of feedback on ID performance and by the increasing interest in joint communication and sensing, the contribution studies joint identification and sensing (JIDAS) over state-dependent discrete memoryless channels with noisy and strictly causal feedback. In this setting, the transmitter simultaneously conveys identification messages and estimates the channel state, thereby tightly coupling communication and sensing functionalities. Upper and lower bounds on the capacity–distortion function are derived, providing a characterization of the fundamental performance limits of JIDAS under noisy feedback. The framework integrates task-oriented communication with state estimation through feedback and supports perception–communication–decision loops envisioned for AI-native 6G systems. As an illustrative example, in vehicle-to-vehicle communication it may be more critical to reliably identify control or warning messages while estimating inter-vehicle distance than to decode high-rate payload data, highlighting JIDAS as a promising approach for future 6G networks.

\vspace*{0.5cm}

\newpage
\subsection{Yaning Zhao: "Hands-On Molecular Communication Testbed for Undergraduate Education"}
\begin{floatingfigure}[r]{6cm}
\mbox{\includegraphics[width=5.5cm, height=9.78cm]{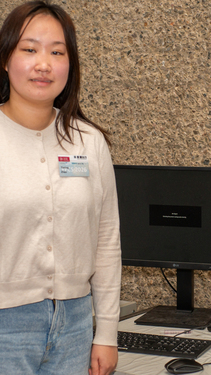}}
\caption{Yaning Zhao}\label{Yaning2}
\end{floatingfigure}
Yaning Zhao (Fig.~\ref{Yaning2}) presented the molecular communication (MC) testbed developed at the Institute for Communications Technology (IfN), TU Braunschweig, for the undergraduate Communication Engineering laboratory course. The primary objective of this experiment is to provide students with an intuitive, reproducible, and practically accessible introduction to key MC concepts through a low-cost fluidic setup. 
The system is based on a continuous background water flow into which three different dye colors are injected as information carriers. At the receiver side, a multi-wavelength photosensor detects the optical signals. To separate the overlapping spectral components and reliably identify the transmitted colors, a zero-forcing–based estimator is employed. This signal processing approach enables robust symbol detection despite channel mixing effects.
The laboratory experiment is designed to be completed independently by students within a single session and requires only foundational knowledge from introductory communication engineering courses. A comprehensive lab script guides participants step by step through channel characterization, spectral separation via pseudoinverse computation, color detection, and simple data transmission using on-off keying.
Pilot implementations have demonstrated that students are able to reproduce the complete communication chain—from transmitter control to signal processing and decoding—and achieve stable data rates of up to 0.5 bit/s over a 15 cm fluidic channel. The testbed thus shows that fundamental principles of molecular communication can be effectively taught using a compact, affordable, and experimentally transparent setup. The experiment will be permanently integrated into the undergraduate laboratory curriculum. The description of the full developement can be found in \cite{gaedeken2025handsonmolecularcommunicationtestbed}
\vspace*{0.5cm}

\newpage

\section{Workshop Talks}
Now we present the talks that were held during the conference.


\subsection{Eduard Jorswieck - Welcome}
\begin{floatingfigure}[r]{6cm}
\mbox{\includegraphics[width=5.5cm, height=9.78cm]{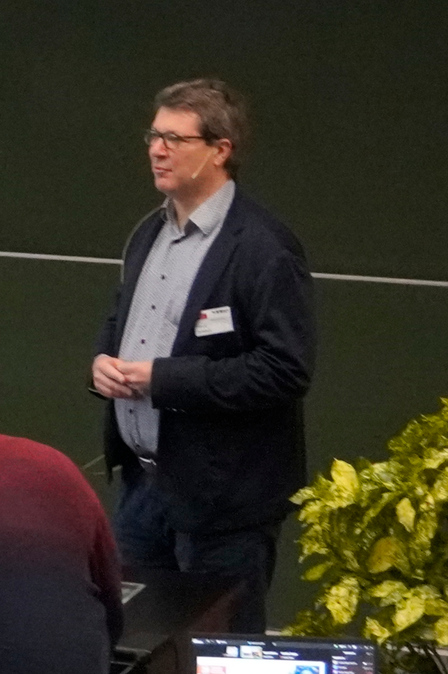}}
\caption{Eduard Jorswieck}\label{EJ}
\end{floatingfigure}
The Foundations of Future Communication Systems (FFCS) conference was officially opened by Eduard Jorswieck (Fig.~\ref{EJ}), who warmly welcomed all participants to Braunschweig and TU Braunschweig. He briefly introduced the city as a historic and innovative research hub in the heart of Germany and presented key facts about the university. TU Braunschweig is a member of the prestigious TU9 Alliance of leading technological universities in Germany, hosts approximately 2,300 researchers and 242 professors, and is structured into six faculties, with the Faculty of Electrical Engineering and Physics (Faculty 5) playing a central role in communication systems research. Currently serving as Dean of Faculty 5, Eduard introduced the Institute for Communications Technology (IFN), which comprises three professors, one research group leader, 52 doctoral researchers, and 2 postdoctoral researchers, and outlined its four research groups and their scientific focus areas. He explained that the idea for FFCS originated from Christian Deppe and that the conference is jointly organized by Christian Deppe, Boulat Bash, Holger Boche, Eduard Jorswieck, Uzi Pereg, and Shun Watanabe, with special thanks to Pin-Hsun Lin and Ines Richlick for their outstanding organizational support. The conference is generously supported by Siemens, MDPI, Springer, and 6G-life. Eduard concluded with an overview of the scientific program, spanning quantum communication, security, molecular communication, semantic communication, and resilient future networks, thereby emphasizing the interdisciplinary character of FFCS and setting the stage for an intensive and inspiring week.

\newpage
\subsection{Anton Trushechkin - Conference key and GHZ state distillation in quantum networks: Scenarios without quantum memory}
\begin{floatingfigure}[r]{6cm}
\mbox{\includegraphics[width=5.5cm, height=9.78cm]{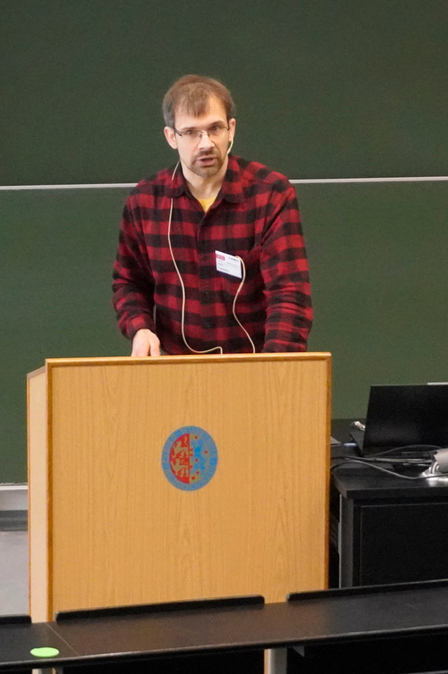}}
\caption{Anton Trushechkin}\label{Anton}
\end{floatingfigure}
Anton Trushechkin (Fig.~\ref{Anton}) presented recent progress on conference secret key agreement and GHZ state distillation in quantum networks, focusing on practically relevant scenarios where intermediate nodes cannot store quantum information. In this setting, participants are connected by bipartite entanglement sources that generate pairwise correlations, from which a multipartite conference key—i.e., a common secret string shared by all participants—must be distilled. The work considers protocols restricted to LOSR (local operations and shared randomness), meaning that quantum systems must be measured immediately and classical communication is allowed only afterwards during the processing of measurement outcomes.

Two main questions are addressed: deriving upper bounds on achievable conference key rates and analyzing the optimality of a simple protocol. This protocol measures each bipartite quantum state separately, extracts bipartite classical keys, and then generates a conference key using classical multipartite information-theoretic methods~\cite{Csiszar2004,PIN2010}. While general quantum protocols may employ collective measurements, the analysis shows that in certain cases the simple strategy is optimal. Upper bounds on the conference key rates are derived using Holevo’s theorem on accessible information, and the simple protocol is shown to achieve these bounds in specific scenarios.

Unlike previous bounds based on bipartitions of the network graph~\cite{Pirandola2020,Das2021}, the derived bounds rely on more general partitions and can reveal more complex structural bottlenecks. The results indicate that further progress will require genuinely multipartite quantum information tools beyond reductions to bipartite measures. Preliminary results have appeared in~\cite{Trushechkin2026}.

\newpage
\subsection{Daniel Kilper - Trends in Quantum Network Systems Research: Findings of the Optica Incubator}
\begin{floatingfigure}[r]{6cm}
\mbox{\includegraphics[width=5.5cm, height=9.78cm]{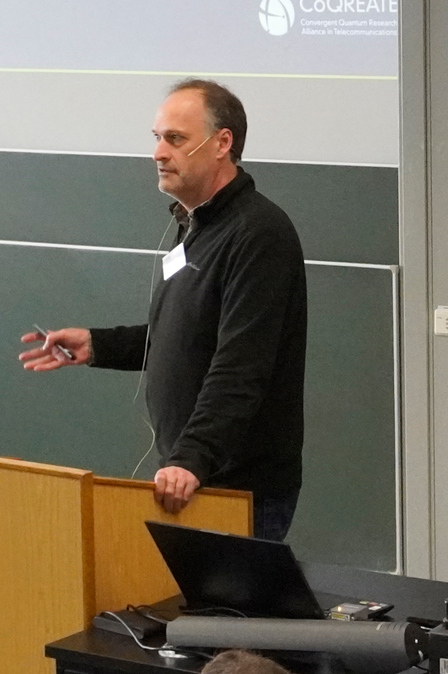}}
\caption{Daniel Kilper}\label{Kilper}
\end{floatingfigure}
Daniel Kilper (Fig.~\ref{Kilper}) presented an overview of current trends in quantum network systems research, focusing on the architectural and engineering challenges that arise when moving from point-to-point quantum communication experiments toward large-scale quantum networks~\cite{kilper2014optical}. While many theoretical results in quantum networking originate from physics and information theory, the development of practical network systems requires a complementary systems-engineering perspective.

The talk emphasized that future quantum networks will rely heavily on optical communication infrastructure. Optical fiber systems already form the backbone of classical communication networks and are expected to provide the physical transport layer for quantum communication technologies such as quantum key distribution and entanglement distribution. However, research on optical networking aspects of quantum networks remains relatively limited. One major obstacle is the absence of widely accepted system architectures that could guide engineering-oriented investigations and enable consistent performance evaluation.

To address this gap, the presented work proposes the development of early system architectures for quantum communication networks. Such architectures provide a framework for studying issues such as resource management, entanglement distribution, routing strategies, and network control mechanisms. Establishing architectural models also enables researchers to formulate meaningful engineering questions and evaluate trade-offs between different technological approaches.

The talk further highlighted the importance of defining common performance metrics and operational practices for quantum network systems. Achieving consensus on such metrics would help accelerate progress in both academic research and industrial development while supporting future standardization efforts. Overall, the presentation emphasized that bridging the gap between theoretical advances and practical network design will be essential for realizing scalable quantum communication infrastructures.

\newpage
\subsection{Janis N\"otzel - Simulations in Quantum Communication}
\begin{floatingfigure}[r]{6cm}
\mbox{\includegraphics[width=5.5cm, height=9.78cm]{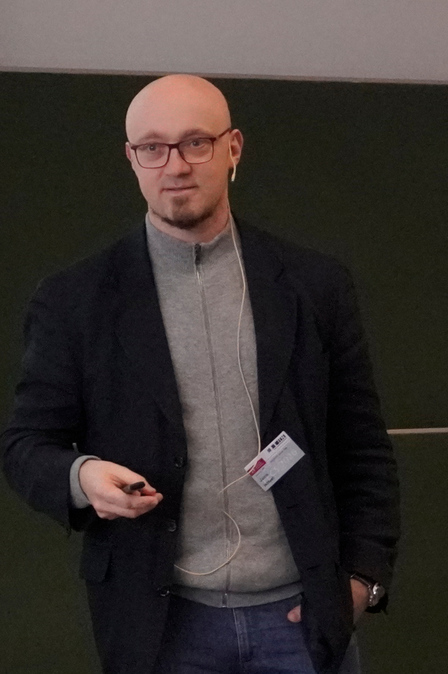}}
\caption{Janis N\"otzel}\label{Janis}
\end{floatingfigure}
Janis N\"otzel (Fig.~\ref{Janis}) first discussed a series of possible applications~\cite{leoSat,qkd,wiretap,jammedWiretap,futureNetworks,operating,jcas} with clear near-term benefits. Starting from plain data transmission, the concept of the joint detection receiver~\cite{saikat} was introduced and discussed. He then moved on to its use for sensing~\cite{infinite} and finally outlined aspects of a novel quantum optical bit commitment protocol~\cite{phaseBC}. 

Well-written software simplifies and accelerates specific aspects of collaboration. The design of software therefore separates tasks that benefit from such acceleration from those for which software offers no advantage. In turn, only software that provides a viable advantage over existing tools is adopted and used in the development of new technologies. In times where AI systems generate software at an unprecedented pace, these boundary conditions enforce a sharp focus on specific areas of quantum communication during software development. 

The common theme connecting the presented examples is the heterogeneity of teams required to properly realize such systems. After discussing these exemplary systems, he proceeded to outline a software system~\cite{qureed,photonWeave} that is suitable for bridging the knowledge gaps. A key aspect of the software is the introduction of clearly defined interfaces, which help to separate concerns. Interestingly, this design decision reflects the layering of information processing systems.

\newpage
\subsection{Ren\'e Schwonnek - Overview on Device-Independent QKD}
\begin{floatingfigure}[r]{6cm}
\mbox{\includegraphics[width=5.5cm, height=9.78cm]{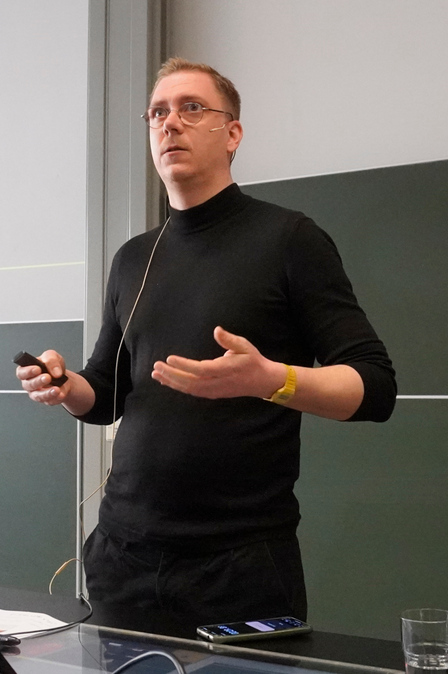}}
\caption{Ren\'e Schwonnek}\label{Schwonnek}
\end{floatingfigure}
Ren\'e Schwonnek (Fig.~\ref{Schwonnek}) gave an overview of device-independent quantum key distribution, describing security guarantees derived from observed statistics, the assumptions that can be removed compared to standard QKD, and the practical barriers and opportunities on the path toward implementations.
Device-independent quantum key distribution (DI-QKD) is widely regarded as the most stringent form of quantum cryptography, because its security is derived directly from observed nonlocal correlations rather than from a detailed characterization of the devices used in the implementation~\cite{Lim2013}. Its central promise is that a Bell violation can be converted into certified randomness and secrecy even under very weak assumptions on the internal functioning of the devices. At the same time, this strong security paradigm comes with substantial practical challenges, in particular with respect to loss, noise, and finite statistics~\cite{Lim2013,Schwonnek2021,TanWolf2024}.
An overview of the basic functioning of DI-QKD is provided, together with a discussion of the main mechanisms by which Bell-test data are converted into secret key. Particular emphasis is placed on the main bottlenecks for realistic implementations, including loopholes and finite-size effects~\cite{Lim2013,Schwonnek2021,TanWolf2024}.
In recent years, major advances have been achieved in both photonic and atom-based platforms, and several key ingredients required for DI-QKD have been demonstrated with steadily increasing experimental control~\cite{Nadlinger2022,Liu2022,Zhang2022,Lu2026}. While fully practical implementations remain challenging, these results have significantly sharpened the picture of what is already experimentally feasible~\cite{Nadlinger2022,Zhang2022,Lu2026}.
Recent developments in protocol design are also highly relevant. These include local-Bell-test architectures, random key-basis approaches, routed Bell-test protocols, and improved entropy-based security analyses~\cite{Lim2013,Schwonnek2021,Lobo2024,LeRoyDeloison2025,TanWolf2024}. Related network-based ideas for extending loophole-free nonlocal correlations over larger distances further strengthen the practical outlook for DI-QKD~\cite{Chaturvedi2024,Lobo2024}. Altogether, these developments suggest that DI-QKD should no longer be viewed only as a foundational possibility, but increasingly as a technology for which concrete paths toward practical implementations can be identified~\cite{Zhang2022,Lu2026,TanWolf2024,LeRoyDeloison2025}.

\newpage
\subsection{Ilja Gerhardt - From Molecules to Entanglement}
\begin{floatingfigure}[r]{6cm}
\mbox{\includegraphics[width=5.5cm, height=9.78cm]{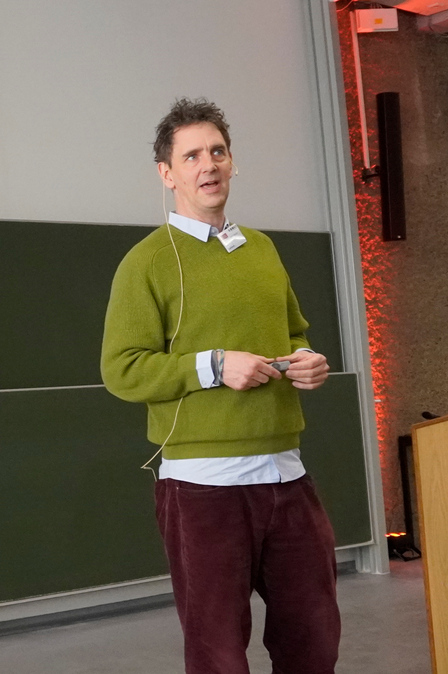}}
\caption{Ilja Gerhardt}\label{Ilya}
\end{floatingfigure}
Ilya Gerhardt (Fig.~\ref{Ilya}) presented the talk discussing how individual molecules can act as quantum optical emitters and building blocks for quantum communication experiments. Although molecules are typically associated with chemistry, the presentation showed that many of their optical properties closely resemble those of well-known quantum emitters such as atoms, ions, or quantum dots.
The talk began with an introduction to fluorescent molecules and their optical level structure. When excited by a laser, these systems emit photons through electronic transitions accompanied by vibrational states. At cryogenic temperatures, the vibrational contributions are largely frozen out, revealing narrow zero-phonon lines with linewidths limited by the excited-state lifetime. These transitions exhibit strong optical coherence and enable molecules to operate as bright and spectrally narrow single-photon sources with emission rates reaching millions of detected photons per second.
Gerhardt then discussed how such molecular emitters can be used in quantum optical experiments. Individual molecules can both emit and absorb single photons with a large effective optical cross section, enabling photon–molecule interactions at the single-emitter level. Experiments have demonstrated photon-mediated communication between distant molecules, Hong–Ou–Mandel interference of photons emitted by independent molecules, and the generation of entangled photonic states through interference and measurement.
A key challenge highlighted in the talk is that many fluorescent molecules lack internal degrees of freedom, such as spin states, that would allow the emitters themselves to become entangled after photon interference. Recent work therefore explores engineered molecular systems that provide additional controllable quantum states. Examples include strongly coupled molecular pairs with dipole–dipole interactions as well as molecular systems with intrinsic spin degrees of freedom, such as carbene-based molecules exhibiting triplet ground states.
Overall, the presentation illustrated how single molecules can serve as versatile quantum optical systems that combine the advantages of atomic-like optical coherence with the scalability of chemical synthesis. These developments open new perspectives for molecular quantum optics.

\newpage
\subsection{Konrad Banaszek - Quantum Noise in Optical Communication Systems: How to Stop Worrying and Start Lovin' it}
\begin{floatingfigure}[r]{6cm}
\mbox{\includegraphics[width=5.5cm, height=9.78cm]{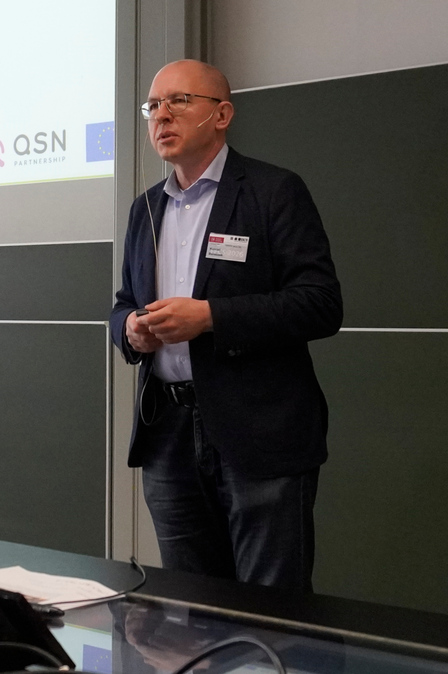}}
\caption{Konrad Banaszek}\label{Konrad}
\end{floatingfigure}
Konrad Banaszek (Fig.~\ref{Konrad}) presented a talk on quantum noise in optical communication systems, discussing the fundamental role of quantum noise and how it shapes the ultimate performance limits of optical receivers. Optical communication systems operate with electromagnetic fields whose fluctuations are governed by quantum mechanics. As optical powers decrease and detection approaches the few-photon regime, these fluctuations become a dominant factor that cannot be eliminated but must instead be properly modeled and exploited.

The talk explained how quantum noise arises from the fundamental properties of light and detection processes, leading to shot noise and other quantum-limited effects that constrain the achievable signal-to-noise ratio. Rather than viewing quantum noise solely as an obstacle, the presentation emphasized that it provides a well-defined theoretical framework for analyzing communication limits and designing optimal receivers. In particular, the quantum description of optical signals allows the derivation of fundamental capacity bounds and clarifies the performance gap between conventional receivers and theoretically optimal quantum measurements.

Several receiver architectures were discussed from this perspective, including coherent detection techniques and more advanced quantum measurement strategies. These approaches illustrate how physical noise models connect directly to communication-theoretic quantities such as error probability and achievable information rates. The presentation highlighted that understanding quantum noise is essential for future high-sensitivity optical communication systems, including deep-space links and quantum communication networks.

Overall, the talk emphasized that quantum noise should not merely be regarded as a limitation but as a fundamental design principle that guides the development of optimal optical communication technologies and receiver architectures.
\newpage

\subsection{Frank Fitzek - 6G-life\textsuperscript{2} - Digital transformation and sovereignty of future communication networks for networked robotics}
\begin{floatingfigure}[r]{6cm}
\mbox{\includegraphics[width=5.5cm, height=9.78cm]{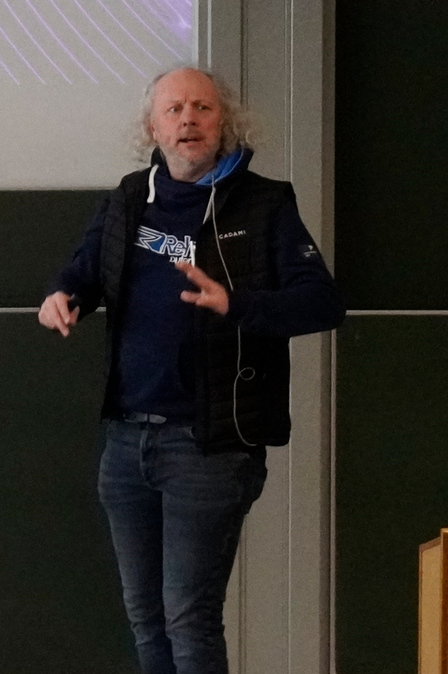}}
\caption{Frank Fitzek}\label{Frank}
\end{floatingfigure}
Frank Fitzek (Fig.~\ref{Frank}) delivered a presentation on the evolution of the 6G-life initiative toward its successor program, 6G-life$^{2}$. In his talk, he outlined that whereas 6G-life primarily focused on foundational research and architectural concepts for next-generation communication systems, 6G-life$^{2}$ places stronger emphasis on structured technology transfer and implementation pathways. A comprehensive summary of the scientific and strategic results of the first phase of 6G-life is documented in the recently published volume~\cite{Fitzek2026}.

A central component of this evolution is the establishment of a “Research to Transfer Hub,” designed to translate scientific results into demonstrators, pilot deployments, and operational testbeds. Particular attention was given to networked robotics as an application domain with stringent requirements in terms of latency, reliability, and security.

The presentation further addressed the notion of digital sovereignty in the context of future communication networks. These networks were characterized as strategic infrastructures whose development requires competencies across the entire innovation chain—from theoretical foundations to deployment and operation. In this framework, the strengthening of European technological capabilities was highlighted as a long-term objective.

In addition, the interplay between artificial intelligence and communication technologies was discussed. A conceptual distinction was made between “Communication for AI,” where advanced network infrastructures enable distributed and collaborative AI systems, and “AI for Communication,” where AI methods are employed to optimize and autonomously manage communication networks. The talk concluded with remarks on the importance of a coherent strategic positioning of European research and industry in the global landscape.

\newpage
\subsection{Yanling Chen - Non-Adaptive Coding for Two-Way Wiretap Channel With or Without Cost Constraints}
\begin{floatingfigure}[r]{6cm}
\mbox{\includegraphics[width=5.5cm, height=9.78cm]{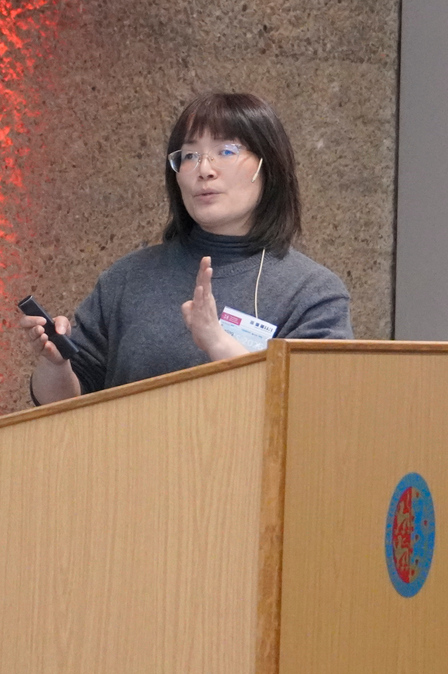}}
\caption{Yanling Chen}\label{Yanling}
\end{floatingfigure}
Yanling Chen (Fig.~\ref{Yanling}) presented results on non-adaptive coding strategies for the two-way wiretap channel (TW-WC) with an external eavesdropper, based on joint work with Prof.~Masahito Hayashi~\cite{HC2024,CH2025}. The talk addressed reliable and secure communication over the TW-WC and discussed different encoder–decoder configurations, both with and without cost constraints, as well as various criteria used to measure reliability and secrecy.

Focusing on a simple non-adaptive coding approach, it was shown that there exist coding schemes for which both the information leakage to the eavesdropper and the average decoding error probability decay exponentially with the number of channel uses. This leads to achievable secrecy rate regions for the TW-WC under a strong joint secrecy constraint. To characterize secrecy and reliability exponents in a compact form, several versions of R\'enyi mutual information were employed. In particular, the R\'enyi mutual information introduced in~\cite{Renyi1961} was used to characterize the secrecy exponent for the TW-WC without cost constraint, while Sibson’s R\'enyi mutual information and its conditional version~\cite{Sibson1969,Csiszar1995,TH2018} were used to describe the error exponent in the same scenario. For channels with cost constraints, another variant of R\'enyi mutual information~\cite{CG2022} was used to characterize both secrecy and error exponents.

The employed techniques build on channel resolvability arguments for multi-user multiple-access channels~\cite{HC2019} and extend classical results such as Gallager’s error exponent~\cite{Gallager1968} for point-to-point communication. Together with random coding arguments based on constant-composition codes for channels with cost constraints, these tools provide a systematic framework for analyzing secrecy and reliability trade-offs in multi-user communication systems.

\newpage
\subsection{Ugo Vaccaro - An Overview of Superimposed Codes and Their Role in Cryptography and Communication}
\begin{floatingfigure}[r]{6cm}
\mbox{\includegraphics[width=5.5cm, height=9.78cm]{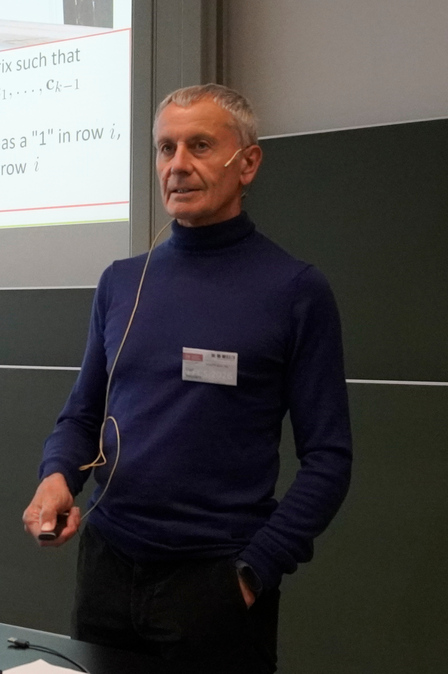}}
\caption{Ugo Vaccaro}\label{Ugo}
\end{floatingfigure}
Ugo Vaccaro (Fig.~\ref{Ugo}) presented recent results on Superimposed Codes (SCs) and their applications to synchronization and security problems in communication systems. A $(t,n,k)$-superimposed code can be represented as a $t\times n$ binary matrix with the property that, for every column and every set of $k-1$ other columns, there exists a row in which the selected column has a $1$ while the other $k-1$ columns have $0$. 

In communication settings such as multiple-access channels (MACs), this matrix interpretation provides a deterministic transmission schedule. Assigning each of the $n$ users a column guarantees that if up to $k$ users transmit simultaneously, each active user has at least one dedicated collision-free time slot among the $t$ available slots. A central challenge in the theory of superimposed codes is therefore the minimization of the code length $t$. Currently, there remains a gap between the best known upper bound $O(k^2\log n)$ and the lower bound $\Omega\!\left(\frac{k^2}{\log k}\log n\right)$.

To address more realistic communication scenarios, the talk also discussed several variants of superimposed codes. In particular, \emph{Robust Shift-Invariant Superimposed Codes (RSISC)}~\cite{BRV} remove the assumption of global synchronization, allowing users to start transmission at arbitrary times while tolerating relative misalignment and noise. Furthermore, \emph{frameproof codes}~\cite{DalaiFRV25} were presented as $q$-ary generalizations of SCs with applications in digital fingerprinting, where they prevent coalitions of up to $k$ users from forging the fingerprint of an innocent user. For this setting, a randomized construction based on the Moser--Tardos algorithmic version of the Lov\'asz Local Lemma was described, producing such codes in expected $O(tn^2)$ time.

Overall, the presentation highlighted that superimposed codes and their variants provide versatile combinatorial tools for decentralized communication and cryptographic applications. At the same time, determining optimally short constructions remains an important open problem, with current research aiming to close the asymptotic gap between known upper and lower bounds.

\newpage
\subsection{Hannes Bartz - Security of Code-Based PIR Schemes}
\begin{floatingfigure}[r]{6cm}
\mbox{\includegraphics[width=5.5cm, height=9.78cm]{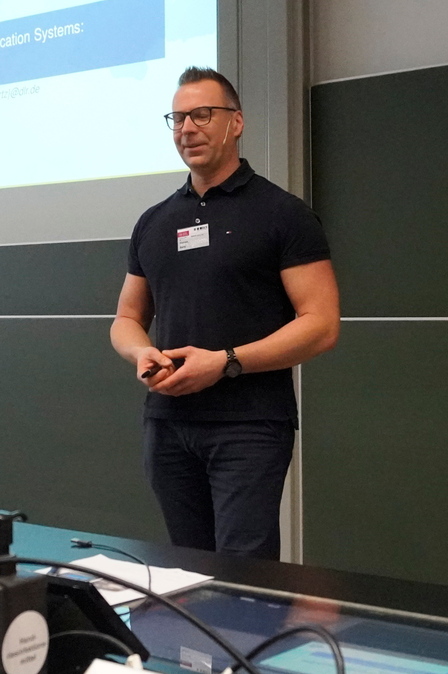}}
\caption{Hannes Bartz}\label{Hannes}
\end{floatingfigure}
Hannes Bartz (Fig.~\ref{Hannes}) presented recent work on the security of code-based private information retrieval (PIR) schemes, based on joint work with Svenja Lage~\cite{LageBartz2026}. Private information retrieval allows a user to obtain a file from a database without revealing to the server which file is requested. While information-theoretic PIR requires multiple non-colluding servers, single-server solutions typically rely on computational assumptions, motivating the study of post-quantum approaches such as code-based cryptography.

The talk discussed how linear code-based encryption schemes can be used to construct PIR protocols exploiting their homomorphic properties. In a generic construction, the client generates encrypted query components that indicate the desired file index, while the server computes a linear combination of the database entries with the encrypted query and returns the result, which the client can decrypt to recover the requested file. However, implementing this approach with code-based cryptography introduces challenges, in particular when many ciphertext additions are required.

One proposed technique is to restrict the support of the error vectors outside an information set of the underlying code. Although this enables efficient ciphertext aggregation, it introduces additional algebraic structure that weakens the security guarantees. The presentation illustrated how such structural properties can be exploited by attacks beyond classical information-set decoding, including linear-algebraic attacks that recover the queried file index.

A concrete example based on previously proposed code-based PIR constructions was analyzed, demonstrating how so-called subquery attacks can identify the requested index by exploiting rank differences in the query matrices. Even revised schemes designed to mitigate these attacks remain vulnerable to similar techniques. The talk concluded with several lessons learned: decoding hardness alone does not guarantee query privacy, linear structures tend to leak information, and restricting the error support can introduce exploitable weaknesses. These findings indicate that future secure code-based PIR constructions may require fundamentally different design approaches.

\newpage
\subsection{Matteo Nerini - Analog Computing for Future Communication Systems}
\begin{floatingfigure}[r]{6cm}
\mbox{\includegraphics[width=5.5cm, height=9.78cm]{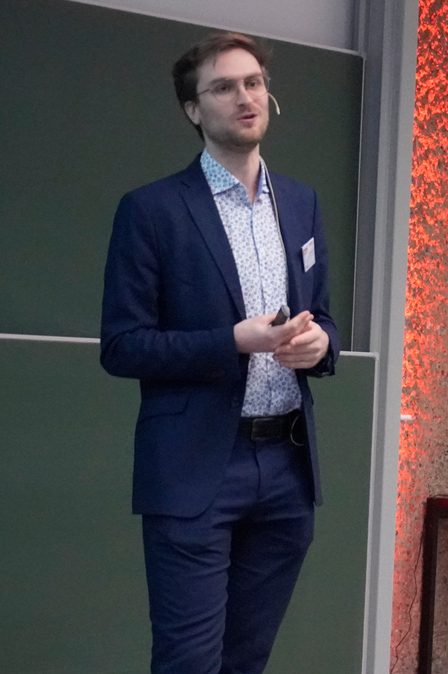}}
\caption{Matteo Nerini}\label{Matteo}
\end{floatingfigure}
Matteo Nerini (Fig.~\ref{Matteo}) presented analog computing, which has recently been revived due to its potential for energy-efficient and highly parallel computations. In this talk, analog computers that linearly process microwave signals, referred to as microwave linear analog computers (MiLACs), and their applications in signal processing and communications were discussed~\cite{ner1,ner2}.

In the first part, a MiLAC is modeled as a multiport microwave network with tunable impedance components, enabling the execution of mathematical operations by reconfiguring the microwave network and applying input signals at its ports~\cite{ner1}. It is demonstrated that a MiLAC can efficiently compute the linear minimum mean square error (LMMSE) estimator and perform matrix inversion with remarkably low computational complexity. Specifically, a matrix can be inverted with complexity growing quadratically with its size.

In the second part, the applications of MiLACs in wireless communications are investigated, highlighting their potential to enable future wireless systems by executing computations and beamforming in the analog domain~\cite{ner2}. MiLAC-aided beamforming enables the flexibility and performance of digital beamforming while significantly reducing hardware costs by minimizing the number of radio-frequency (RF) chains and relying only on low-resolution analog-to-digital converters (ADCs) and digital-to-analog converters (DACs). In addition, it eliminates per-symbol operations by avoiding digital-domain processing and substantially reduces the computational complexity of zero-forcing (ZF), which scales quadratically with the number of antennas instead of cubically.

\newpage
\subsection{Prakash Narayan - Multiterminal Secrecy}
\begin{floatingfigure}[r]{6cm}
\mbox{\includegraphics[width=5.5cm, height=8.78cm]{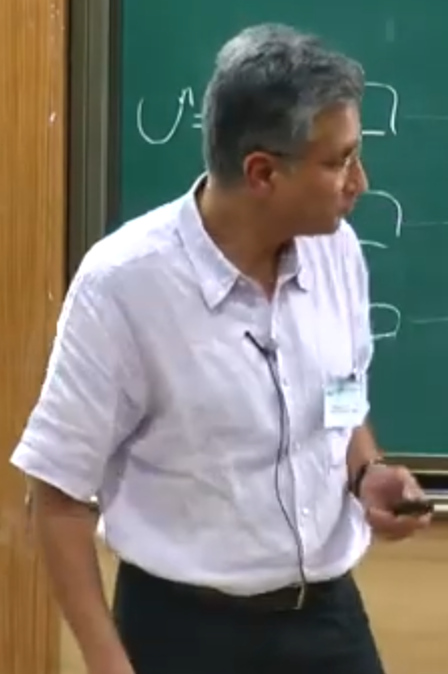}}
\caption{Prakash Narayan}\label{Prakash}
\end{floatingfigure}

Prakash Narayan (Fig.~\ref{Prakash}) presented results on multiterminal secrecy, based on collaborations with Imre Csisz'ar and Sagnik Bhattacharya \cite{BhattacharyaNarayan2024SharedInformationTree}, \cite{CsiszarNarayan2004SecrecyMultipleTerminals}, \cite{CsiszarNarayan2008SecrecyMultiterminal}, \cite{CsiszarNarayan2013SecrecyGenerationMAC}, \cite{NarayanTyagi2016MultiterminalSecrecy}.

He first considered Shannon-theoretic secret key generation (SKG) for a multiterminal source model with arbitrarily many terminals, each observing one component of a discrete memoryless multiple source, with unrestricted interactive public communication. The terminals cooperate to generate a secret key for later encrypted communication, while secrecy is required against an eavesdropper observing only the public discussion. The secrecy-capacity characterization reveals close links to multiterminal source coding without secrecy constraints.

Second, SKG by several parties was studied for a multiterminal channel model with one input terminal, multiple output terminals, a secure noisy channel, and an unlimited-capacity public noiseless channel. The key is generated jointly by the input and output terminals through public communication. A single-letter secrecy-capacity characterization is obtained against an eavesdropper observing only the public communication. This capacity is achieved by source emulation with noninteractive public communication, where the input terminal sends no public message and each output terminal sends at most one deterministic public message.

Third, SKG by several parties was studied for multiterminal channel models with multiple input and output terminals connected by a secure noisy channel and assisted by an unlimited-capacity public noiseless channel. The terminals generate the key with the help of public communication. Single-letter lower and upper bounds on secrecy capacity are obtained; these coincide in special cases, but not in general. Links are also established between secrecy capacity and the transmission-capacity region of the corresponding multiple access channel without secrecy constraints.

Common randomness plays a central role. Shared information and conditional shared information arise as measures of mutual and conditionally mutual dependence among multiple random variables, with important operational meanings.

This work was supported by the U.S. National Science Foundation.

\newpage


\subsection{Falko Dressler - Molecular Communication in Different Fluids}
\begin{floatingfigure}[r]{6cm}
\mbox{\includegraphics[width=5.5cm, height=9.78cm]{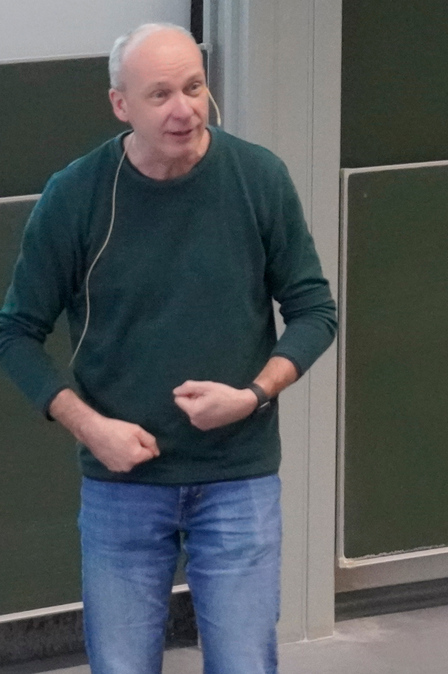}}
\caption{Falko Dressler}\label{Falko}
\end{floatingfigure}
Falko Dressler (Fig.~\ref{Falko}) presented the talk \emph{“Molecular Communication in Different Fluids”}, discussing how the physical properties of the transmission medium affect molecular communication systems. The presentation was based on recent experimental work evaluating molecular communication channels in realistic biological environments, particularly blood \cite{Debus2025}.

Molecular communication has been proposed as a key technology for the Internet of Bio-Nano Things, where nanoscale devices exchange information through the human circulatory system. While many theoretical studies assume simple fluids such as water, real in-body environments involve blood, a complex non-Newtonian fluid whose flow properties significantly influence signal propagation.

The talk presented experimental results obtained from a molecular communication testbed using superparamagnetic iron oxide nanoparticles (SPIONs) as signaling particles. Measurements were performed for water, a blood substitute, and real porcine blood, enabling a systematic comparison of the resulting channel impulse responses.

The results show that viscosity, flow velocity, and injection geometry strongly affect signal propagation. In particular, blood introduces additional channel effects that are not observed in simpler fluids. To capture these phenomena, the authors extended an existing channel model to account for non-Newtonian flow behavior, enabling more accurate modeling of molecular communication in biological environments.
\newpage
\subsection{Massimiliano Pierobon - Molecular Communication for the Living: Subjective Information and Emergent Communication Paradigms}
\begin{floatingfigure}[r]{6cm}
\mbox{\includegraphics[width=5.5cm, height=9.78cm]{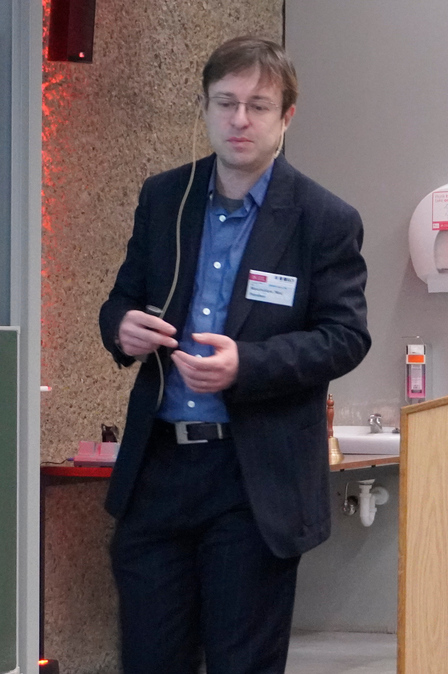}}
\caption{Massimiliano Pierobon} \label{Massimiliano}
\end{floatingfigure}
Massimiliano Pierobon (Fig.~\ref{Massimiliano}) presented a talk exploring how information-theoretic concepts can be extended to better describe communication processes in living systems. Molecular communication (MC) studies how information is encoded, transmitted, and decoded using molecules as physical carriers. In contrast to classical electromagnetic communication, molecules are discrete physical entities whose propagation is governed by stochastic processes such as Brownian motion and diffusion, which fundamentally shape the resulting communication channels.

The talk reviewed the foundations of MC theory, emphasizing that individual molecules can be viewed as atomic information units whose transport and interaction processes determine the behavior of biological communication systems. Building on this perspective, the presentation introduced the concept of \emph{subjective information}, which evaluates information not only by statistical correlation but by its usefulness for an agent interacting with its environment. In biological contexts, information is meaningful only insofar as it influences the internal state and decision-making processes of living organisms.

Using models of interacting agents and simulated biological systems, the framework quantifies how environmental signals contribute to the survival and adaptive behavior of an organism. This viewpoint connects molecular communication with broader questions of biological signaling and the emergence of communication protocols in living systems. In particular, communication mechanisms may emerge as evolutionary strategies that allow organisms to acquire and process information relevant for maintaining viability.

Overall, the talk highlighted how integrating molecular communication theory with concepts from biology, statistical physics, and information theory may lead to new communication paradigms where performance is evaluated not only in terms of transmission efficiency but also in terms of the functional role that information plays in enabling adaptive behavior in living systems~\cite{Barker2022}.

\newpage
\subsection{Liubov Bakhchova - Molecular Miscommunication in Biological Signaling}
\begin{floatingfigure}[r]{6cm}
\mbox{\includegraphics[width=5.5cm, height=9.78cm]{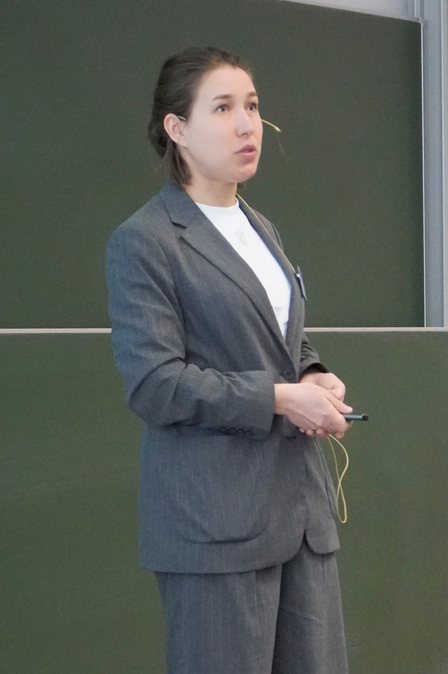}}
\caption{Liubov Bakhchova} \label{Liubov}
\end{floatingfigure}
Liubov Bakhchova (Fig.~\ref{Liubov}) presented a talk related to the characterization of organic mixed ionic–electronic conductors (OMIECs) and their role in bioelectronic devices. The presentation focused on experimental methods for understanding the internal electrochemical and structural processes occurring in organic electrochemical transistors (OECTs), which are promising components for next-generation bioelectronics, neuromorphic systems, and flexible sensors \cite{Mukhin2026}.

OECTs operate through the coupling of ionic and electronic transport processes. When a voltage is applied, ions from the electrolyte penetrate the organic semiconductor channel and modulate its electronic conductivity through electrochemical doping and dedoping. Understanding the microscopic processes governing this interaction—such as ion injection, water uptake, and morphological changes of the polymer film—is essential for improving device performance, stability, and reproducibility.

The talk highlighted the use of electrochemical quartz crystal microbalance with dissipation monitoring (EQCM-D) as a powerful in-situ characterization technique. EQCM-D allows the real-time measurement of mass changes at the device surface together with information about the viscoelastic properties of the material. By monitoring shifts in resonance frequency and dissipation of a quartz sensor, the method provides insight into ion transport, swelling effects, and structural dynamics of OMIEC layers during operation.

These measurements enable researchers to link electrochemical activity with physical changes in the material and thereby better understand the mechanisms governing OECT behavior. Such insights are crucial for designing more reliable bioelectronic interfaces and improving the performance of devices used in biosensing, neural interfaces, and soft electronics. Overall, the presentation illustrated how advanced experimental techniques can bridge the gap between materials science, electrochemistry, and communication-inspired bioelectronic systems.

\newpage
\subsection{Jessica Bariffi - Coding Theory for DNA Storage}
\begin{floatingfigure}[r]{6cm}
\mbox{\includegraphics[width=5.5cm, height=9.78cm]{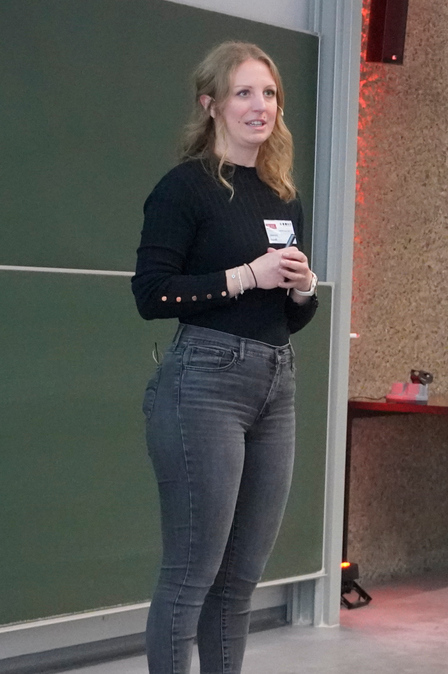}}
\caption{Jessica Bariffi} \label{Jessica}
\end{floatingfigure}

Jessica Bariffi (Fig.~\ref{Jessica}) presented an introduction to the coding-theoretic foundations of DNA-based data storage, focusing on challenges arising from the underlying channel model. DNA has emerged as a promising storage medium due to its extremely high data density and long-term durability compared to conventional magnetic or optical storage technologies. After outlining the DNA storage pipeline—including binary-to-quaternary encoding, strand synthesis, PCR amplification, and sequencing—the talk focused on the resulting error models.

Unlike classical communication channels dominated by substitution errors, DNA synthesis and sequencing introduce synchronization errors such as insertions and deletions. These errors disrupt positional alignment and require alternative metrics such as the Levenshtein distance, making insertion–deletion correcting codes central to reliable DNA storage systems. A key example is the family of Varshamov–Tenengolts (VT) codes~\cite{VTCodes1965}, which can correct a single insertion or deletion with asymptotically optimal redundancy of order $\log(n)$. Extending such constructions to multiple deletions remains a major open problem, as current explicit codes do not yet match known existential bounds.

The talk also discussed the retrieval problem, where stored DNA strands are reconstructed from multiple noisy copies. Modern approaches combine classical coding techniques with clustering and sequence reconstruction algorithms to exploit redundancy more efficiently. Overall, DNA storage represents not only a biochemical innovation but also a rich coding-theoretic challenge involving synchronization errors, constrained strand lengths, and scalable decoding methods.

\newpage
\subsection{Martin Korte - Microlia signalling: the good, the bad and the ugly side of neuroinflammation}
\begin{floatingfigure}[r]{6cm}
\mbox{\includegraphics[width=5.5cm, height=9.78cm]{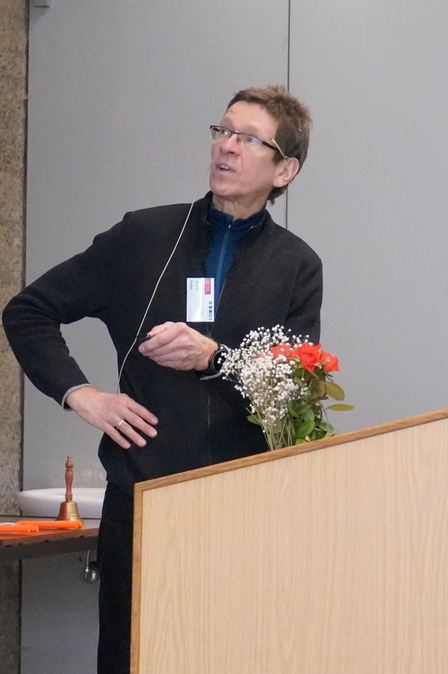}}
\caption{Martin Korte} \label{Martin}
\end{floatingfigure}
Martin Korte (Fig.~\ref{Martin}) presented a talk discussing how molecular signalling processes in the brain regulate neural function, protection, and disease. The presentation highlighted the central role of microglia—the resident immune cells of the central nervous system—in sensing and processing molecular signals within neural tissue.

In the brain, communication does not occur solely through electrical synapses between neurons but also through a complex network of molecular signals, including ATP, cytokines, chemokines, metabolites, and damage-associated molecular patterns. Microglia detect and interpret these signals and adjust their behavior accordingly. Through this process, they contribute to synaptic pruning, immune surveillance, tissue repair, and the maintenance of neural homeostasis.

The talk emphasized the dual role of microglial signalling. Under healthy conditions, microglia support neural plasticity and protect neural circuits from damage. However, dysregulated signalling pathways can lead to chronic neuroinflammation, which is associated with neurodegenerative diseases and cognitive decline. Understanding these mechanisms is therefore essential for explaining how molecular communication in the brain can shift from beneficial regulation to harmful pathological processes.

Overall, the presentation illustrated how neuroimmune signalling provides a natural example of highly complex molecular communication systems. Studying these biological communication processes may offer valuable insights for the development of theoretical models and engineered molecular communication systems that operate in similarly complex biochemical environments.

\newpage


\subsection{Erik G. Larsson - Decentralized learning over unreliable communication networks}
\begin{floatingfigure}[r]{6cm}
\mbox{\includegraphics[width=5.5cm, height=9.78cm]{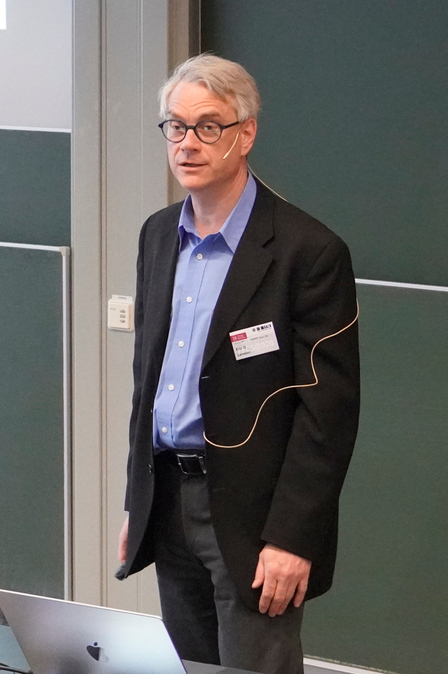}}
\caption{Erik G. Larsson}\label{Erik}
\end{floatingfigure}
Erik G. Larsson (Fig.~\ref{Erik}) discussed the decentralized gradient descent (DGD) algorithm and its sibling diffusion, which are workhorses in decentralized machine learning, distributed inference and estimation, and multi-agent coordination. He presented a novel, principled framework for the analysis of DGD and diffusion for strongly convex, smooth objectives and arbitrary undirected topologies, using contraction mappings coupled with a result known as the mean Hessian theorem (MHT). The use of these tools yields tight convergence bounds, both in the noise-free and noisy regimes. While these bounds are qualitatively similar to results found in the literature, the approach based on contractions together with the MHT decouples the algorithm dynamics (how quickly the algorithm converges to its fixed point) from its asymptotic convergence properties (how far the fixed point is from the global optimum). This leads to a simple and intuitive analysis that is accessible to a broader audience.

The resulting analyses extend directly to multiple local gradient updates, time-varying step sizes, sampling noise in the gradients, noisy communication links, and random topologies. In the noisy case, the approach avoids assumptions on bounded gradient noise variance. When applied to random topologies, it is sufficient that the mixing matrix is symmetric \emph{in expectation}.

These analyses unify, considerably simplify, and extend results in the literature that were obtained using more complicated proof techniques. This makes the framework an attractive way of describing, using, and teaching decentralized convex optimization. Many of the presented results can be used in a plug-and-play fashion by practitioners who want to establish convergence of their algorithms without reproducing intricate technical arguments~\cite{larsson2025unified}.

\newpage
\subsection{Bho Matthiesen - Resilient Radio Access Network: AI and the Unknown Unknowns}
\begin{floatingfigure}[r]{6cm}
\mbox{\includegraphics[width=5.5cm, height=9.78cm]{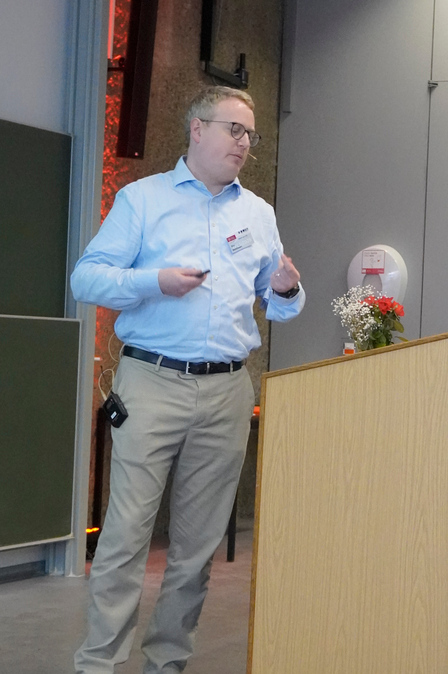}}
\caption{Bho Matthiesen} \label{Bhoma}
\end{floatingfigure}
Bho Matthiesen (Fig.~\ref{Bhoma}) presented a talk addressing the challenge of designing wireless networks that remain reliable under highly dynamic and unpredictable operating conditions. Future radio access networks (RANs) will increasingly rely on artificial intelligence for tasks such as resource allocation, interference management, and network control. However, AI-driven systems face a fundamental limitation: models are typically trained on known data distributions and may fail when confronted with previously unseen scenarios, often referred to as “unknown unknowns.”

The presentation discussed how such uncertainties arise in practical wireless systems due to changing environments, user behavior, hardware impairments, and adversarial conditions. When AI-based controllers encounter situations outside their training domain, their decisions may become unreliable or even harmful for network operation. Ensuring resilience therefore requires mechanisms that allow the network to detect, interpret, and safely react to such unexpected events.

To address these challenges, the talk highlighted approaches that combine machine learning with robust system design principles. In particular, methods for uncertainty detection, anomaly identification, and fallback strategies were discussed as ways to maintain reliable network performance when AI predictions become uncertain. These techniques enable networks to recognize when their learned models are no longer trustworthy and to adapt their behavior accordingly.

Overall, the presentation emphasized that resilient next-generation wireless systems must integrate AI with robust control and monitoring mechanisms. By explicitly accounting for unknown unknowns, future radio access networks can maintain reliable operation even in highly complex and evolving communication environments~\cite{Matthiesen2025}.

\newpage

\subsection{Hossein Ahmadi - Semantic Overt--Covert Splitting with Post-Subtraction KL Guarantees and Ambiguity Constraints}
\begin{floatingfigure}[r]{6cm}
\mbox{\includegraphics[width=5.5cm, height=9.78cm]{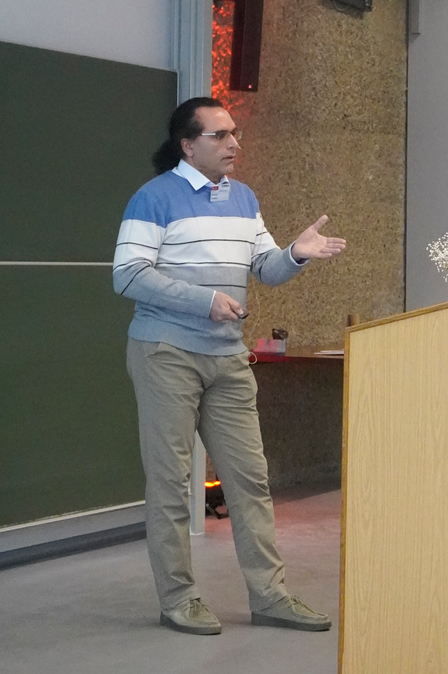}}
\caption{Hossein Ahmadi}\label{Hossein}
\end{floatingfigure}
Hossein Ahmadi (Fig.~\ref{Hossein}) proposed a semantic perspective on overt–covert splitting in which the overt layer is interpreted not merely as “public traffic,” but as a deliberate semantic disclosure interface that reveals only coarse meaning, while the covert layer provides a minimal refinement necessary to complete the intended task output. Building on the task-oriented/semantic communication framework \cite{Gunduz_JSAC23}, semantic content is modeled as a pair consisting of (i) a content/meaning class (i.e., what the message is about) and (ii) an intent/action label (i.e., what should be done). The receiver’s objective is to reliably recover this semantic pair, while an adversary observing the overt layer remains uncertain about the underlying meaning.
The core semantic principle follows a coarse-to-fine representation: the overt layer conveys a coarse semantic index (e.g., a cluster identifier), whereas the covert layer transmits a compact refinement that disambiguates the intended semantic content within that cluster. To formalize the notion of “semantic privacy” in the overt layer, an ambiguity requirement is introduced, ensuring that multiple semantic meanings remain plausible to an adversary even after observing the overt content. This ambiguity constraint is imposed at the semantic level—limiting inference about meaning—and is explicitly decoupled from physical-layer detectability.
The refinement is carried by a covert micro-payload that must remain difficult to detect by a warden. This leads to a two-constraint design: (i) an ambiguity constraint governing semantic leakage through the overt channel, and (ii) a detectability constraint limiting the warden’s ability to detect the covert refinement, typically benchmarked against the square-root-law regime for covert communication \cite{Bash2013LPDAWGN}. The resulting design parameters include the granularity of the overt semantic partition, the amount of refinement required to achieve the desired task accuracy, and the strictness of the ambiguity and detectability constraints. Overall, semantic splitting introduces an additional, actionable design layer on top of covert communication mechanisms: the overt layer can be tuned to remain semantically “safe,” while a minimal covert refinement suffices to enable reliable task recovery at the receiver.

\newpage
\subsection{Armin Dekorsy - Semantic Communications: Human Decision-Making and Multi-Tasking}
\begin{floatingfigure}[r]{6cm}
\mbox{\includegraphics[width=5.5cm, height=9.78cm]{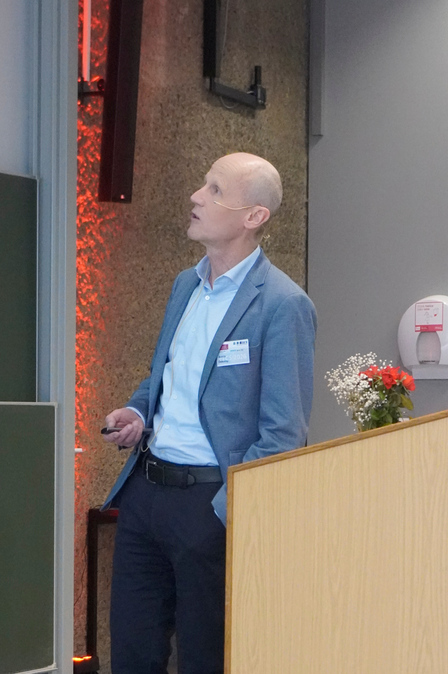}}
\caption{Armin Dekorsy}\label{Armin}
\end{floatingfigure}
Armin Dekorsy (Fig.~\ref{Armin}) presented recent perspectives on semantic communication (SemCom) from the viewpoint of human decision-making and multi-tasking. Semantic communication shifts the focus of communication system design from transmitting raw data reliably to transmitting task-relevant meaning, thereby aligning communication with application goals and human cognition. Building on Shannon’s information theory, modern SemCom systems construct compact, task-specific representations of observation signals (e.g., images) and transmit semantic variables such as object classes or text. This is typically formulated using information-theoretic principles such as the information bottleneck or InfoMax criterion, which maximize the mutual information between the received signal and the relevant semantic variables~\cite{beck2023semantic,beck_advancing_2025}. 

The talk highlighted that this paradigm introduces a fundamental trade-off between semantic fidelity, communication resources, and cognitive constraints. Experimental results show that semantic communication can outperform classical Shannon-based designs in several aspects. By tightly coupling application and communication layers, systems such as the open-source SINFONY framework demonstrate significant bandwidth savings and robustness to channel noise~\cite{beck2023semantic,beck2023software,beck2024model,beck_advancing_2025}. When integrated into human decision-support systems, semantic communication can also adapt to cognitive limitations by prioritizing task-relevant information while reducing communication overhead~\cite{beck2026integrating,beck_advancing_2025}. Furthermore, extensions to semantic multi-task communication using shared encoder architectures enable simultaneous support for multiple tasks and can improve performance compared to separate task-specific encoders~\cite{halimi2024multitask,halimi2025cooperative,halimi2025implicit}. 

Overall, the presentation emphasized that semantic communication offers a promising framework for connecting artificial and human intelligence systems, where communication is optimized not for raw bit rates but for the relevance and utility of the transmitted information.
\newpage
\subsection{Neha Sangwan - Byzantine Channels and Systems}
\begin{floatingfigure}[r]{6cm}
\mbox{\includegraphics[width=5.5cm, height=9.78cm, height=9.78cm]{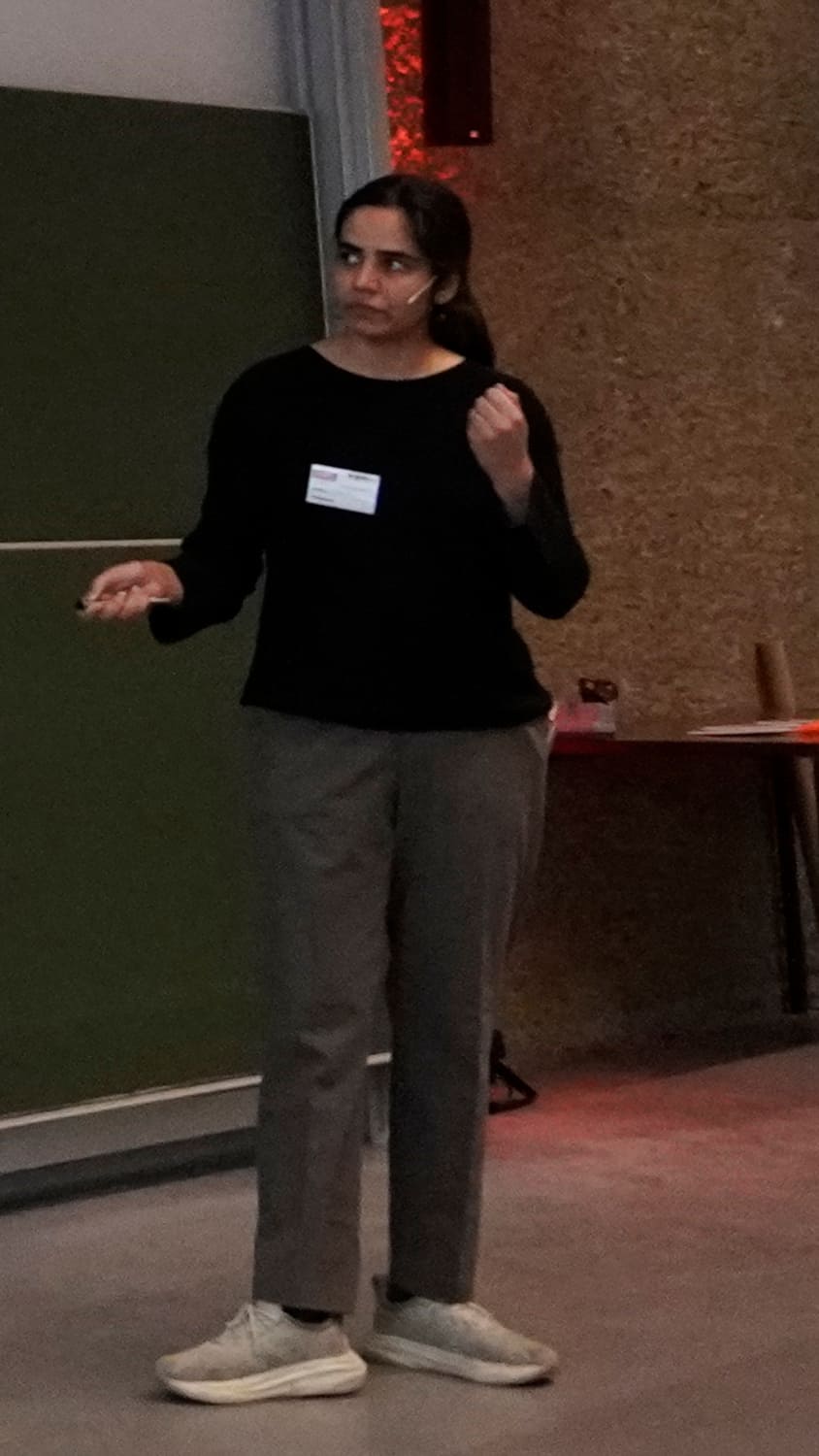}}
\caption{Neha Sangwan} \label{Neha}
\end{floatingfigure}
Neha Sangwan (Fig.~\ref{Neha}) presented models of Byzantine channels and systems, focusing on communication and inference when components may behave adversarially, and discussing coding and system-level strategies that provide robustness against worst-case manipulations.

5G networks offer exceptional reliability and availability, ensuring consistent performance and user satisfaction. Yet, they may still fail when confronted with unexpected conditions. A resilient system is able to adapt to real-world complexity, including operating conditions that were not anticipated during system design. This makes resilience a vital attribute for communication systems that must sustain service in scenarios where models are absent or too complex to provide statistical guarantees. Such considerations indicate that artificial intelligence (AI) will play a major role in enabling resilience.

The talk examines the challenges of designing AI for resilient radio access networks, highlighting the limitations of current statistical learning methods in the presence of rare and unforeseen disruptions. This underscores that the development of truly resilient networks will extend well beyond the 6G era. Ultimately, achieving such resilience requires a paradigm shift in how AI systems are engineered to handle uncertainty and unpredictability.
\newpage
\subsection{Marc Geitz - Some ideas for identification codes in practice}
\begin{floatingfigure}[r]{6cm}
\mbox{\includegraphics[width=5.5cm, height=9.78cm]{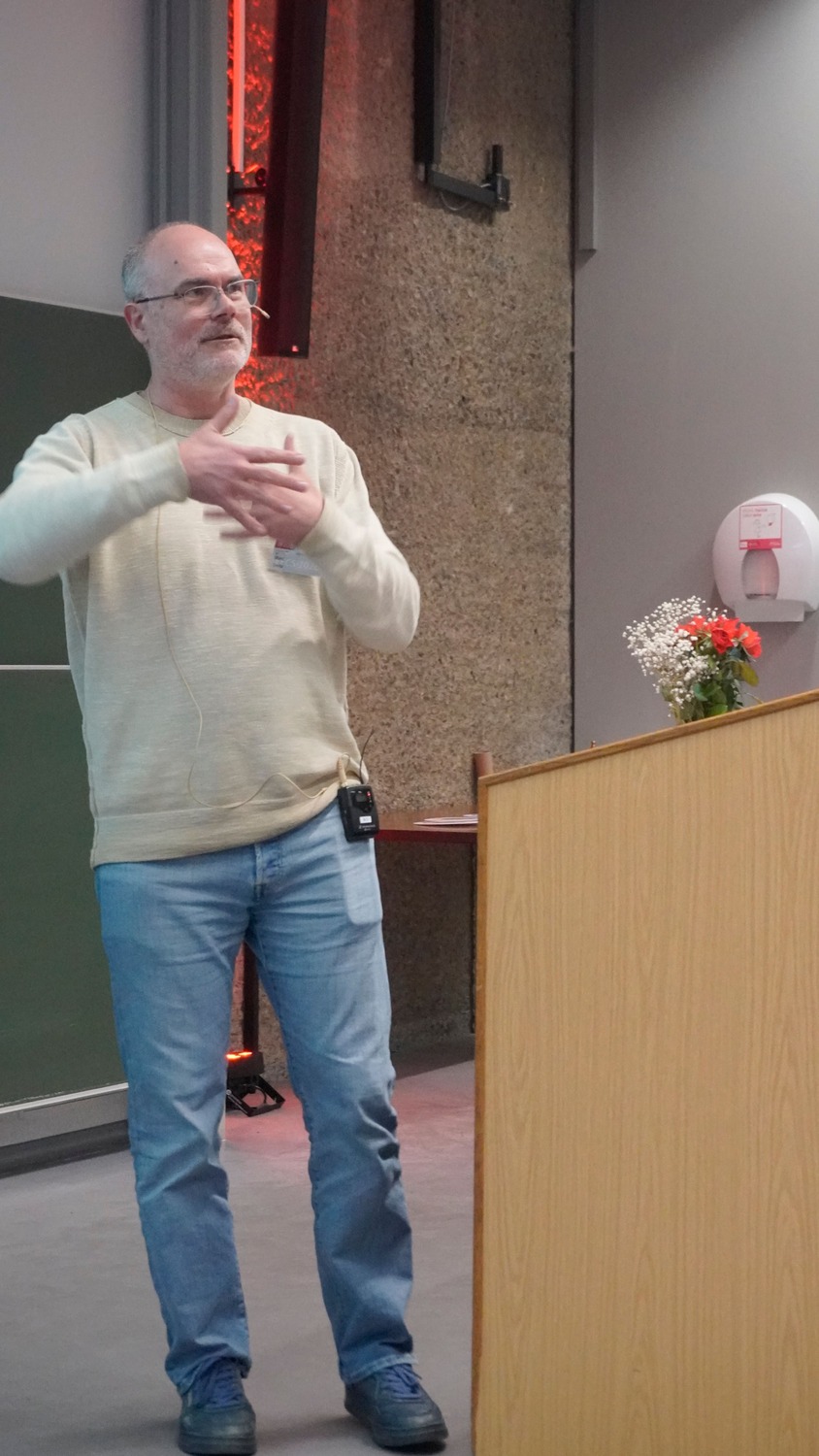}}
\caption{Marc Geitz} \label{Marc}
\end{floatingfigure}
Marc Geitz (Fig.~\ref{Marc}) presented ideas on bringing identification codes closer to practical communication systems. Identification via channels differs fundamentally from classical message transmission: instead of decoding a full message, the receiver only decides whether a specific message was sent. This change in the decoding task allows the number of identifiable messages to grow doubly exponentially with the blocklength, making identification particularly attractive for large-scale systems such as authentication, access control, and massive machine-type communication.

The talk explored how identification concepts could be incorporated into modern communication architectures. One direction considers combining identification with practical channel coding techniques such as polar or LDPC codes in order to obtain reliable and efficiently implementable schemes. Another perspective investigates the integration of identification mechanisms with physical-layer security and authentication protocols, where the objective is not only reliable identification but also protection against impersonation and eavesdropping.

Further discussion addressed possible system architectures in which identification replaces classical data transmission for certain control tasks. Examples include device discovery, access requests in massive networks, and lightweight authentication procedures. In such scenarios, identification codes could significantly reduce communication overhead while supporting an extremely large number of potential users.

Overall, the presentation highlighted that while identification theory offers remarkable theoretical gains in the number of distinguishable messages, realizing these advantages in practice requires new code constructions, efficient decoding algorithms, and integration with existing communication protocols.

\newpage
\subsection{Stefan Wegele - Quantum Annealer for automated dispatching in railway operations}
\begin{floatingfigure}[r]{6cm}
\mbox{\includegraphics[width=5.5cm, height=9.78cm]{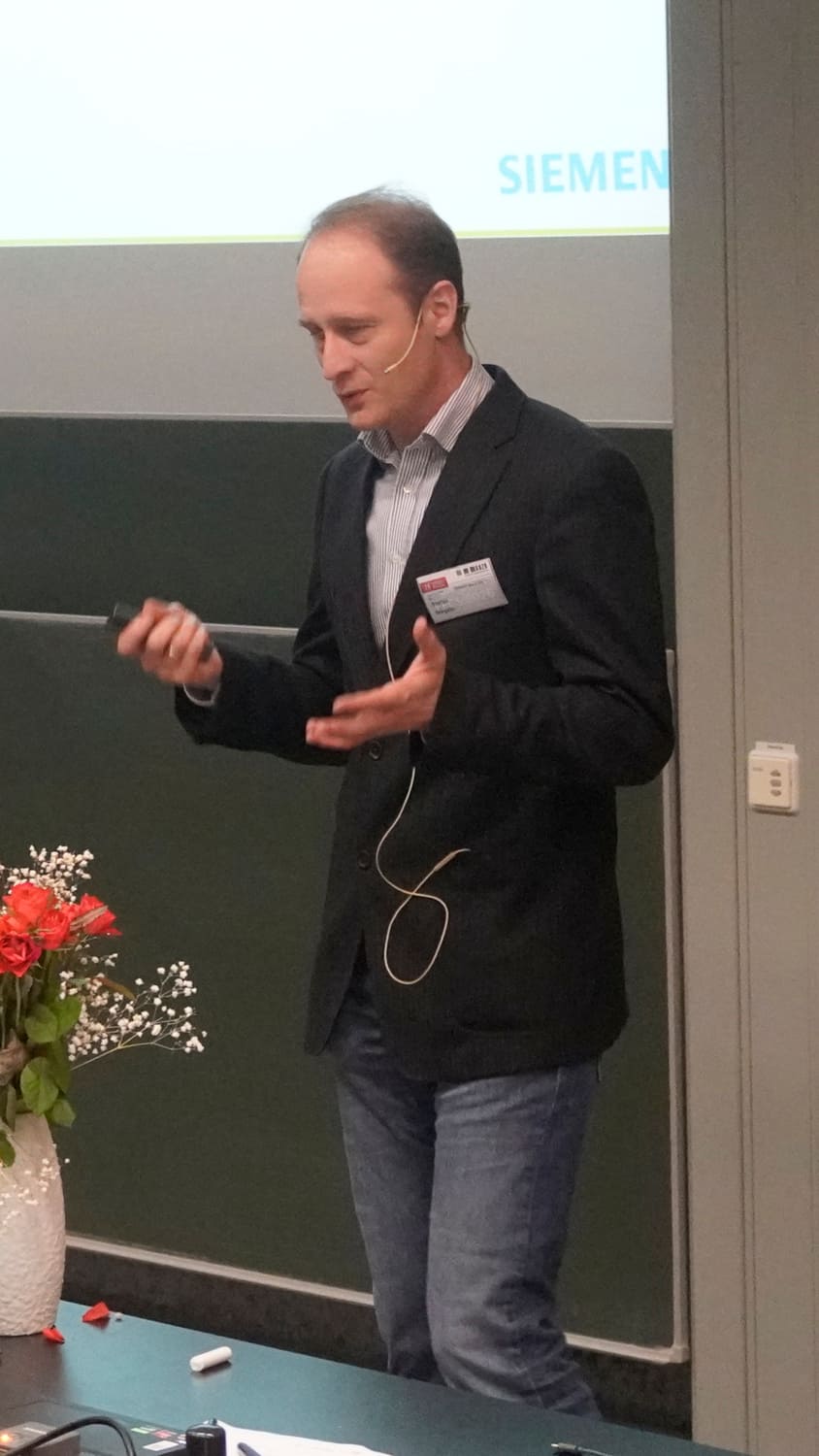}}
\caption{Stefan Wegele} \label{stefan}
\end{floatingfigure}
Stefan Wegele (Fig.~\ref{stefan}) presented recent work on the use of quantum annealing for automated dispatching in railway operations, focusing on how large-scale scheduling conflicts can be formulated and solved using quadratic unconstrained binary optimization (QUBO) models \cite{Wegele2026}. The railway dispatching problem arises when delays propagate through a network and conflicts occur between trains competing for limited infrastructure such as tracks or stations. In real railway systems with thousands of kilometers of tracks and many daily train movements, these conflicts must be resolved quickly in order to maintain operational stability. :contentReference[oaicite:0]{index=0}

The presentation showed how dispatching decisions can be represented as binary variables describing alternative routing or scheduling choices for each train. Conflicts between alternatives are translated into penalty terms, leading to a QUBO objective that balances several criteria, including avoiding simultaneous track usage and minimizing total delay. This formulation allows the dispatching problem to be mapped to quantum annealing hardware or specialized optimization accelerators.

Two heuristic optimization strategies were discussed. The first approach performs conflict resolution after generating candidate schedules, assigning high penalties to collisions and iteratively refining the solution. The second integrates conflict resolution directly into the optimization process by assigning penalties to both conflicts and delays, thereby enabling a joint optimization of feasibility and performance.

Computational experiments on simplified railway networks demonstrate that the QUBO-based framework can efficiently evaluate thousands of alternatives and resolve conflicts within seconds when implemented in optimized C++ environments or on dedicated annealing hardware. The results indicate that quantum-inspired optimization techniques provide a promising direction for real-time dispatching support in complex railway networks.
\newpage


\subsection{Stefano Buzzi - Overcoming hardware impairments in cell-free massive MIMO through differential space time block coding}
\begin{floatingfigure}[r]{6cm}
\mbox{\includegraphics[width=5.5cm, height=9.78cm]{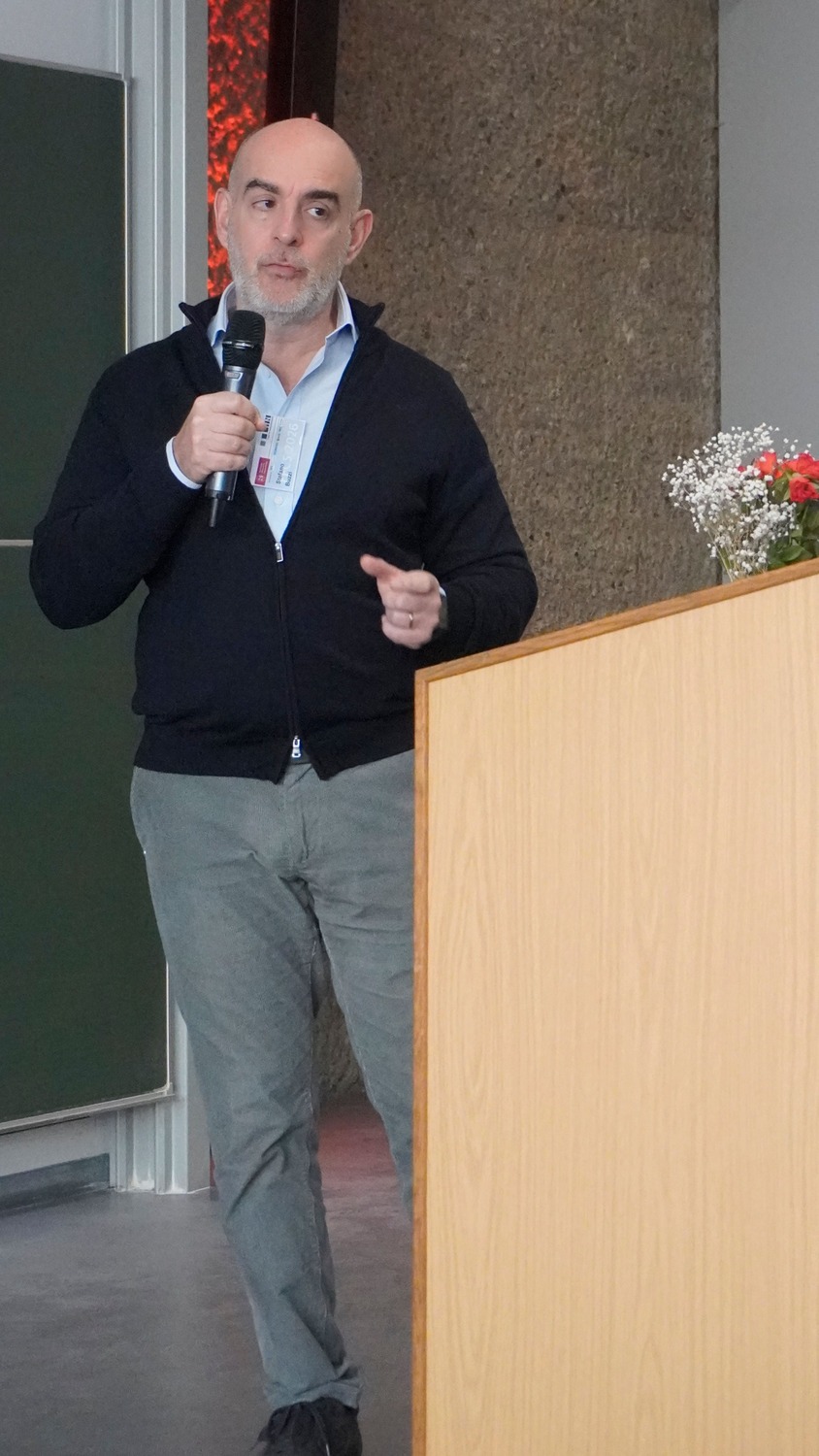}}
\caption{Stefano Buzzi} \label{stefano}
\end{floatingfigure}
Stefano Buzzi (Fig.~\ref{stefano}) presented recent advances on overcoming hardware impairments in cell-free massive MIMO (CF-mMIMO) systems using differential space–time block coding (DSTBC), addressing practical limitations caused by phase misalignments across distributed access points \cite{FreitasBuzziInterdonato2025}. CF-mMIMO is considered a promising architecture for future wireless networks because it replaces conventional cellular deployments with many distributed access points jointly serving users, thereby providing uniform service quality, improved macro-diversity, and better support for emerging 6G applications such as integrated sensing and communication and mobile edge computing. :contentReference[oaicite:0]{index=0}

A major practical challenge in CF-mMIMO arises from hardware impairments and oscillator phase mismatches between distributed access points. Perfect coherent joint transmission requires accurate phase synchronization among the access points, which in practice is difficult to achieve. Existing solutions rely on over-the-air synchronization procedures or optical fronthaul links, both of which introduce additional complexity and overhead.

The presentation proposed differential space–time block coding as a practical alternative that avoids the need for channel state information at the receiver. In this approach, information is encoded in the phase difference between consecutive transmission blocks rather than relying on absolute phase knowledge. While uplink channel estimation is still performed at the network side to compute beamformers and suppress multiuser interference, the user equipment can decode the downlink signal without explicit CSI. Simulation results demonstrate that the proposed approach mitigates phase misalignment effects and enables reliable communication even in the presence of hardware impairments.

Overall, the results illustrate that differential transmission techniques can significantly simplify the implementation of CF-mMIMO systems while preserving the advantages of distributed massive MIMO architectures for future wireless networks.

\newpage
\subsection{Maurizio Magarini - Semantic/molecular Communication}
\begin{floatingfigure}[r]{6cm}
\mbox{\includegraphics[width=5.5cm, height=9.78cm]{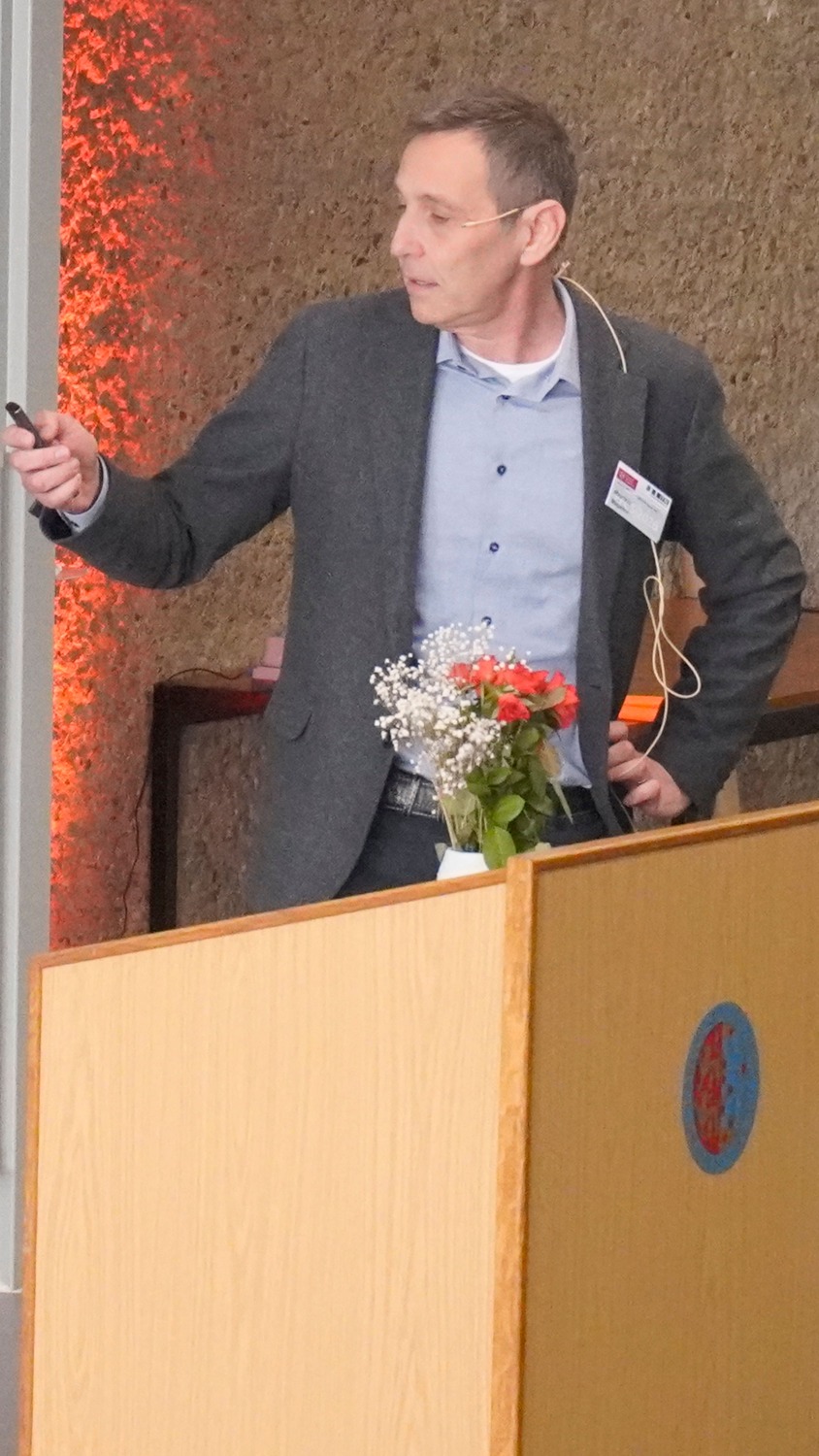}}
\caption{Maurizio Magarini} \label{Mourizio}
\end{floatingfigure}
Maurizio Magarini (Fig.~\ref{Mourizio}) presented recent advances on semantic information in molecular communication systems, focusing on how meaningful information transfer can be quantified in diffusion-based biological networks \cite{MagariniStano2021}. Unlike classical Shannon information theory, which captures syntactic information as statistical dependence between signals, the presented approach emphasized task-oriented communication scenarios where the objective is not reliable bit transmission but achieving functional goals such as survival, adaptation, or targeted intervention.

To address this, the talk adopted the semantic information framework of Kolchinsky and Wolpert, where semantic information is defined as the portion of information that is causally necessary for an agent to maintain its viability in an environment. In this context, viability is quantified through the negative Shannon entropy of the agent’s state distribution, representing the system’s ability to remain in a self-organized, far-from-equilibrium state.

Several case studies were discussed to illustrate this concept. In synthetic cell communication models, agents were shown to extract relevant environmental information, enabling a comparison between Shannon and semantic information measures \cite{MagariniStano2021}. In bacterial chemotaxis scenarios, it was demonstrated that maximizing mutual information is not required for optimal survival; instead, viability is more closely related to transfer entropy under environmental constraints \cite{Brand2024GLOBECOM}. Furthermore, a hybrid navigation strategy for nanoscale medical agents operating in tumor microenvironments achieved improved reliability and reduced detection time compared to purely chemotactic or random strategies \cite{Hedayati2025NANOCOM}.

Overall, the presentation highlighted that incorporating semantic information provides a powerful framework for analyzing and designing molecular communication systems in which meaning, adaptation, and task effectiveness are central.

\newpage
\subsection{Nam Tran - Integrated Communication and Reconfigurable Intelligent Surfaces (RIS)}
\begin{floatingfigure}[r]{6cm}
\mbox{\includegraphics[width=5.5cm, height=9.78cm]{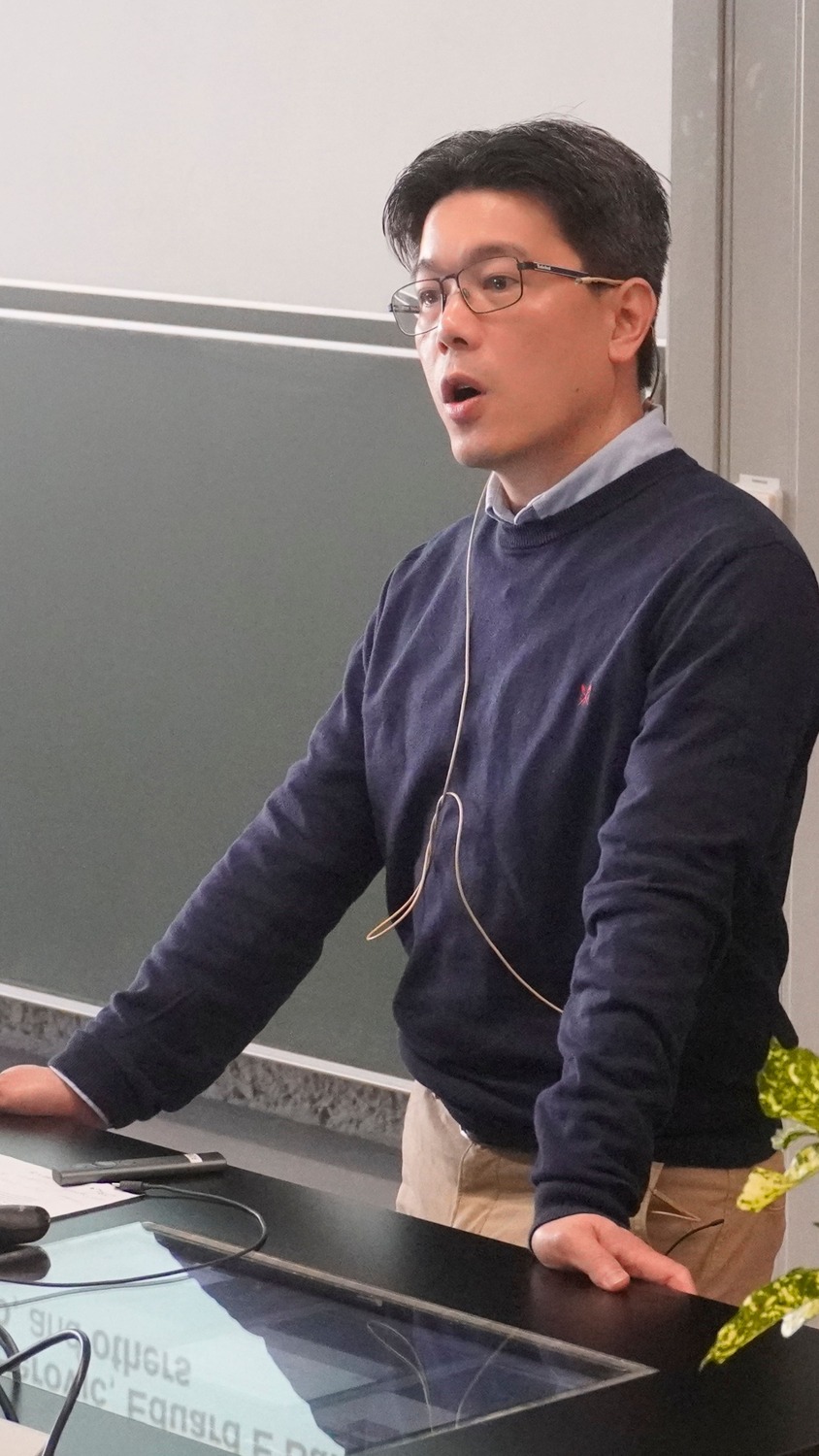}}
\caption{Nam Tran} \label{Nam}
\end{floatingfigure}
Nam Tran (Fig.~\ref{Nam}) presented recent advances in reconfigurable intelligent surfaces (RIS) and stacked intelligent metasurfaces (SIMs), focusing on how programmable propagation environments can reshape coverage and interference patterns in future wireless networks. The talk first reviewed key lessons learned from current RIS research and deployments, which motivate the development of SIM architectures consisting of multiple layers of intelligent metasurfaces. By exploiting electromagnetic interactions among meta-atoms across different layers, SIMs enable signal processing directly in the wave domain and can combine multiple data streams without relying solely on digital precoding.

A central challenge in SIM-aided communication systems is the large-scale and highly nonconvex nature of the associated optimization problems. Previous studies have suggested that, for a fixed SIM thickness, system performance quickly saturates as the number of layers increases. The presentation introduced several optimization formulations for SIM-aided systems in both single-user (SU) and multi-user (MU) MIMO scenarios and discussed algorithmic techniques that overcome this limitation.

For SU-MIMO systems, hybrid optimization approaches were presented that maximize achievable rate and enable performance to scale with the number of SIM layers~\cite{Bahingayi:SUMIMO:2025}. The talk further examined the interaction between linear digital precoding and wave-domain precoding under practical modulation schemes such as QPSK and QAM, showing that even small amounts of digital precoding can significantly improve performance~\cite{Stefan:SIM:HolographicMIMO:2024}. For MU-MIMO systems, sum-rate maximization was studied within an alternating optimization framework, where prioritizing the optimization of metasurface phase shifts can mitigate the performance saturation effect as the number of layers increases~\cite{Bahingayi:MUMIMO:2026}. 

Overall, the presentation highlighted the strong potential of stacked intelligent metasurfaces to enhance wireless system performance when combined with suitable optimization methods and system-level design.
\newpage
\subsection{Mohammad Soleymani - Toward 6G Networks: Tradeoff-Aware Multi-Objective Optimization for RIS-Assisted Communications}
\begin{floatingfigure}[r]{6cm}
\mbox{\includegraphics[width=5.5cm, height=9.78cm]{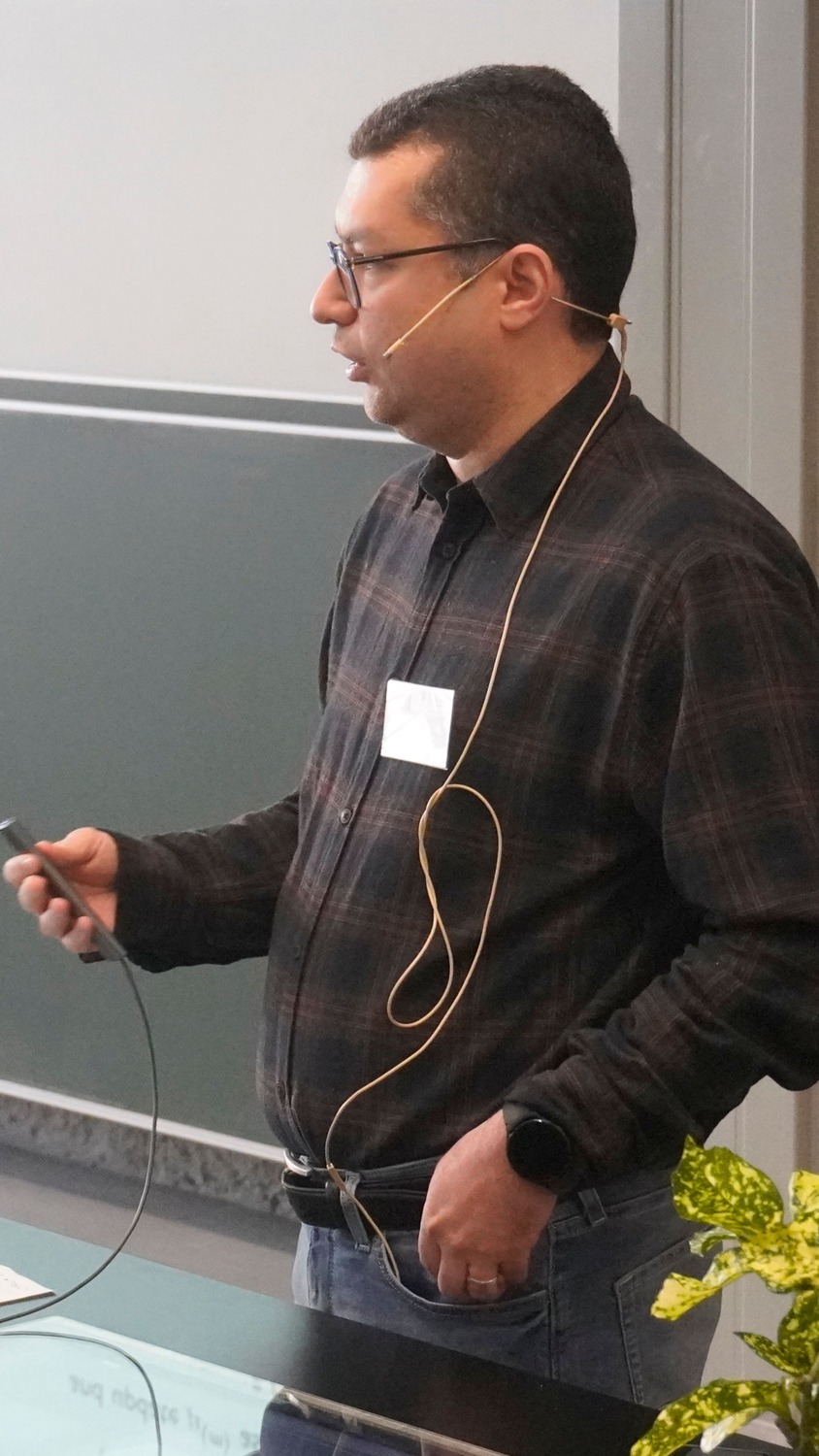}}
\caption{Mohammad Soleymani} \label{Mohammad}
\end{floatingfigure}
Mohammad Soleymani (Fig.~\ref{Mohammad}) presented recent work on multi-objective optimization for emerging 6G wireless systems, with a focus on reconfigurable intelligent surface (RIS)-assisted communications. Future wireless networks must simultaneously improve multiple key performance indicators such as spectral efficiency, energy efficiency, latency, and reliability. Since these objectives are often conflicting, the talk emphasized the importance of optimization frameworks that explicitly capture trade-offs between different performance metrics.
A general methodological framework based on fractional matrix programming was introduced for solving optimization problems that frequently arise in wireless communication systems, particularly in short-packet and finite blocklength regimes. The approach provides a unified method for handling ratio-type performance metrics, such as energy efficiency, and enables efficient optimization of fractional matrix expressions~\cite{soleymani2025framework}.
Building on this framework, the presentation examined RIS-assisted multi-user MIMO systems operating under ultra-reliable low-latency communication (URLLC) constraints. In particular, the spectral–energy efficiency trade-off of nearly passive RIS architectures was analyzed, comparing conventional diagonal RIS designs with more general beyond-diagonal structures~\cite{soleymani2025spectral}. The results reveal fundamental trade-offs between spectral efficiency and energy efficiency and demonstrate how RIS architecture choices significantly influence system performance.
A further case study considered the joint optimization of energy efficiency and latency in RIS-assisted multi-user MIMO downlink systems employing rate-splitting multiple access (RSMA). The results indicate that RSMA can significantly improve energy efficiency and reduce latency compared to conventional spatial division multiple access (SDMA), while also benefiting from synergy with RIS-based propagation control~\cite{soleymani2026ris}. Overall, the talk highlighted the importance of trade-off-aware optimization methods for understanding the design space of future wireless communication systems.

\newpage
\subsection{S\'andor Fekete - Communication and Connectivity in Parallel Robot Swarm Reconfiguration}
\begin{floatingfigure}[r]{6cm}
\mbox{\includegraphics[width=5.5cm, height=9.78cm]{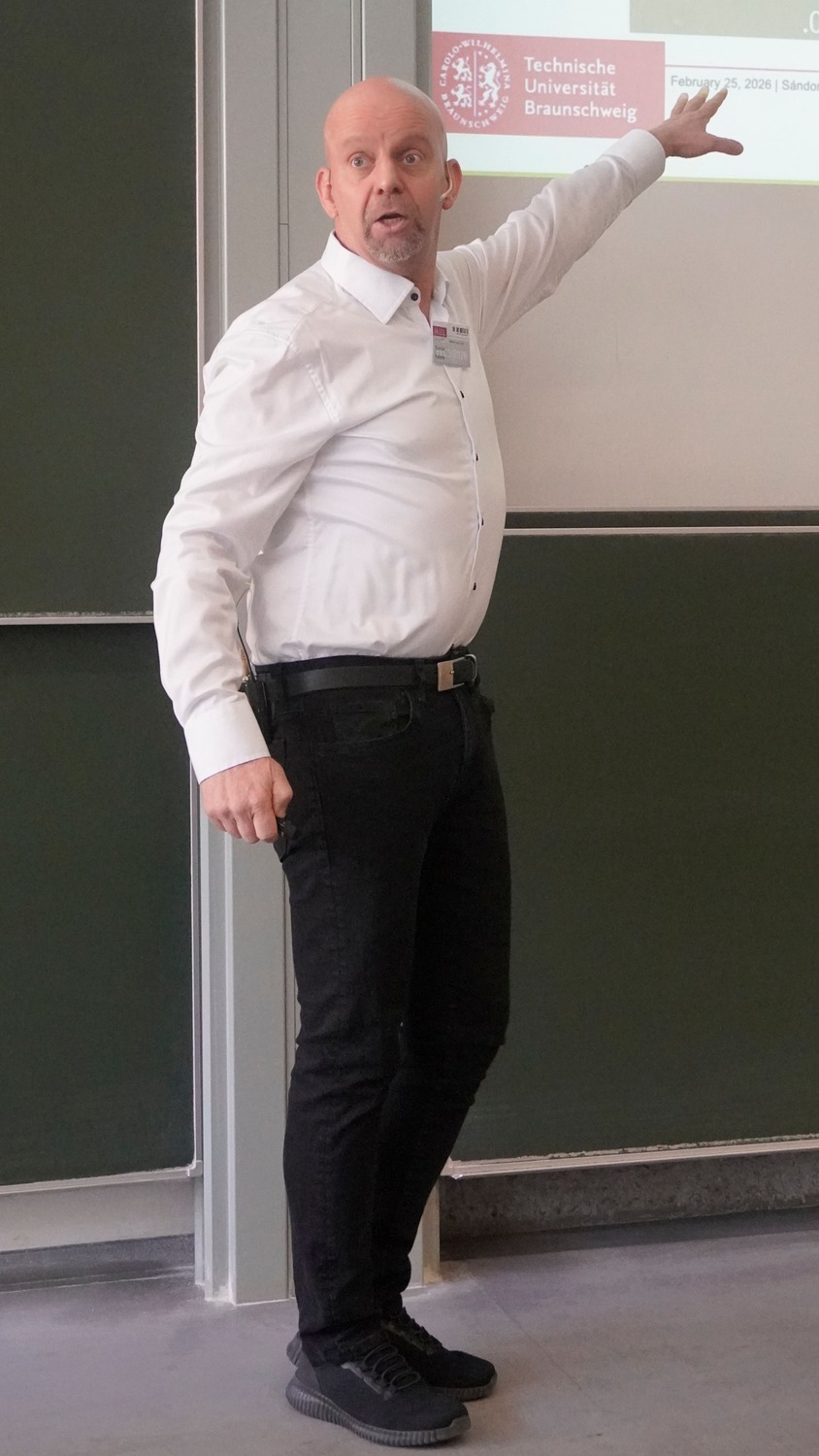}}
\caption{S\'andor Fekete} \label{sandor}
\end{floatingfigure}
S\'andor Fekete (Fig.~\ref{sandor}) presented an overview of algorithmic methods for reconfiguring large swarms of objects under constraints on coordination, communication, and connectivity. The talk addressed a broad spectrum of scenarios, ranging from small-scale systems to extremely large swarms, and highlighted the challenges of achieving efficient collective behavior in high-dimensional settings \cite{fekete2020coordinating}.

Starting from fundamental issues in coordinating swarms at extreme dimensions, the presentation introduced coordinated motion planning for particle systems, where the goal is to minimize reconfiguration time while avoiding collisions \cite{demaine19coordinated,becker18coordinatedvideo,demaine18coordinated}. A practical application was then discussed in the context of dense traffic systems, where coordinated vehicle motion can improve flow efficiency and reduce fuel consumption \cite{fekete2014method}.

The talk further explored scalable approaches for constructing large structures in space using swarms of simple robots, referred to as *Space Ants*, emphasizing decentralized control and robustness \cite{abdel20spaceants,fekete21cadbots}. Another key focus was on particle swarms controlled by uniform global inputs, where coordinated behavior emerges through symmetry-breaking mechanisms such as environmental obstacles or boundary friction \cite{becker14particle,becker15tilt,becker19particle,baez20friction,schmidt20friction}. These techniques enable applications such as targeted drug delivery and micro-scale assembly \cite{becker20targeted,becker20assembly}.

Subsequently, the presentation addressed scenarios requiring persistent connectivity within the swarm, introducing algorithmic strategies that ensure structural integrity during reconfiguration \cite{bourgeois22spaceants2,fekete23connected}. Finally, a novel approach to energy-efficient tracking of moving objects was presented, based on primal-dual optimization methods from mathematical programming \cite{fekete2026tracking,loi2026droneairtrafficcontrol}.

Overall, the talk demonstrated how algorithmic insights and optimization techniques can enable scalable, efficient, and robust control of large swarms across a wide range of applications.

\newpage


\subsection{Slawomir Stanczak - ISAC in 6G}
\begin{floatingfigure}[r]{6cm}
\mbox{\includegraphics[width=5.5cm, height=9.78cm]{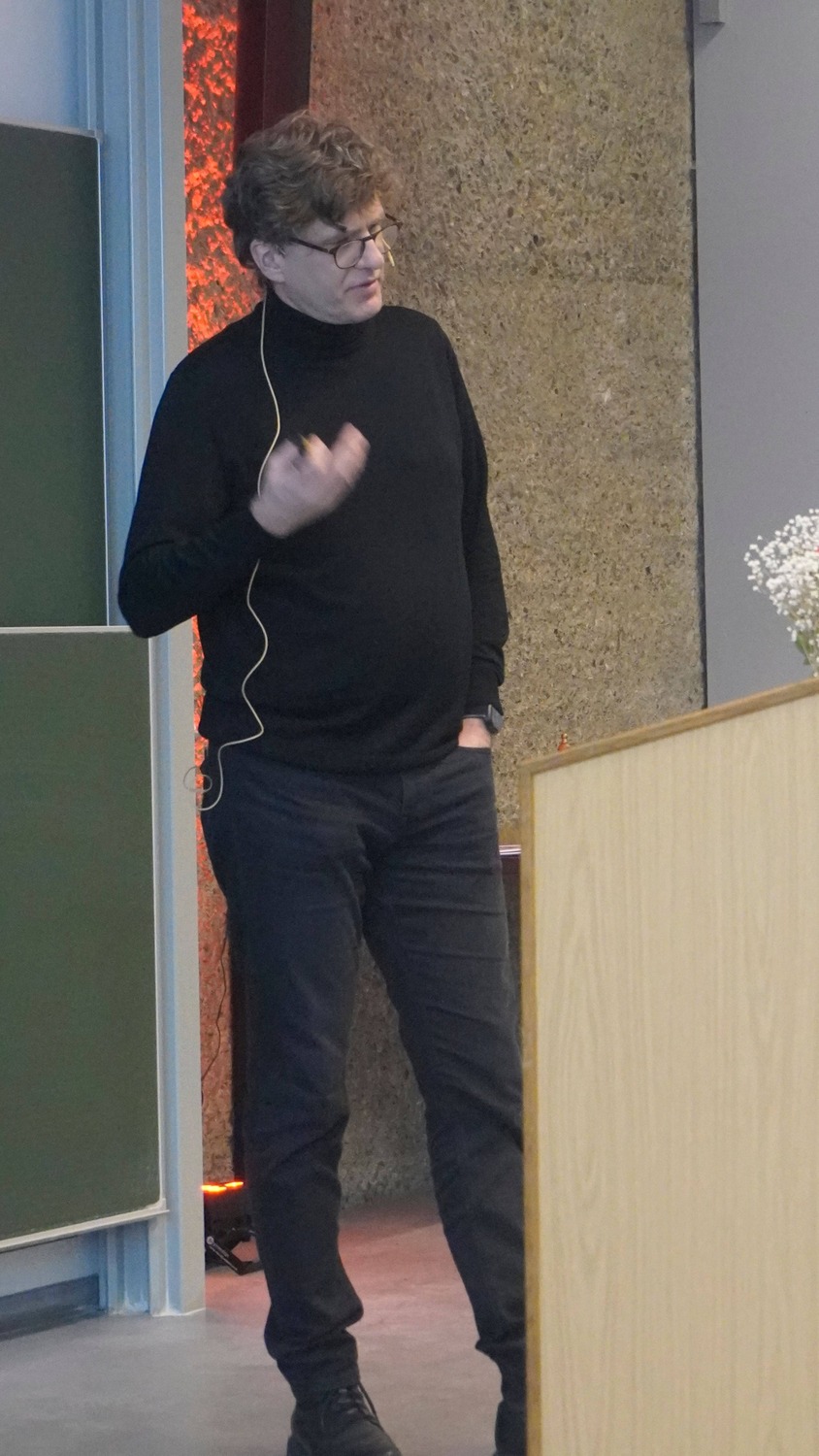}}
\caption{Slawomir Stanczak} \label{slawomir}
\end{floatingfigure}

Slawomir Stanczak (Fig.~\ref{slawomir}) presented the talk \emph{“Integrated Sensing and Communication (ISAC) in 6G,”} discussing how future wireless systems can jointly perform data transmission and environmental sensing using the same radio signals. In emerging 6G networks, communication infrastructure will not only provide connectivity but will also act as a distributed sensing platform capable of detecting objects, estimating positions, and monitoring the surrounding environment.

The talk focused on the integration of sensing functionality into wireless communication systems, where transmitted communication signals can simultaneously be exploited for radar-like sensing tasks. This joint design introduces new challenges, as communication and sensing objectives often impose different requirements on waveform design, beamforming, and signal processing.

A key part of the presentation addressed detection and estimation problems in monostatic ISAC systems employing analog beamforming. In such systems, the same antenna array is used for both transmitting and receiving signals, enabling efficient hardware implementations but introducing additional constraints on signal design and processing. The analysis showed how optimized beamforming and detection strategies can significantly improve sensing performance while maintaining reliable communication capabilities.

The results highlight the importance of jointly optimizing communication and sensing objectives in future wireless networks. Integrated sensing and communication is expected to become a fundamental building block of 6G systems, enabling applications such as environmental awareness, autonomous systems, and intelligent transportation \cite{Hernangomez2024}.

\newpage


\subsection{Arun Padakandla - Rate Regions for Multi-terminal Quantum Channels}
\begin{floatingfigure}[r]{6cm}
\mbox{\includegraphics[width=5.5cm, height=9.78cm]{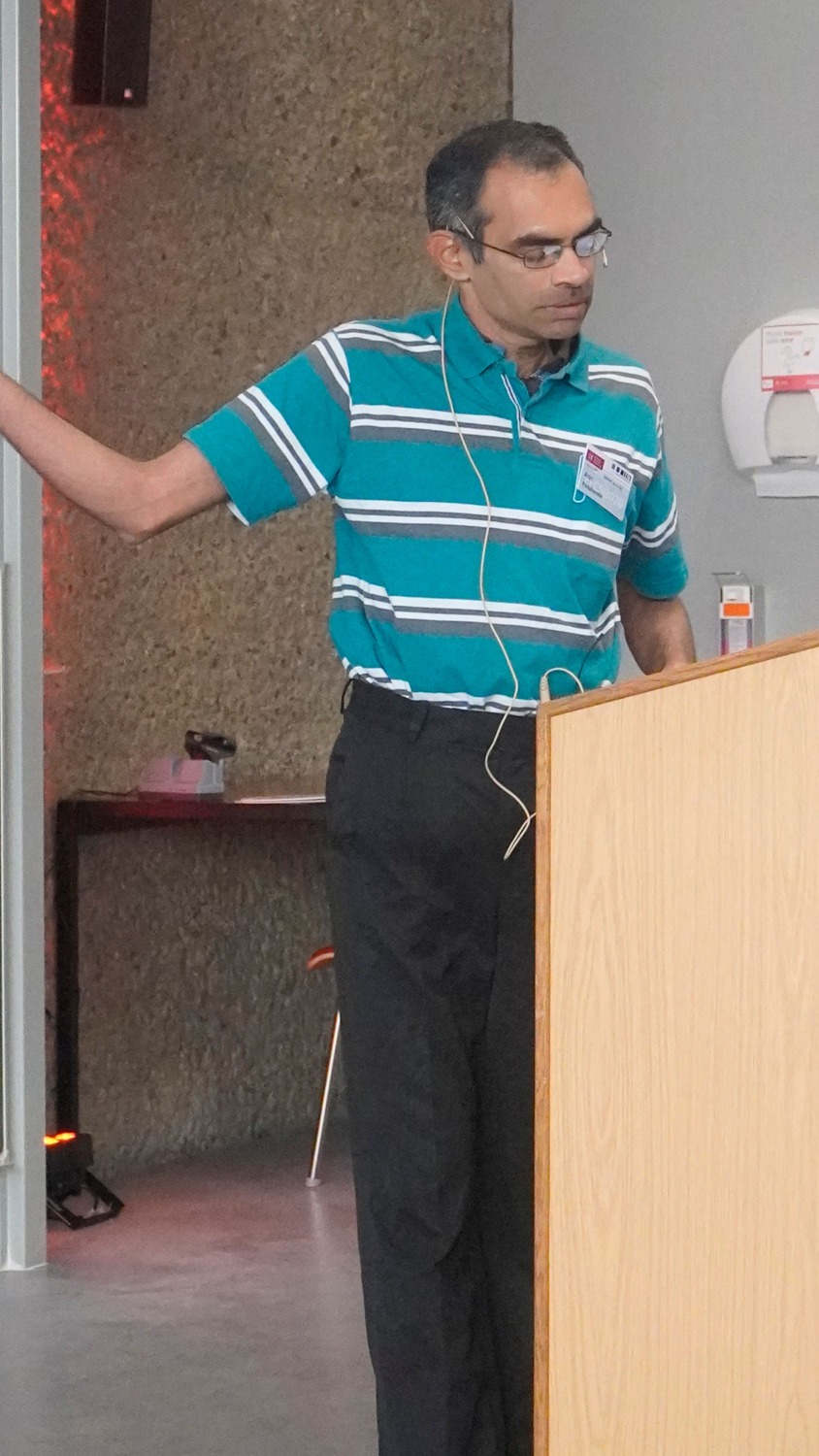}}
\caption{Arun Padakandla} \label{arun}
\end{floatingfigure}
Arun Padakandla (Fig.~\ref{arun}) presented recent advances on the problem of characterizing the bit carrying (CQ) capacity (regions) of multi-terminal quantum channels including $3-$user quantum broadcast and $3-$user quantum interference channels. The addressed problems consider the scenario of communicating multiple bit streams over three user quantum broadcast (QBC) and interference channels (QIC) and focuses specifically on characterizing new inner bounds to the corresponding capacity regions. 

Classical–quantum interference channels are traditionally studied using extensions of the Han–Kobayashi strategy, where each transmitter splits its message into components and receivers decode parts of the interfering signals. Existing constructions rely on unstructured i.i.d. random codes and simultaneous decoding. However, in multi-user settings such as the three-user interference channel, the interference observed at each receiver is inherently a combination of signals from multiple transmitters. Decoding such interference involving bivariate functions of transmitted signals cannot be accomplished via unstructured IID codes. The talk demonstrated that jointly designed coset codes possessing inter- and intra-structural properties can enable efficient decoding of bivariate interference and thereby significantly improve  achievable rate regions by enabling receivers to decode algebraic combinations of interfering transmissions. In particular, by carefully coordinating the codebooks across users, it becomes possible to reliably decode sums of interfering codewords even when the individual signals remain uncertain. This structured decoding approach yields strictly larger inner bounds for certain classical–quantum interference and broadcast channels compared to previously known schemes.

A key technical contribution is the extension of Sen’s simultaneous decoding framework—based on tilting, smoothing, and augmentation techniques to decoding simultaneously into structured coset code ensembles. This allows the analysis of simultaneous decoding with coset codes and leads to new achievable rate regions for multi-terminal classical–quantum networks.

\newpage
\subsection{Christoph Hirche - Differential Privacy}
\begin{floatingfigure}[r]{6cm}
\mbox{\includegraphics[width=5.5cm, height=9.78cm]{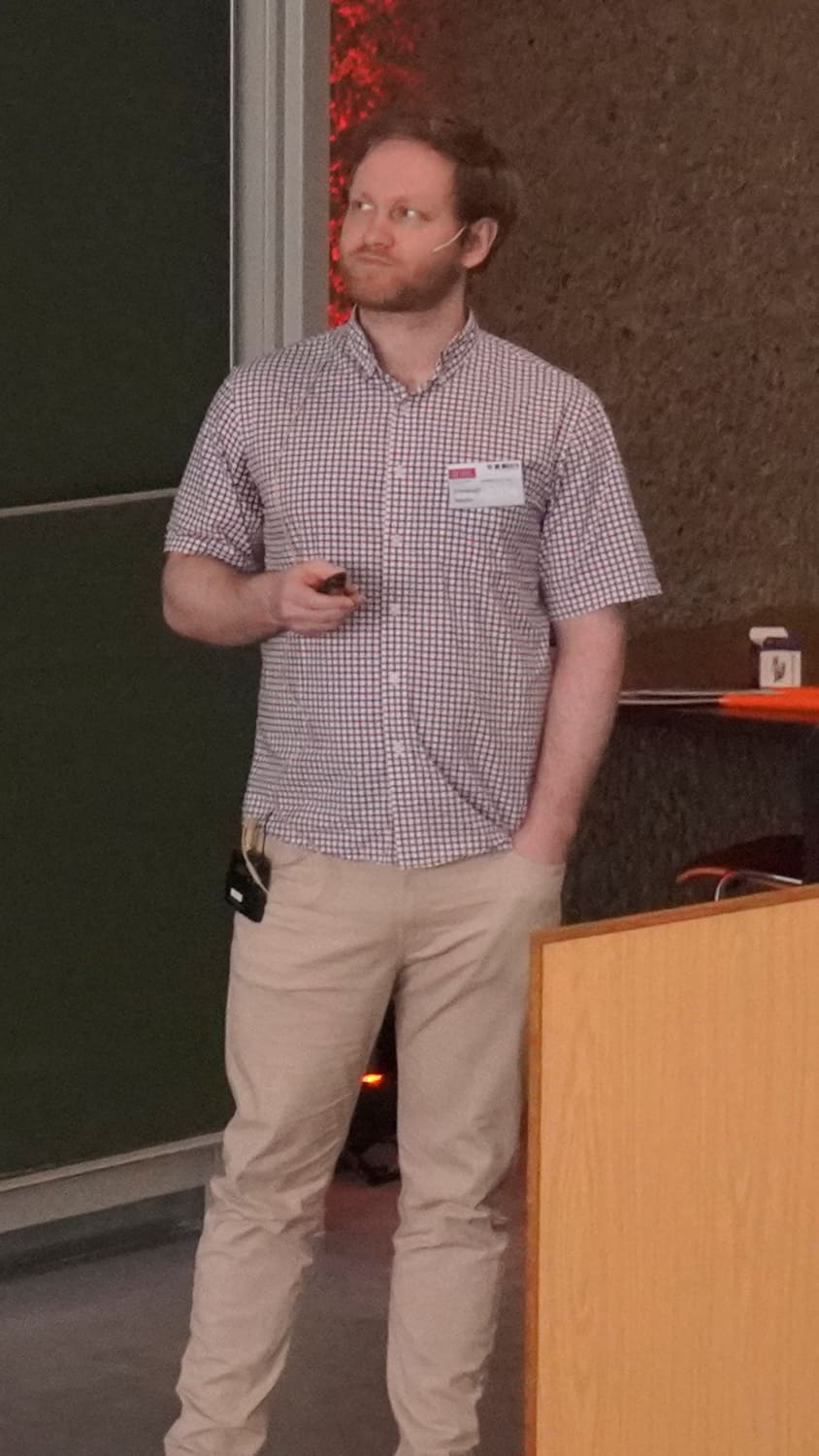}}
\caption{Christoph Hirche} \label{Christoph}
\end{floatingfigure}
Christoph Hirche (Fig.~\ref{Christoph}) presented the talk \emph{“Differential Privacy: Classical and Quantum Perspectives,”} discussing information-theoretic approaches to privacy guarantees in both classical and quantum settings. Differential privacy has emerged as a central framework for protecting sensitive information when statistical queries are performed on datasets. Informally, it ensures that the output distribution of a mechanism changes only slightly when a single individual’s data is modified, thereby limiting the information that can be inferred about any specific participant.

The talk reviewed the classical formulation of differential privacy and then extended the discussion to quantum information processing. In the quantum setting, datasets and queries may involve quantum states and quantum channels, which requires generalizing privacy definitions and analytical tools. From an information-theoretic perspective, privacy guarantees can be characterized using distinguishability measures between probability distributions or quantum states.

The presentation introduced a quantum generalization of differential privacy and analyzed its properties using concepts from quantum information theory such as trace distance and relative entropy. These tools allow one to quantify the privacy leakage of quantum mechanisms and to study the fundamental trade-offs between privacy, accuracy, and information extraction.

Overall, the talk highlighted that differential privacy provides a unifying framework for analyzing privacy guarantees across classical and quantum data-processing scenarios. The information-theoretic viewpoint not only clarifies the operational meaning of privacy guarantees but also provides a foundation for designing privacy-preserving protocols in future quantum information systems \cite{Hirche2022}.

\newpage
\subsection{Andreas Winter - Playing competitive games better with separable quantum states than with any classical correlation}
\begin{floatingfigure}[r]{6cm}
\mbox{\includegraphics[width=5.5cm, height=9.78cm]{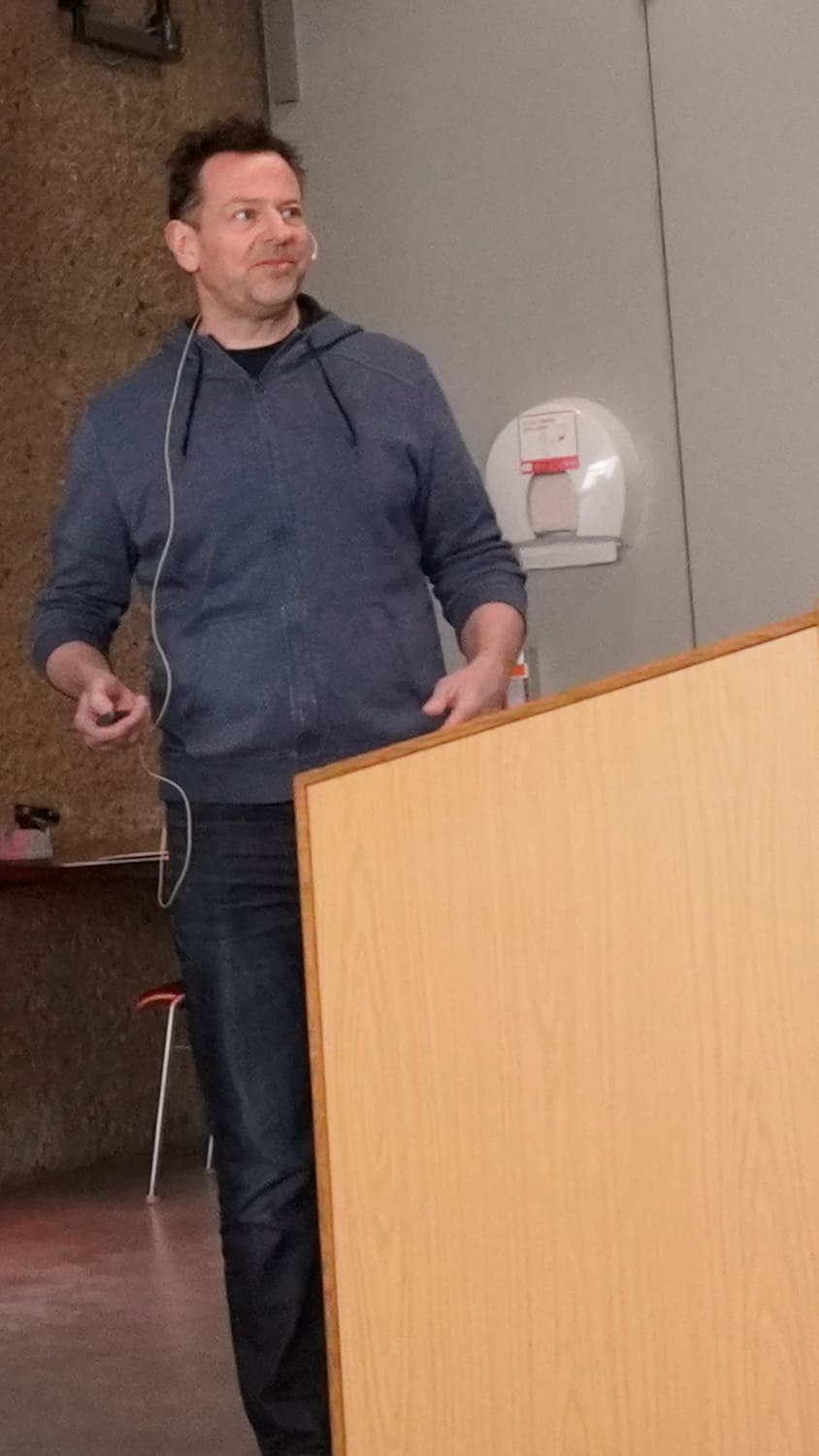}}
\caption{Andreas Winter} \label{andreas}
\end{floatingfigure}
Andreas Winter (Fig.~\ref{andreas}) presented a talk on quantum equilibria in games, discussing how quantum information concepts modify the theory of strategic games by extending the notion of classical correlated equilibria to settings in which players share quantum states and perform local measurements before choosing their actions.

In classical game theory, correlated equilibria allow players to coordinate their strategies using shared classical randomness. The work presented in the talk shows that quantum correlations can enlarge this set of possibilities. In particular, even \emph{separable} quantum states, which do not contain entanglement, can generate correlations that outperform any purely classical shared-randomness strategy in certain competitive games. This demonstrates that the operational power of quantum resources in games is not limited to entanglement alone.

The talk further connected these observations to the broader framework of quantum equilibria. In this setting, players follow equilibrium strategies derived from local measurements on a shared quantum state, and the resulting outcomes can exceed the performance of all classical correlated equilibria. This provides a new perspective on the role of quantum states in strategic interaction and clarifies how quantum information can influence competitive decision-making.

Overall, the presentation highlighted that quantum game theory reveals advantages of quantum correlations beyond the standard entanglement paradigm. The concept of quantum equilibria offers a natural framework for understanding these effects and for studying how quantum resources can enhance distributed strategic tasks and information-processing games.

\newpage
\subsection{Saikat Guha - Measurement-based entanglement distillation with atomic qubits and constant-rate quantum repeaters}
\begin{floatingfigure}[r]{6cm}
\mbox{\includegraphics[width=5.5cm, height=9.78cm]{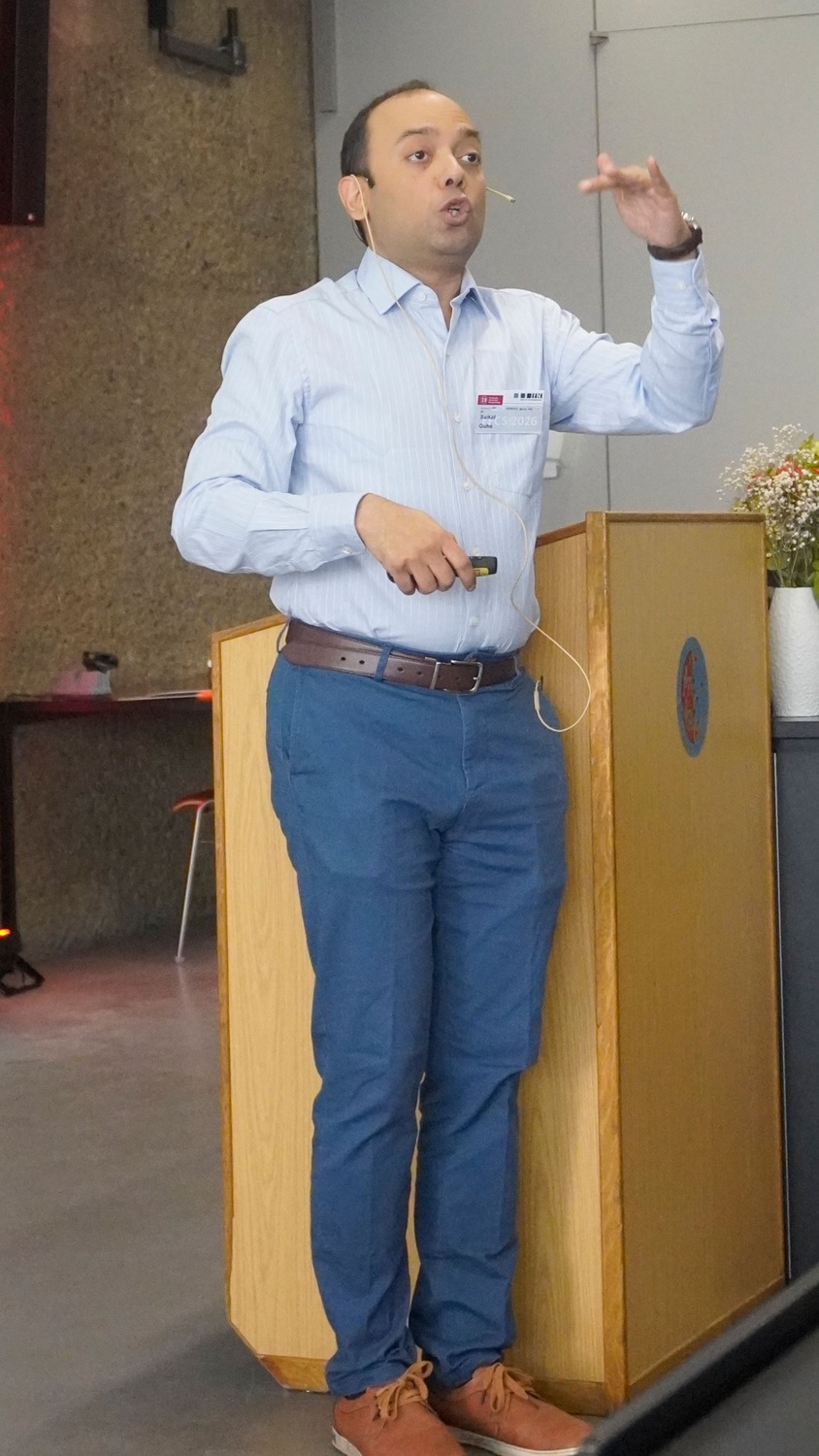}}
\caption{Saikat Guha} \label{saikat}
\end{floatingfigure}
Saikat Guha (Fig.~\ref{saikat}) presented a talk addressing the challenge of distributing high-quality entanglement over long distances in quantum communication networks. Entanglement distribution is a fundamental resource for quantum key distribution, distributed quantum computing, and quantum sensing. However, optical losses and noise in long-distance channels significantly degrade entanglement and limit the achievable communication rates.

The talk discussed a measurement-based approach to entanglement distillation implemented with atomic qubits acting as quantum memories. In this architecture, photonic links are used to establish noisy entanglement between remote nodes, while local measurements and classical communication are employed to distill higher-quality entangled states. By combining measurement-based operations with atomic quantum memories, the protocol enables efficient purification of entanglement generated over lossy optical channels.

A key result of the work is the possibility of achieving \emph{constant-rate} quantum repeater architectures. Unlike conventional repeater schemes whose rates decrease rapidly with distance, the proposed approach can maintain a constant entanglement generation rate over long distances under suitable conditions. This is achieved through a scalable repeater design that integrates entanglement generation, storage, and measurement-based processing at intermediate nodes.

Overall, the presentation highlighted how measurement-based protocols and atomic-qubit memories can enable practical quantum repeater networks capable of supporting long-distance quantum communication. Such architectures are expected to play a central role in the development of future quantum internet infrastructure~\cite{Auletta2021}.
\newpage


\subsection{Rawad Bitar - Low-Communication Secure and Private Federated Learning}
\begin{floatingfigure}[r]{6cm}
\mbox{\includegraphics[width=5.5cm, height=9.78cm]{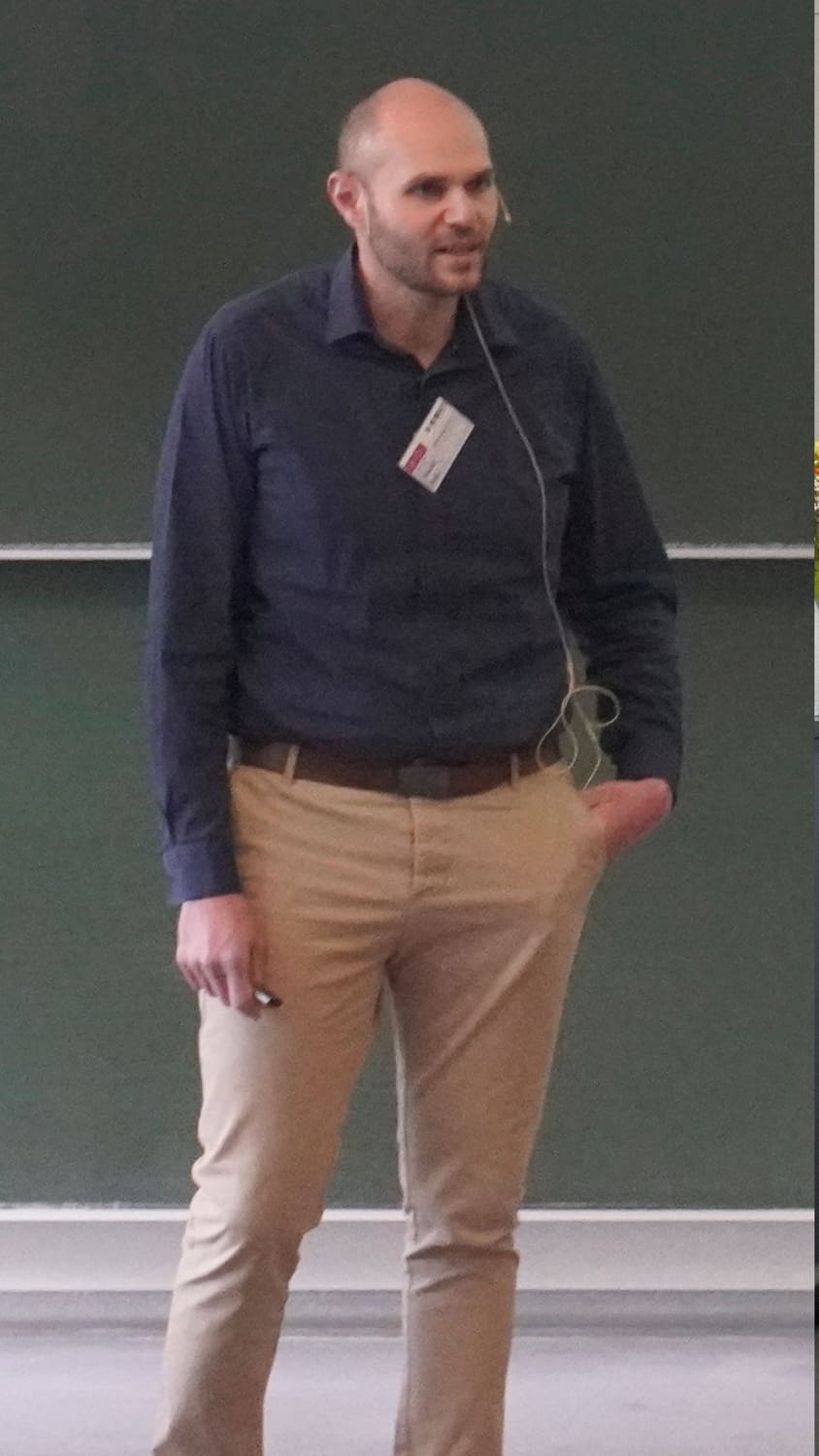}}
\caption{Rawad Bitar} \label{rawad}
\end{floatingfigure}
Rawad Bitar (Fig.~\ref{rawad}) discussed secure and private federated learning with low communication overhead, focusing on cryptographic and information-theoretic techniques that reduce the number of transmitted bits while maintaining privacy, robustness, and model utility.

Federated learning has emerged as a ubiquitous paradigm for training a central model on data generated and owned by multiple participating users. Despite its advantages, achieving trustworthy federated learning faces several challenges, including preserving the privacy of users’ data, ensuring robustness against malicious users, and reducing the required communication cost.

While jointly ensuring privacy and robustness is desirable, a fundamental tension exists between these two requirements. On the one hand, privacy guarantees require hiding users’ data from the federator, while on the other hand, detecting malicious users requires computing statistics of the users’ data.

In this talk, coding-theoretic techniques, in particular secret sharing and private information retrieval schemes, are used to jointly achieve both requirements. As a result, the proposed scheme enables theoretical guarantees for robustness and privacy even in settings where the users’ data are drawn from different distributions~\cite{egger2025private}. However, the price to pay is an exacerbated communication cost scaling polynomially with the number of users. To reduce the communication cost, zeroth-order optimization tools are incorporated into the developed techniques~\cite{egger2025byzantine}.

\newpage
\subsection{Laura Luzzi - Covert communication over additive noise channels}
\begin{floatingfigure}[r]{6cm}
\mbox{\includegraphics[width=5.5cm, height=9.78cm]{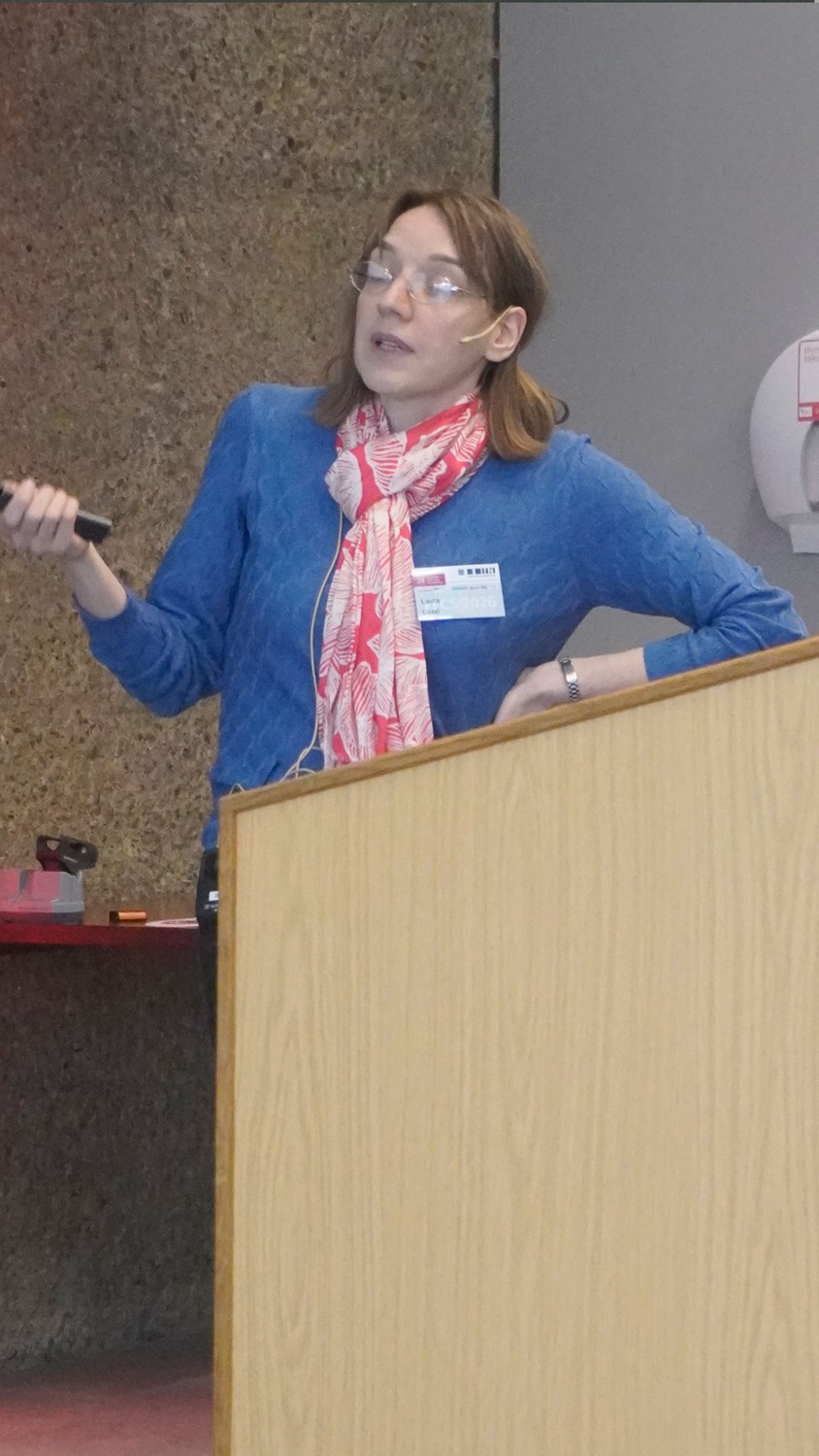}}
\caption{Laura Luzzi} \label{laura}
\end{floatingfigure}
Laura Luzzi (Fig.~\ref{laura}) presented recent advances in covert communication over additive noise channels, addressing the fundamental limits of hiding the very existence of communication from an adversary~\cite{BouetteLuzziWang2023,BouetteLuzziWang2025}. In this talk, a covert communication setting is considered, where a transmitter and a receiver aim to prevent an eavesdropper from detecting that communication is ongoing. It is known that the channel capacity in this setting is zero, and that the amount of information that can be sent reliably and covertly scales with the square root of the number of channel uses.

The scaling constant of the square-root law is studied for a general class of memoryless additive noise channels, and it is shown that it is upper bounded by the square root of the varentropy of the noise. Under additional assumptions, this upper bound is tight.

In the final part of the talk, the asymptotic limits of covert communication over an i.i.d. Gaussian channel are considered when allowing a positive average error probability $\epsilon$. In this scenario, the strong converse does not hold, and the scaling constant $C_{\epsilon}$ depends on $\epsilon$. Upper and lower bounds for $C_\epsilon$ are derived, and it is shown that allowing a small positive error probability enables the transmission of additional covert information.

\newpage
\subsection{Igor Bjelakovi\'c - Physical layer security and privacy}
\begin{floatingfigure}[r]{6cm}
\mbox{\includegraphics[width=5.5cm, height=9.78cm]{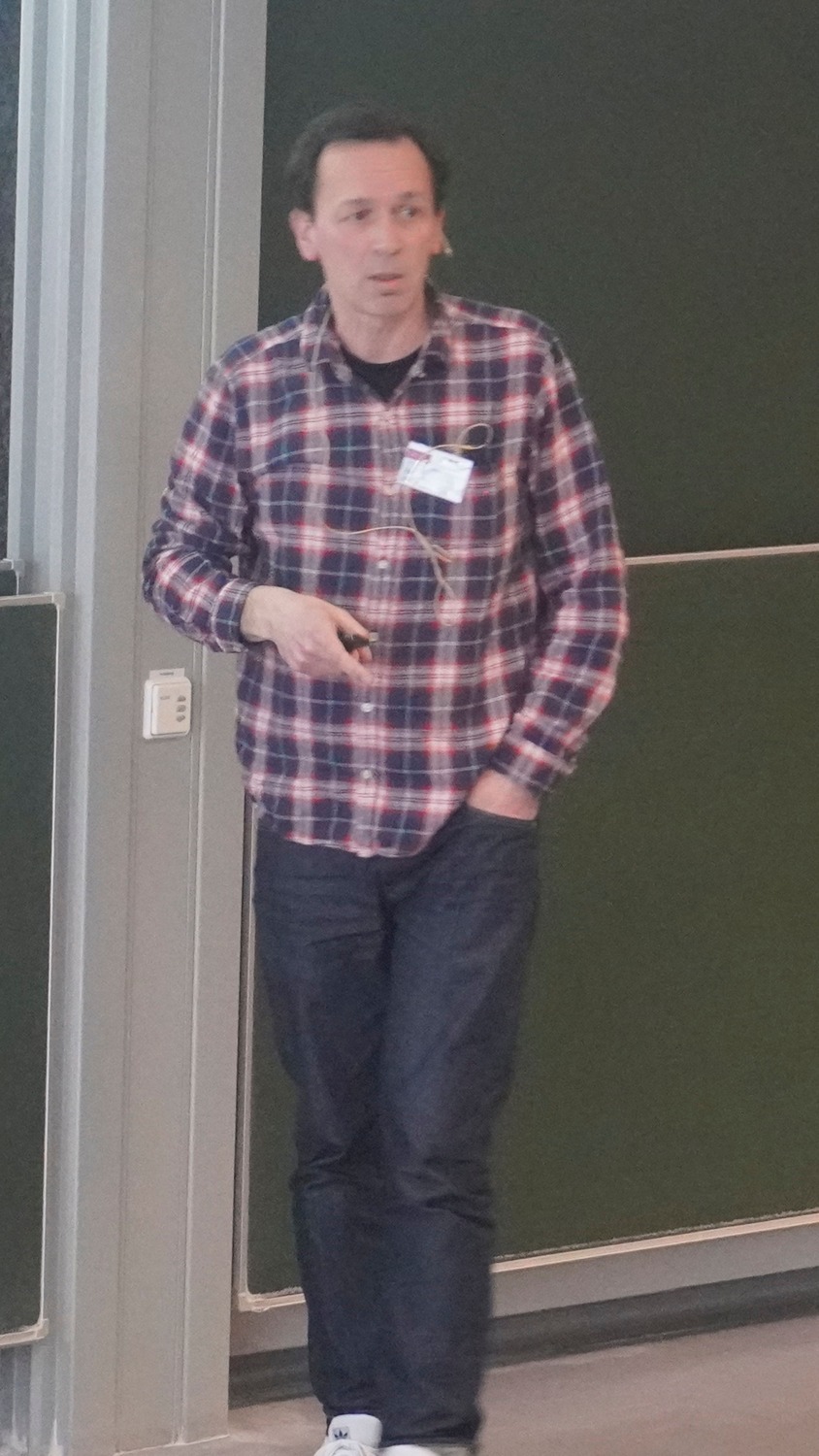}}
\caption{Igor Bjelakovi\'c} \label{Igor}
\end{floatingfigure}
Igor Bjelaković (Fig.~\ref{Igor}) presented recent results on physical-layer security and privacy in quantum and analog communication systems, focusing on classical–quantum (cq) wiretap channels and secure over-the-air computation. The motivation stems from quantum side-channel attacks, where physical processes such as photon emissions from electronic devices can leak information that can be modeled as cq bosonic channels. This allows the integration of wiretap channel theory with cryptographic notions of security. 

In this talk, two approaches to security and privacy in quantum and analog settings were presented. In the first part, a result on keyless communication over channels with classical continuous-alphabet input, infinite-dimensional quantum output, and an infinite-dimensional quantum side channel was introduced. The focus is not primarily on the result itself but rather on the proof technique for establishing security, which exhibits strong parallels to methods from Glivenko–Cantelli theory of empirical processes and PAC bounds in statistical learning. Operationally defined semantic security is taken as the security metric, and an equivalent form—known as distinguishing security—is transformed analytically via duality arguments into a representation directly amenable to Glivenko–Cantelli techniques. This approach enables finite-blocklength results in the infinite-dimensional setting, which are exemplified through numerical evaluations for a noisy cascaded beam splitter quantum-optical channel. The results presented in this part were obtained in collaboration with M. Frey, J. Nötzel, and S. Stanczak, and published in \cite{frey2025}.

The second part of the talk introduces a model for an analog variant of Secure Multi-Party Computation. The starting point is analog over-the-air computation for a class of nomographic functions whose outer functions are sufficiently regular to permit analytical treatment even under correlated fading and noise processes. Suitable security and privacy criteria based on an analog variant of encryption are introduced. Proofs of reliability, security, and privacy rely on lattice-coding methods adapted for use in the analog domain. The initial results were published in \cite{brune2023}.

\newpage
\subsection{Gerhard Wunder - Bilinear Compressive Security}
\begin{floatingfigure}[r]{6cm}
\mbox{\includegraphics[width=5.5cm, height=9.78cm]{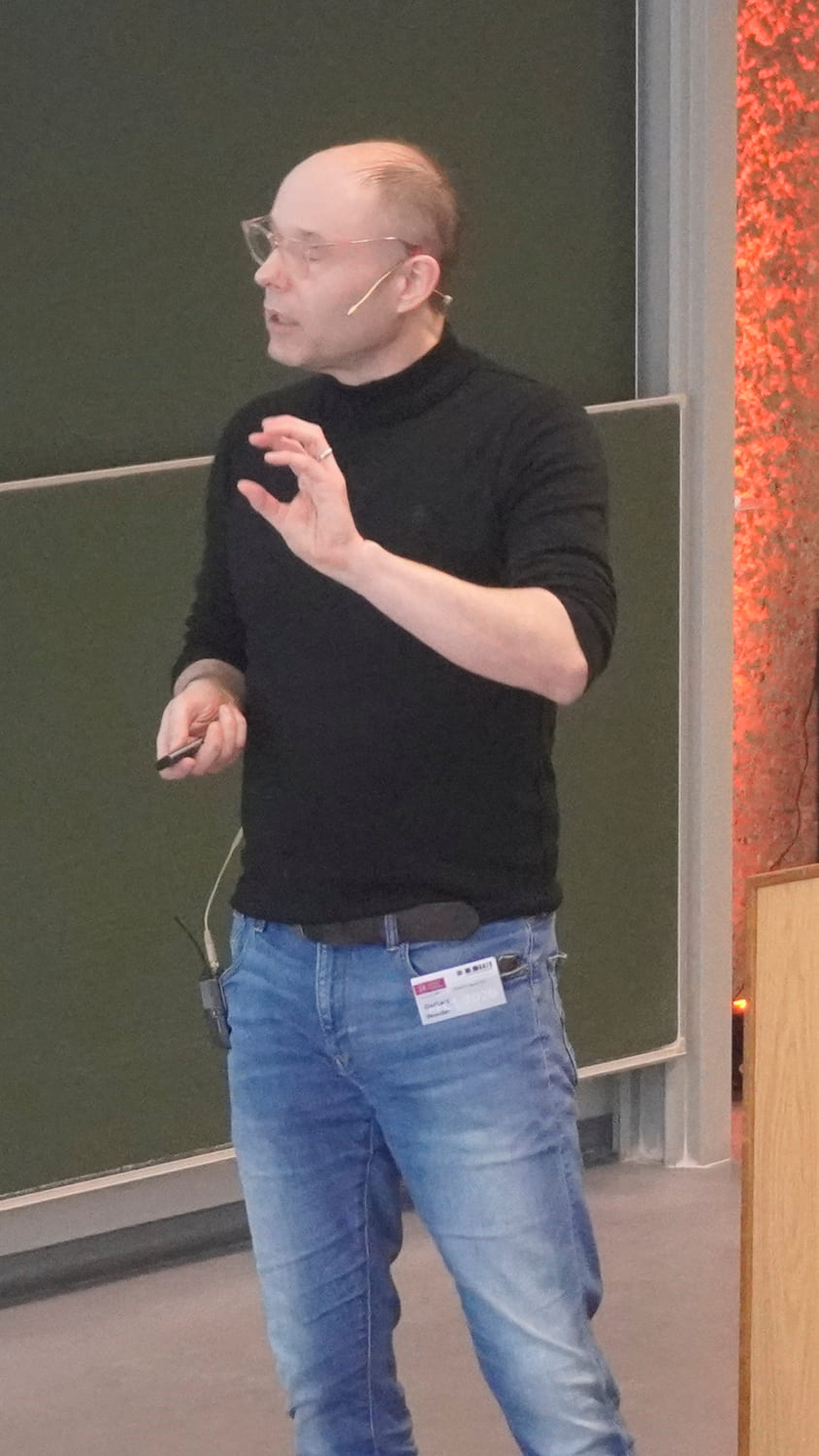}}
\caption{Gerhard Wunder} \label{Gerhard}
\end{floatingfigure}
Gerhard Wunder (Fig.~\ref{Gerhard}) presented the concept of \emph{Bilinear Compressive Security} (BCS), a cryptographic approach that combines compressed sensing with bilinear measurement models to obtain efficient encryption schemes suitable for resource-constrained systems. Compressed sensing exploits the sparsity of signals and allows reconstruction from a small number of random linear measurements, typically modeled as $y = Qx$, where $x$ is a sparse signal and $Q$ is a random measurement matrix \cite{Candes2006,Donoho2006}. \cite{WunderICC2026}
Earlier proposals suggested using compressed sensing directly as a private-key cryptosystem, where the measurement matrix $Q$ serves as the secret key and the ciphertext is given by $y = Qx$. However, such linear constructions are vulnerable to known-plaintext or chosen-plaintext attacks: observing several plaintext–ciphertext pairs allows an adversary to recover the matrix $Q$ by solving a linear system.
To address this weakness, the talk introduced a bilinear measurement model of the form
\(
y = h * Qx ,
\)
where $h$ is a randomly generated sparse filter and $*$ denotes convolution. In this scheme, Alice generates a fresh sparse filter $h$ for each transmission and computes the ciphertext using the secret key matrix $Q$. Bob, who knows $Q$, jointly recovers both the sparse signal $x$ and the filter $h$ by solving a blind deconvolution problem using hierarchical compressed sensing techniques.
The security analysis shows that recovering the secret key from plaintext–ciphertext observations becomes significantly harder in this bilinear setting. In particular, under certain symmetry assumptions on the filter distribution, key recovery reduces to solving multiple phase-retrieval problems, which prevents an adversary from uniquely identifying the measurement matrix. The results demonstrate that the bilinear design removes the critical linear structure exploited in attacks on classical compressive encryption while retaining the efficiency and sparsity-based reconstruction properties of compressed sensing.
Overall, the presentation highlighted bilinear compressive security as a promising direction for lightweight cryptographic schemes, combining sparse signal processing with modern security analysis and offering potential applications in energy-constrained communication systems and secure sensing architectures.

\newpage
\subsection{Aydin Sezgin - Enhancing Physical Layer Security with Reconfigurable Arrays}
\begin{floatingfigure}[r]{6cm}
\mbox{\includegraphics[width=5.5cm, height=9.78cm]{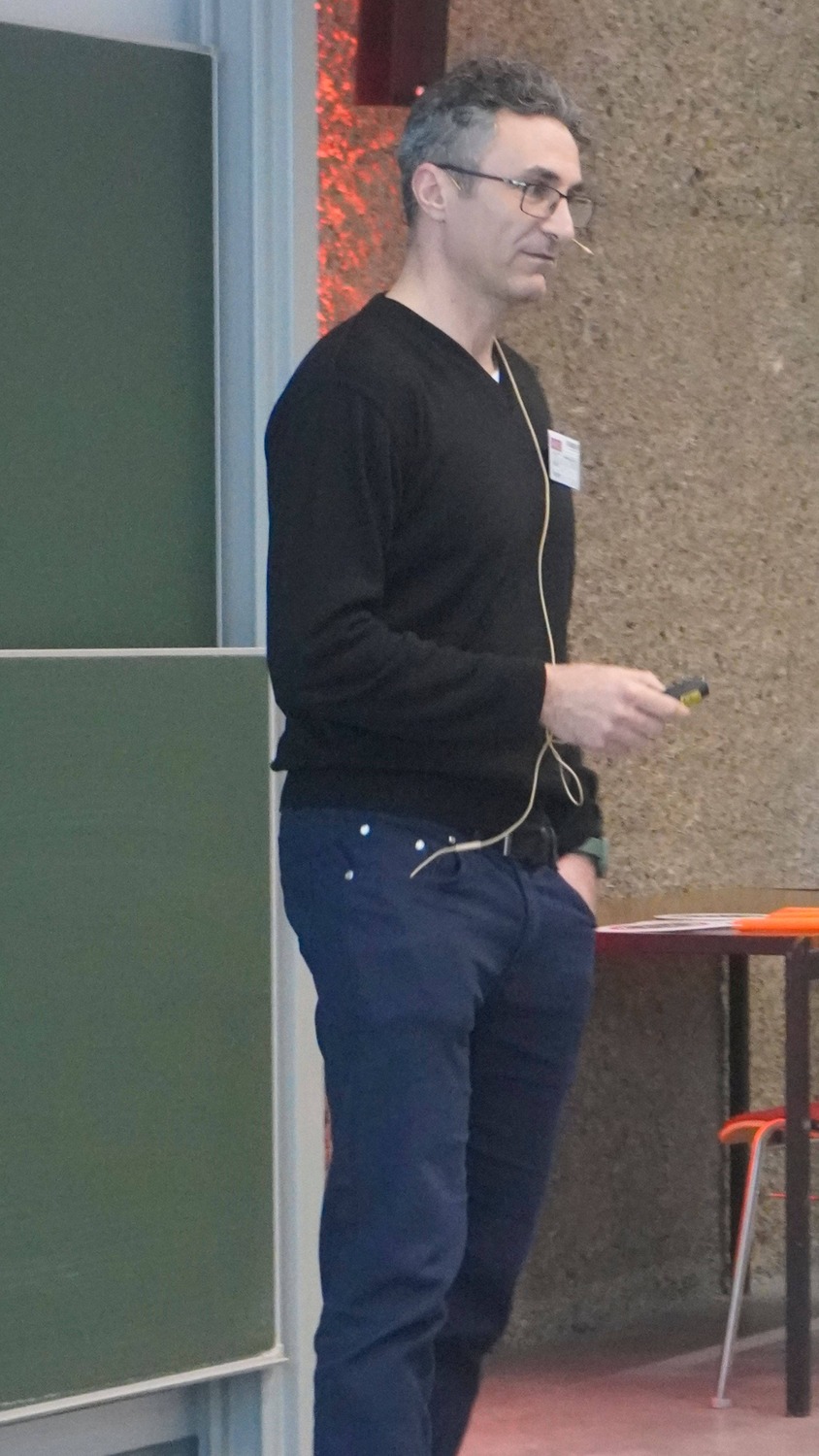}}
\caption{Aydin Sezgin} \label{aydin}
\end{floatingfigure}
Aydin Sezgin (Fig.~\ref{aydin}) presented a talk discussing how emerging programmable radio technologies can strengthen security mechanisms directly at the physical layer of wireless systems. Physical layer security exploits the inherent randomness of wireless channels to protect confidential communication against eavesdropping, providing an alternative or complement to conventional cryptographic methods.

The talk focused on the use of reconfigurable antenna arrays and programmable propagation environments to shape wireless signals in a way that improves secrecy performance. By dynamically adjusting the radiation pattern and spatial properties of transmitted signals, reconfigurable arrays can enhance the signal quality at the intended receiver while simultaneously degrading reception at potential eavesdroppers. Such adaptive beamforming techniques allow the transmitter to exploit spatial channel variations to create favorable conditions for secure communication.

Several signal processing and optimization techniques were discussed for configuring these arrays in practical wireless systems. In particular, the work demonstrates how intelligent control of array parameters and propagation characteristics can significantly increase secrecy rates and robustness against interception. These methods are especially relevant for future wireless networks in which programmable metasurfaces and reconfigurable intelligent surfaces become integral components of the radio infrastructure.

Overall, the presentation highlighted that reconfigurable array technologies provide powerful new tools for strengthening physical layer security in next-generation communication systems. By combining adaptive beamforming with programmable environments, future wireless networks may achieve more resilient and flexible protection against eavesdropping~\cite{Li2024EnhancingSecrecyRate}.
\newpage
\subsection{Gohar Kyureghyan - Cryptography meets Mathematics}
\begin{floatingfigure}[r]{6cm}
\mbox{\includegraphics[width=5.5cm, height=9.78cm]{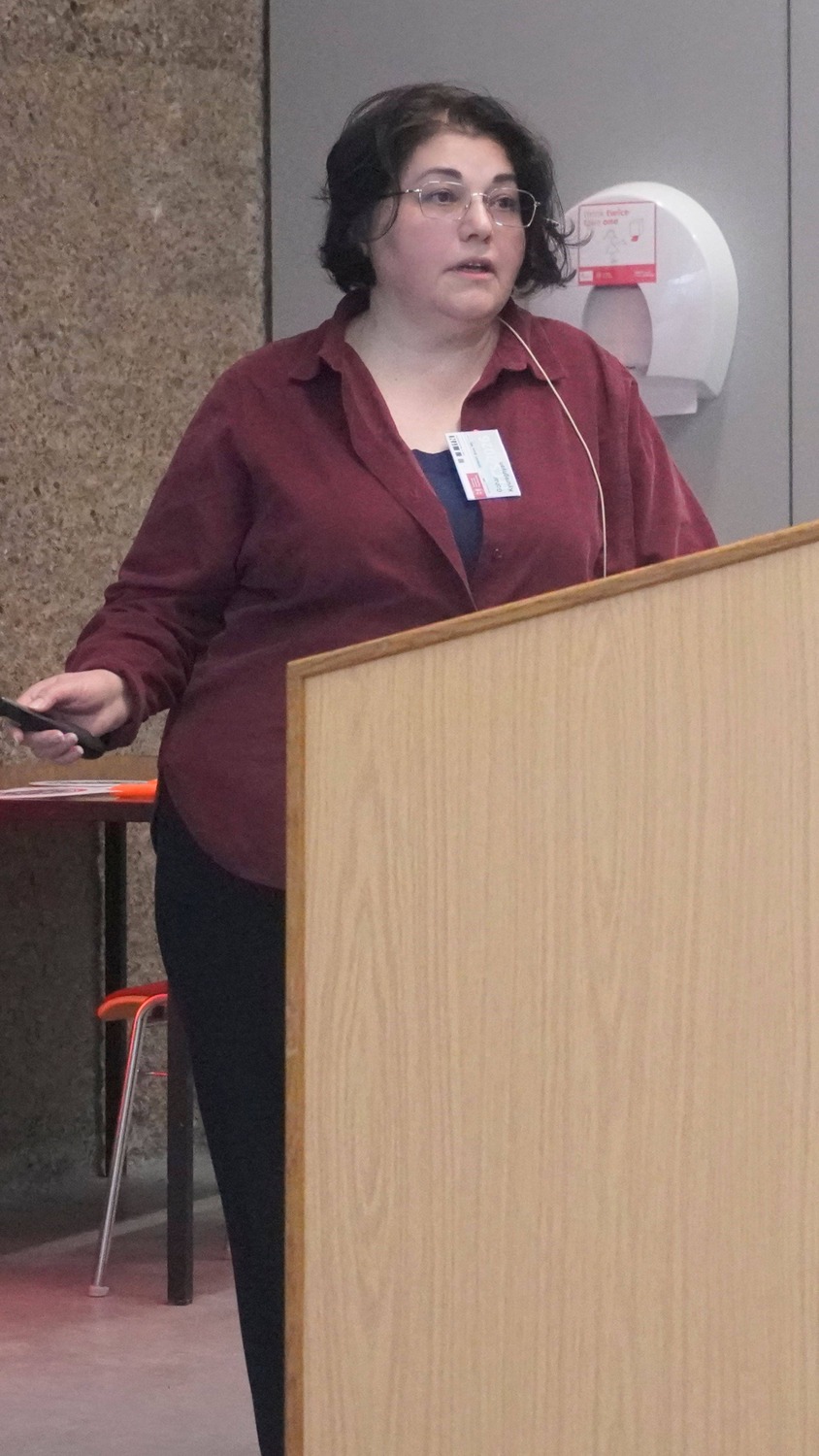}}
\caption{Gohar Kyureghyan} \label{gohar}
\end{floatingfigure}
Gohar Kyureghyan (Fig.~\ref{gohar}) presented recent results on the algebraic structure of a nonlinear mapping over the finite field $\mathbb{F}_2$, which is closely related to cryptographic primitives such as SHA-3 and \textsc{Ascon} \cite{kriepke-kyureg1,kriepke-kyureg2}. The mapping
$\chi : \mathbb{F}_2^n \to \mathbb{F}_2^n$ defined componentwise by
\[
\chi(x)_i = x_i + x_{i+2}(1 + x_{i+1}),
\]
captures essential nonlinear operations used in modern lightweight cryptography.

To analyze its structure, the talk introduced a family of shift-invariant mappings $\gamma_{2k}$ constructed via cyclic shifts and componentwise multiplication. Their linear span $\Gamma_n$ over $\mathbb{F}_2$ serves as a framework for expressing iterates of $\chi$ and understanding their algebraic dependencies.

For odd~$n$, it was shown that every iterate $\chi^j$ can be represented as a linear combination of the mappings $\gamma_{2k}$. In this setting, the space $\Gamma_n$ has dimension $(n+1)/2$, and the subset
\[
G_n = \gamma_0 + \mathrm{span}{\gamma_2,\dots,\gamma_{n-1}}
\]
forms an abelian group under composition. A key result is that this group is isomorphic to the unit group of the residue ring $\mathbb{F}_2[X]/(X^{(n+1)/2})$, establishing a direct link between shift-invariant permutations and invertible binary polynomials.

For even~$n$, the structure changes: $\Gamma_n$ has full dimension $n$, and $G_n$ includes non-bijective mappings. However, the subset of bijections within $\Gamma_n$ still forms an abelian group, now isomorphic to the unit group of $\mathbb{F}_2[X]/(X^n + X^{n/2})$.

Overall, the presentation demonstrated how these algebraic isomorphisms enable both a deeper theoretical understanding and efficient construction of large families of commuting, shift-invariant permutations, reducing complex permutation analysis to the study of corresponding polynomial structures.

\newpage
\subsection{Rafael Schaefer - Semantic Communication -- An Information Theoretic Perspective}
\begin{floatingfigure}[r]{6cm}
\mbox{\includegraphics[width=5.5cm, height=9.78cm]{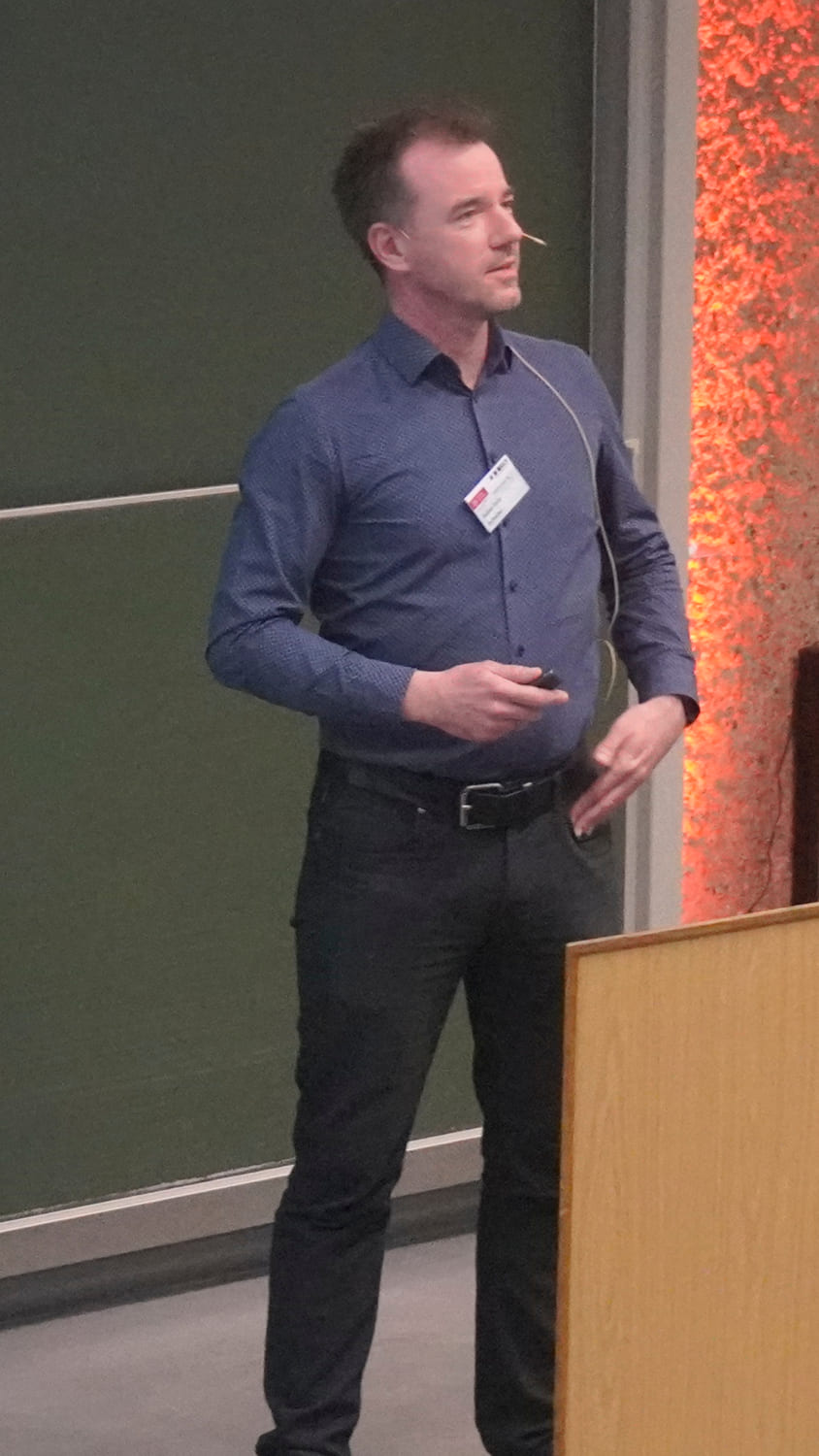}}
\caption{Rafael Schaefer} \label{rafael}
\end{floatingfigure}
Rafael Schaefer (Fig.~\ref{rafael}) presented recent advances on semantic communication from an information-theoretic perspective, emphasizing the transition from reliable symbol transmission to the communication of meaningful information. The talk introduced two complementary mathematical frameworks that extend classical Shannon theory while incorporating semantic aspects.

The first part proposed a probabilistic model inspired by the philosophical concepts of Constraining Affordances and Levels of Abstraction (LoA). Within this framework, communication is described in terms of meanings rather than symbols, with Shannon’s model emerging as a special case in which each message corresponds to exactly one meaning. By integrating semantic content, the notion of channel capacity is extended, showing that the achievable rate of expressing information can exceed the classical Shannon capacity when meanings are properly exploited. Furthermore, the model distinguishes between physical noise and semantic noise, the latter arising from mismatches in context or knowledge between sender and receiver.

The second part focused on semantic communication in the presence of semantic channel noise but assuming an ideal physical channel. A state-dependent channel model was introduced in which the traditional notion of channel state is replaced by context, determining how messages are interpreted. Various scenarios were analyzed, including identical, partially shared, and mismatched contexts between communicating parties. The results demonstrate that contextual knowledge has a significant impact on both message expressibility and achievable communication rates.

Overall, the presentation provided a rigorous information-theoretic foundation for semantic communication, highlighting how meaning and context influence communication limits and bridging classical Shannon theory with emerging context-aware communication paradigms. The results were published in \cite{schaefer-wcnc2026},\cite{schaefer-arxiv2026}.

\newpage
\subsection{Paolo Santini - LDPC and MDPC codes in post-quantum cryptography and the issue of decoding failures}
\begin{floatingfigure}[r]{6cm}
\mbox{\includegraphics[width=5.5cm, height=9.78cm]{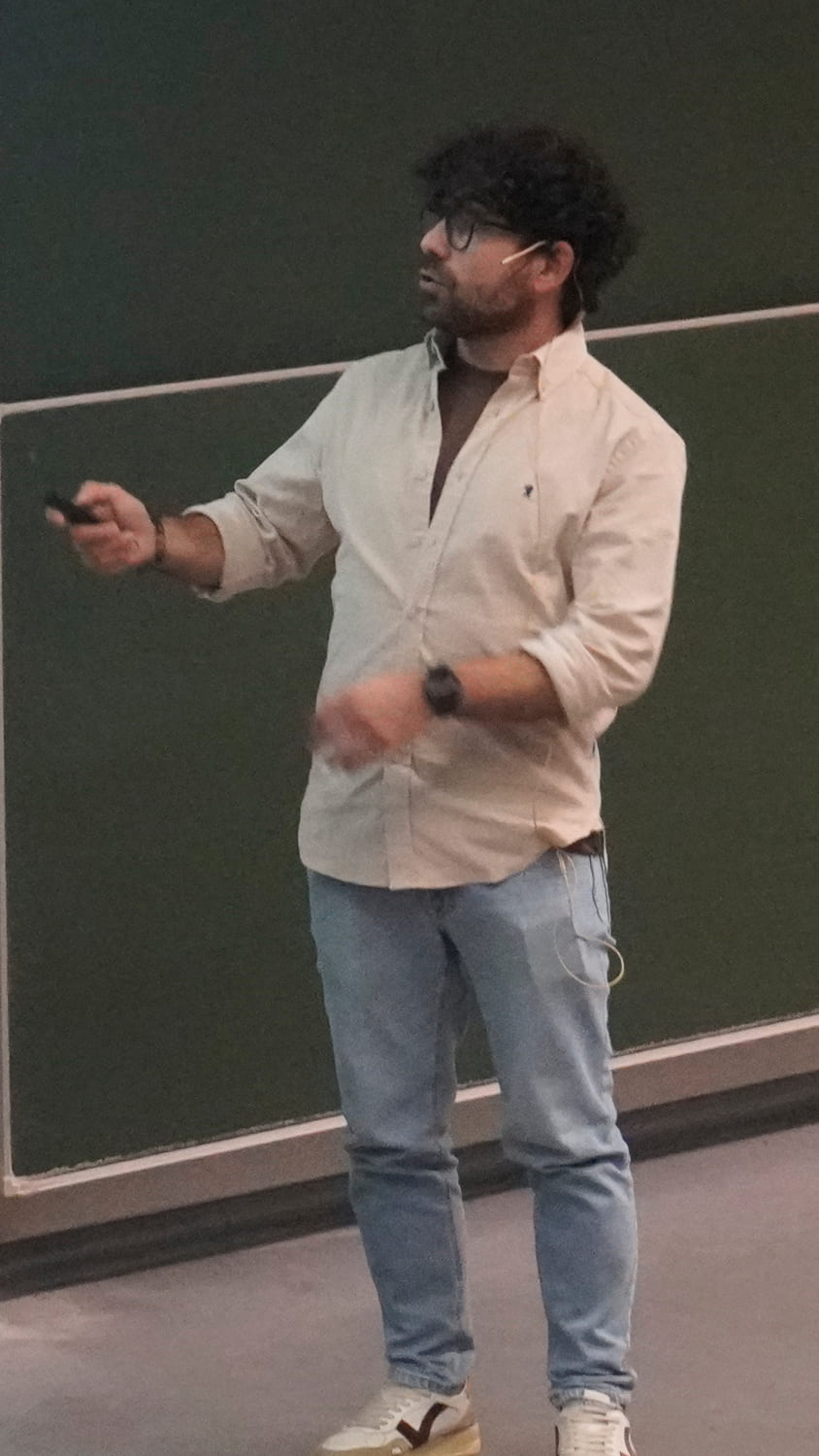}}
\caption{Paolo Santini} \label{paolo}
\end{floatingfigure}
Paolo Santini (Fig.~\ref{paolo}) presented recent advances in code-based post-quantum cryptography, focusing on LDPC and QC-MDPC codes and the challenge of decoding failures in the McEliece framework. Post-quantum cryptography aims to develop cryptographic systems that remain secure against both classical and quantum adversaries, and coding-theoretic constructions constitute one of the main candidate approaches~\cite{McEliece1978,Misoczki2013}.

BIKE is a post-quantum scheme based on codes that has been a candidate in the NIST competition for the standardization of post-quantum schemes. Thanks to the use of Quasi-Cyclic Moderate-Density Parity-Check (QC-MDPC) codes, BIKE offers relatively compact keys and competitive performance. However, as a major drawback (and arguably the main reason hindering its standardization), there are concerns regarding its IND-CCA2 security. Indeed, QC-MDPC codes are decoded using bit-flipping (BF) decoders, which are iterative decoders with a typically nonzero decoding failure rate (DFR). Each decoding failure translates into a decryption failure, and to achieve indistinguishability under adaptively chosen ciphertext attacks (IND-CCA2), it is necessary to ensure that the DFR is negligible (e.g., less than $2^{-\lambda}$). This requires (i) selecting a decoder with strong performance and (ii) deriving a theoretical model to predict the behavior of the DFR. Despite several attempts, no definitive solution has yet been identified, and the problem remains open.

The talk revisits the idea of using error-correcting codes for public-key encryption (dating back to the seminal 1978 work by McEliece) and outlines the most relevant characteristics of BIKE. Particular emphasis is placed on the DFR problem, highlighting open questions and research directions. Recent results on BF-Max, a decoder that has recently been proposed~\cite{baldelli2025bf}, are described. It is shown that, among state-of-the-art decoders, BF-Max achieves a favorable trade-off between error-correction capability and computational complexity.

\newpage


\subsection{Thomas K\"urner - Terahertz communication}
\begin{floatingfigure}[r]{6cm}
\mbox{\includegraphics[width=5.5cm, height=9.78cm]{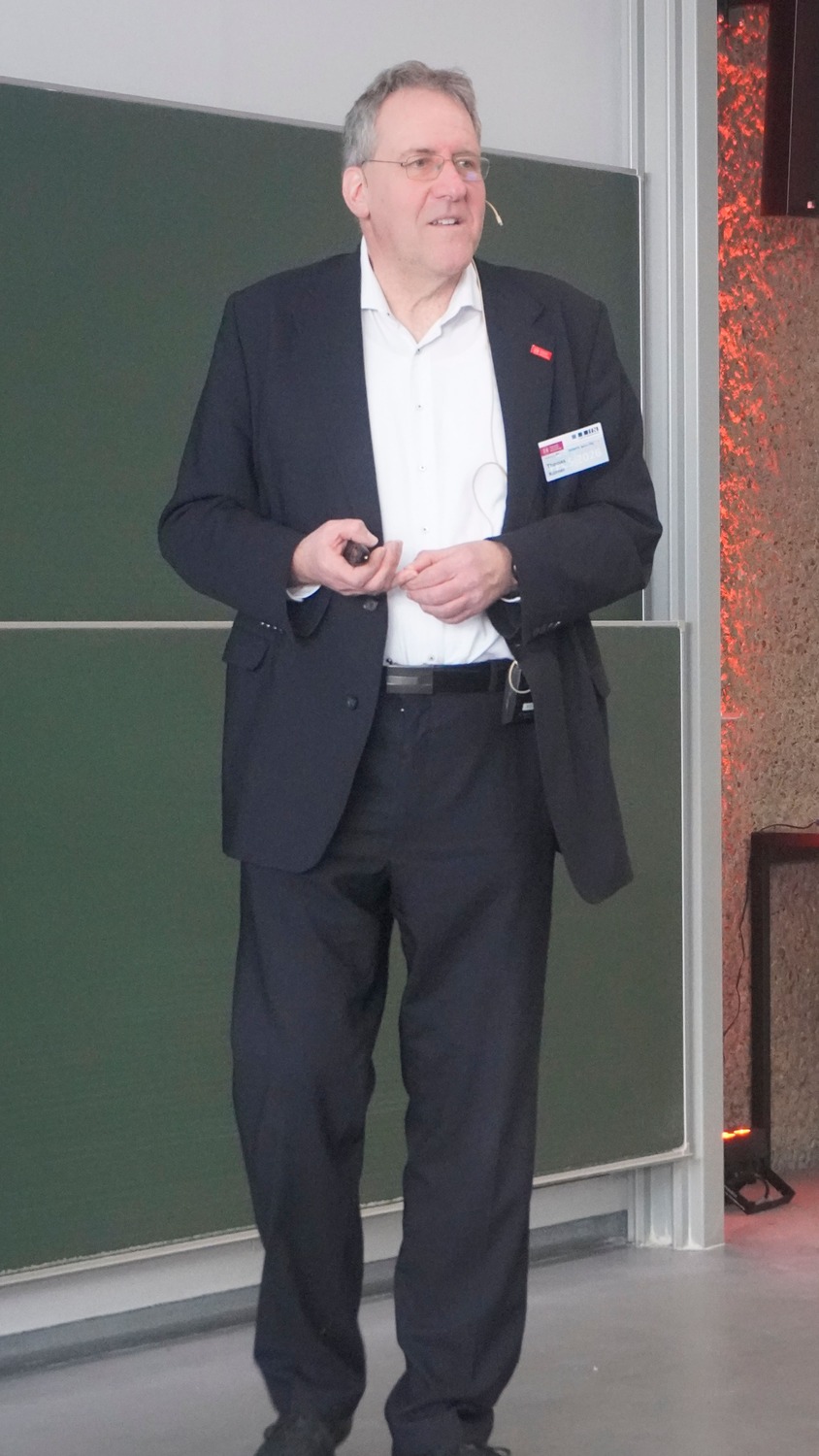}}
\caption{Thomas K\"urner} \label{thomas}
\end{floatingfigure}
Thomas Kürner (Fig.~\ref{thomas}) presented an overview of recent developments in terahertz (THz) communications and discussed the technological and regulatory foundations required to enable wireless systems operating beyond 275\,GHz~\cite{Jornet2024THz}. The motivation for THz communications is the rapidly increasing demand for extremely high data rates in future wireless networks, with long-term visions targeting data rates of up to 1\,Tbps per user. Achieving such rates requires the utilization of the large bandwidth available in the THz spectrum. \cite{Jornet2024THz} \cite{KuernerThoR2024} \cite{Kuerner2020WRC}

The talk highlighted key propagation characteristics of THz channels. Due to the very small wavelengths, highly directional antennas and beamforming techniques are required. While atmospheric attenuation can be significant in outdoor scenarios, it is often negligible in indoor environments. Accurate channel characterization therefore plays a central role. The presentation discussed measurement techniques for THz channel sounding and reported results from recent measurement campaigns, including studies of propagation in aircraft cabins, indoor environments, and data centers. These measurements enable the development of hybrid channel models that combine measurement data with ray-tracing based simulations.

Recent technological advancements were also reviewed. Progress in semiconductor technologies and novel materials has enabled the development of electronic, photonic, and plasmonic THz devices. Experimental demonstrations already show the feasibility of real-time data transmission at 300\,GHz, including a bidirectional backhaul link achieving 2\,$\times$\,20\,Gbit/s over a distance of 150\,m using 16-QAM modulation~\cite{KuernerThoR2024}. 

Finally, the talk discussed ongoing activities in standardization and spectrum regulation. International efforts such as the ETSI ISG THz group and IEEE initiatives are currently preparing the groundwork for future THz communication standards. Although large spectral resources have been identified between 275\,GHz and 450\,GHz, open challenges remain, including coexistence with passive services, channel modeling for mobile scenarios, and the identification of viable large-scale application domains.

\newpage
\subsection{Mahtab Mirmohseni - Semantic Communication}
\begin{floatingfigure}[r]{6cm}
\mbox{\includegraphics[width=5.5cm, height=9.78cm]{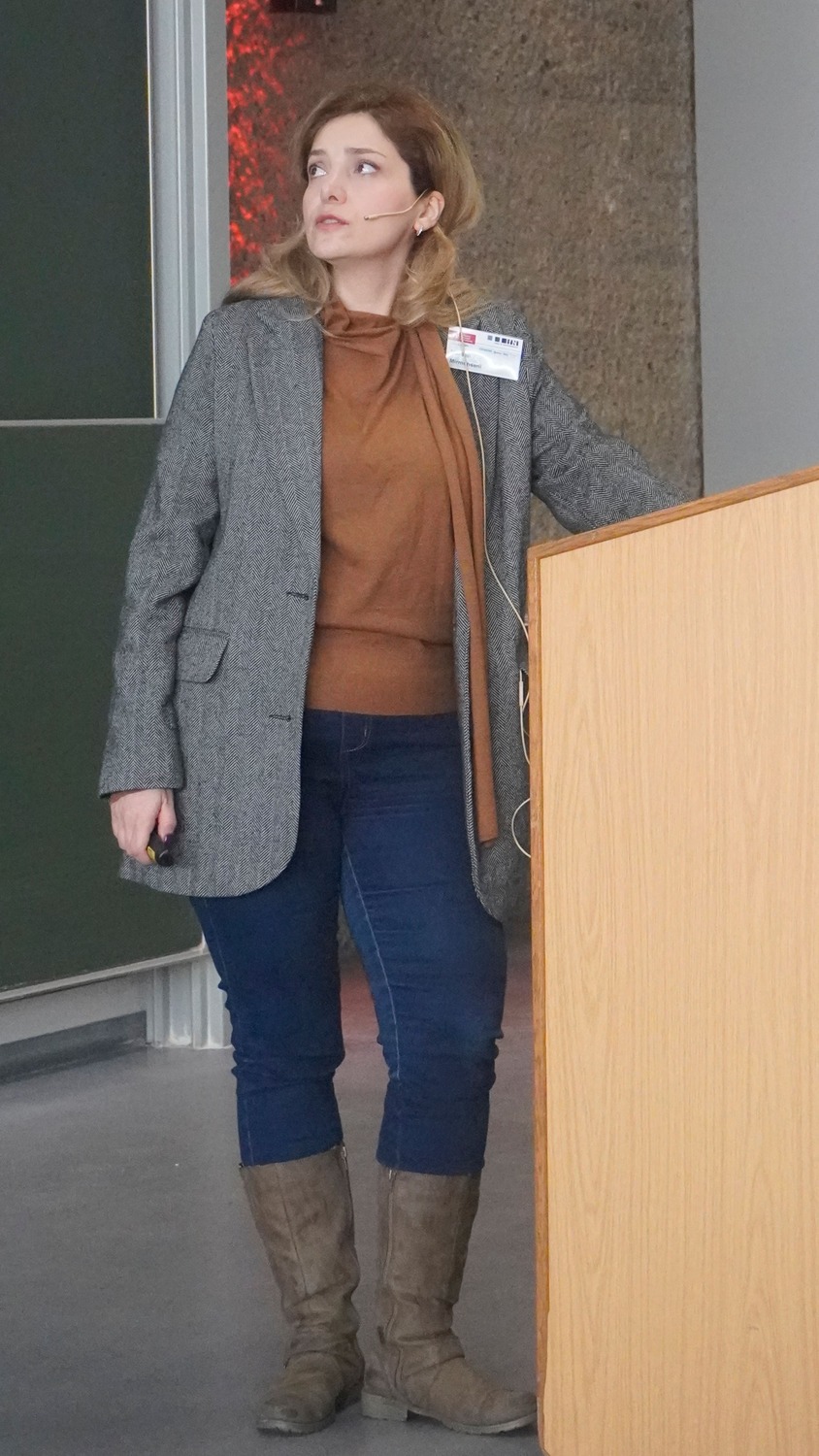}}
\caption{Mahtab Mirmohseni} \label{mahtab}
\end{floatingfigure}
Mahtab Mirmohseni (Fig.~\ref{mahtab}) presented recent work on information-theoretic security for semantic communication systems, based on joint work with Denis Kozlov and Rahim Tafazolli~\cite{kozlov2024secure}. The talk examined how classical information-theoretic tools can be extended to analyze secrecy in semantic communication, where the objective is to transmit task-relevant meaning rather than raw data.

Semantic communication integrates meaning, context, and task objectives into the communication process and can significantly improve the efficiency of wireless networks. The presentation focused on an information-theoretic framework in which a semantic source consists of an intrinsic semantic component and an observed component that are statistically correlated. The communication system must satisfy distortion constraints for both components while guaranteeing secrecy against an eavesdropper. 

The problem is formulated through a rate–distortion–equivocation region that captures the trade-offs between communication rate, reconstruction accuracy, and secrecy. Converse and achievability results were derived for semantic communication over a wiretap channel. The proposed coding scheme combines source coding, encryption, and channel coding using techniques such as superposition coding and message splitting.

Finally, the talk outlined several research directions, including multimodal semantic sources, per-modality secrecy constraints, and extensions to multi-user communication settings.
\newpage
\subsection{Alessio Zappone - Radio resource allocation in metasurface-based wireless networks}
\begin{floatingfigure}[r]{6cm}
\mbox{\includegraphics[width=5.5cm, height=9.78cm]{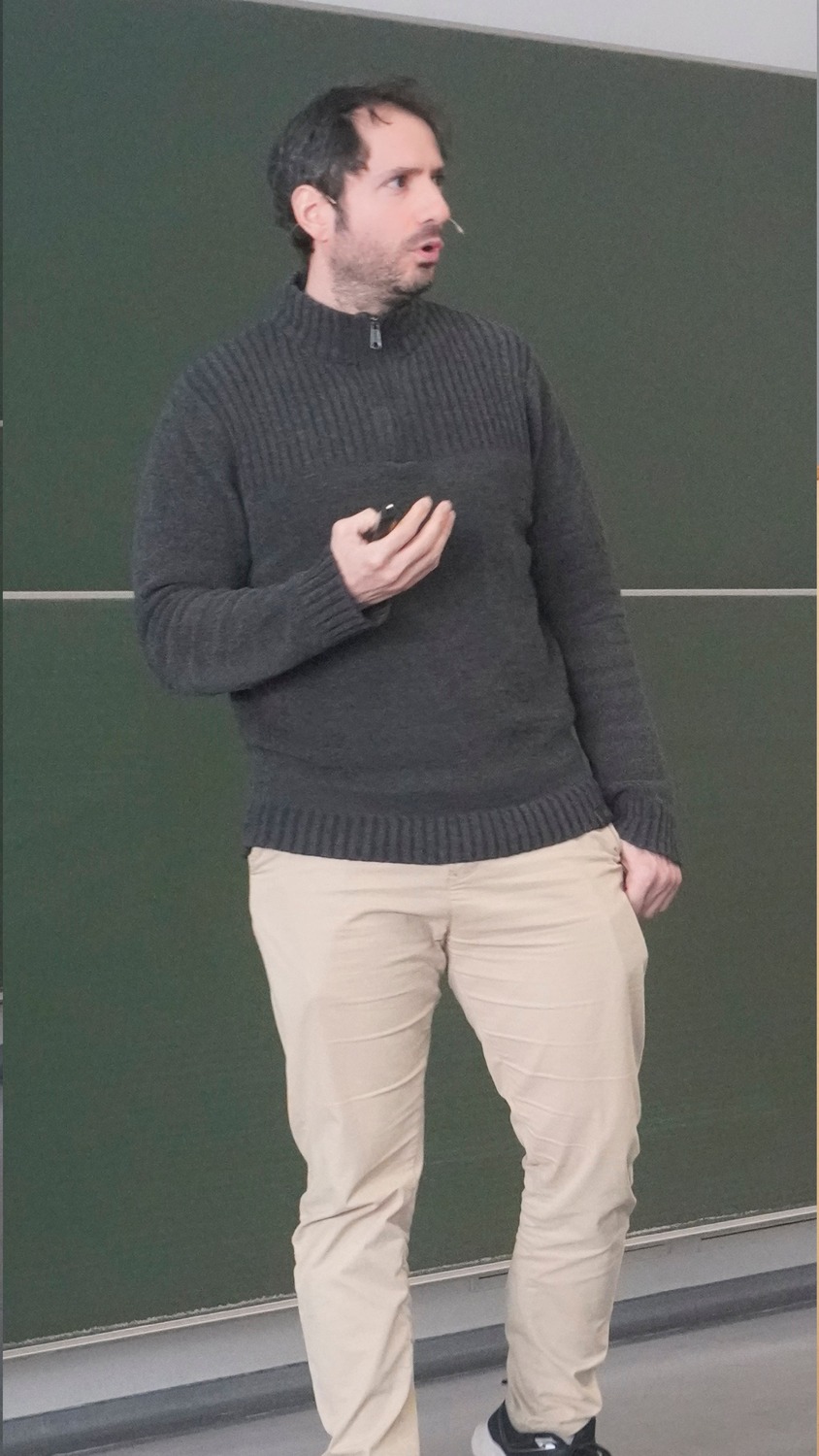}}
\caption{Alessio Zappone} \label{alessio}
\end{floatingfigure}
Alessio Zappone (Fig.~\ref{alessio}) discussed radio resource allocation in metasurface-based networks, presenting optimization formulations and algorithmic approaches that account for reconfigurable propagation, coupled variables, and practical control overhead. The talk is divided into two parts.
In the first part, the focus is on a single-user multiple-input multiple-output system, in
which two nearly-passive metasurfaces are deployed in the near-field of the transmit and receive
antenna arrays, respectively. The two metasurfaces are not wired to the antenna feeders, but are
reached by the transmit signal via wireless propagation. In this context, the problem of optimizing
the system energy efficiency with respect to the two metasurface matrices and the transmit
covariance matrix is tackled. In the special case of a single data-stream transmission, the optimal
reflection matrices of the two metasurfaces are obtained in closed-form, while the beamforming
vector is obtained by means of the sequential fractional programming optimization framework. In
the general case of multi-stream transmission, a numerical algorithm based on sequential
fractional programming is developed to optimize the system radio resources. The results show
that using the metasurfaces close to the transmit and receive antenna arrays allows reducing the
number of digital radio-frequency chains, while at the same time providing large energy efficiency
levels. This is the case in both the single-stream and multi-stream scenarios. More details can be
found in \cite{ZapTCOM}.
In the second part of the talk, a novel method for radio resource allocation is presented,
which aims at reducing the computational complexity of optimization methods based on
traditional optimization theory. The considered approach is based on the cross-entropy
maximization algorithm, and is shown to converge to an optimized energy efficiency level, with
lower complexity than state-of-the-art method. The approach does not require any concavity/
convexity properties of the objective functions and constraints, and neither does it require the
computation of any gradient. It is guaranteed to converge in probability to a solution that provides
a desired level of objective functions, provided the desired level is feasible.

\newpage
\subsection{Stephan ten Brink - Integrated Sensing and Communication}
\begin{floatingfigure}[r]{6cm}
\mbox{\includegraphics[width=5.5cm, height=9.78cm]{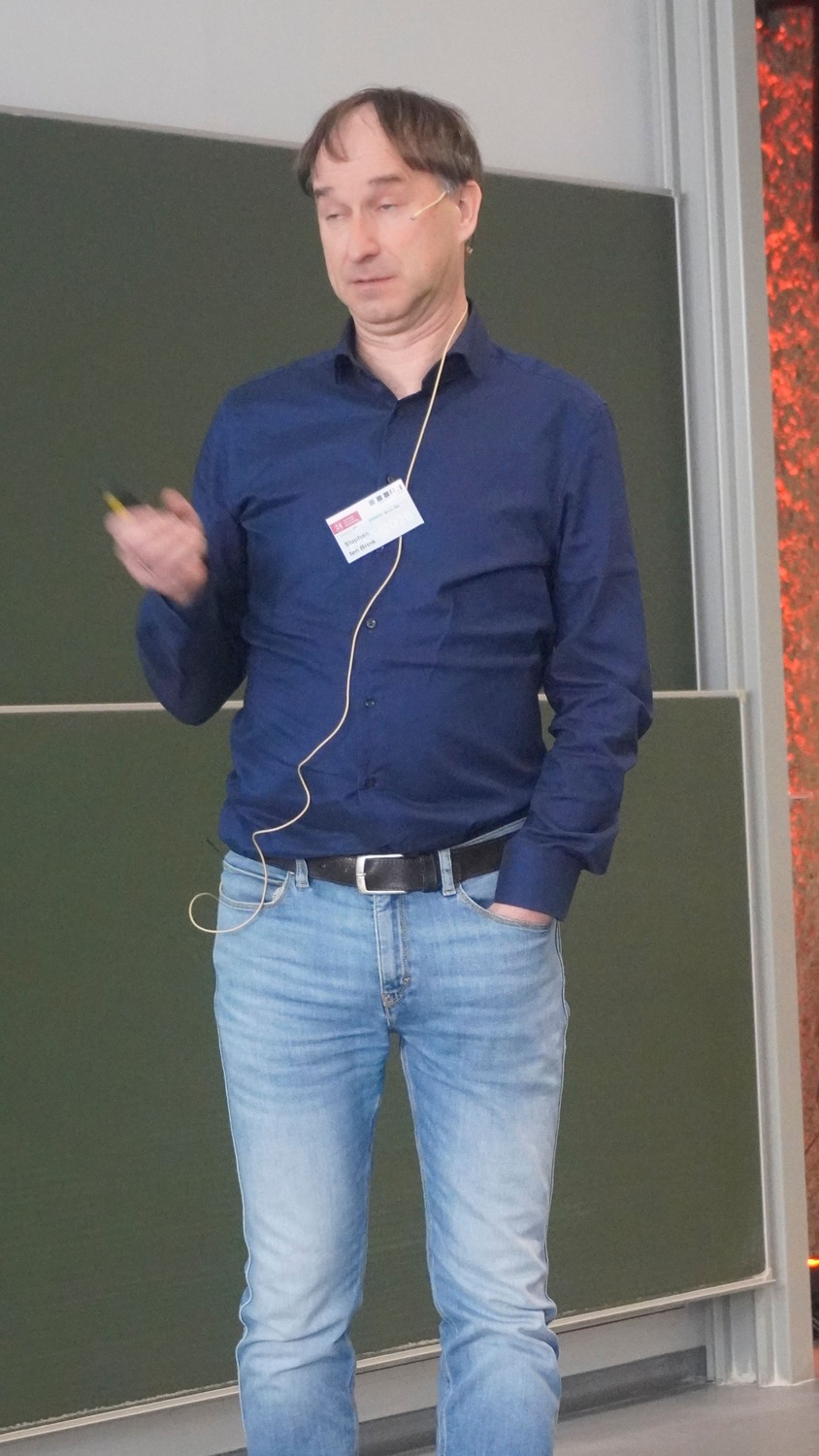}}
\caption{Stephan ten Brink} \label{stephanten}
\end{floatingfigure}
Stephan ten Brink (Fig.~\ref{stephanten}) discussed integrated sensing and communication, highlighting waveform and receiver design considerations, and illustrating how joint design enables new capabilities while creating new trade-offs between sensing accuracy and data throughput.

Recent developments in multiple-input multiple-output (MIMO) technology for both communication and sensing applications are highlighted. Topics include channel measurements, channel modeling, and AI-based post-processing of channel state information, with potential benefits for 6G wireless networks. Distributed yet tightly synchronized MIMO prototypes for coherent channel measurements are shown to be useful in further assessing the potential of the discussed schemes.
\newpage


\subsection{Tadashi Wadayama - Mutual Information Estimation via Score-to-Fisher Bridge for Nonlinear Gaussian Noise Channels}
\begin{floatingfigure}[r]{6cm}
\mbox{\includegraphics[width=5.5cm, height=9.78cm]{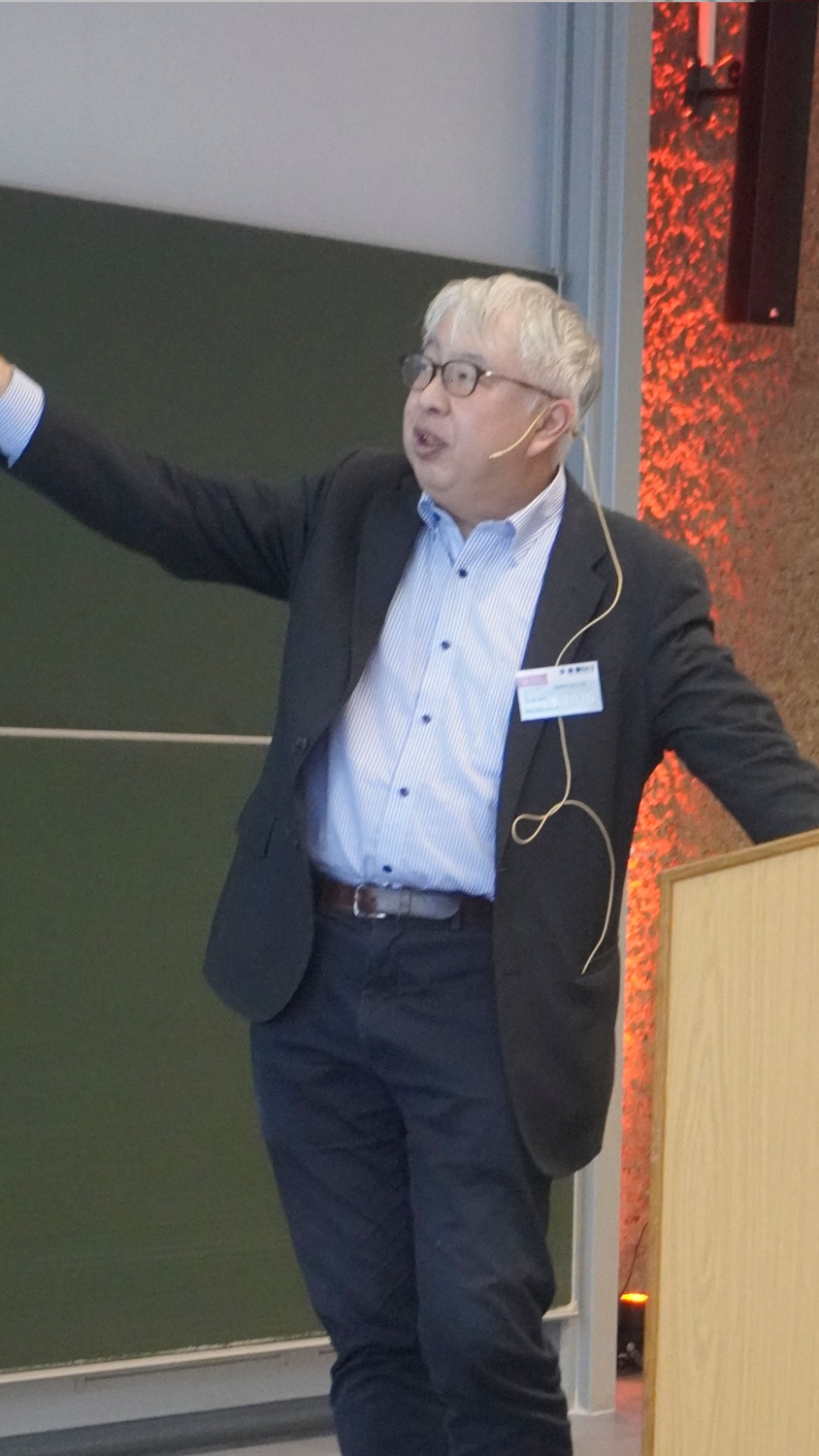}}
\caption{Tadashi Wadayama} \label{tadashi}
\end{floatingfigure}
Tadashi Wadayama (Fig.~\ref{tadashi}) presented a numerical method to evaluate mutual information (MI) in nonlinear Gaussian noise channels by using denoising score matching (DSM)~\cite{vincent2011,song2019} learning to estimate the score function of the channel output. Via de Bruijn's identity~\cite{stam1959}, Fisher information~\cite{fisher1922,cover2006} estimated from the learned score function yields accurate estimates of MI through a Fisher integral representation for a variety of priors and channel nonlinearities.

A comprehensive theoretical foundation for the Score-to-Fisher bridge methodology is developed, along with practical guidelines for its implementation. Extensive validation experiments are conducted, comparing the approach with closed-form solutions and a kernel density estimation~\cite{moon1995,kraskov2004} baseline. The results of the numerical experiments demonstrate that the method is both practical and efficient for MI estimation in nonlinear Gaussian noise channels.

Details of the work are available at \url{https://arxiv.org/abs/2510.05496}.
\newpage
\subsection{Karl-Ludwig Besser - Building Resilience in Wireless Communication Systems With a Secret-Key Budget}
\begin{floatingfigure}[r]{6cm}
\mbox{\includegraphics[width=5.5cm, height=9.78cm]{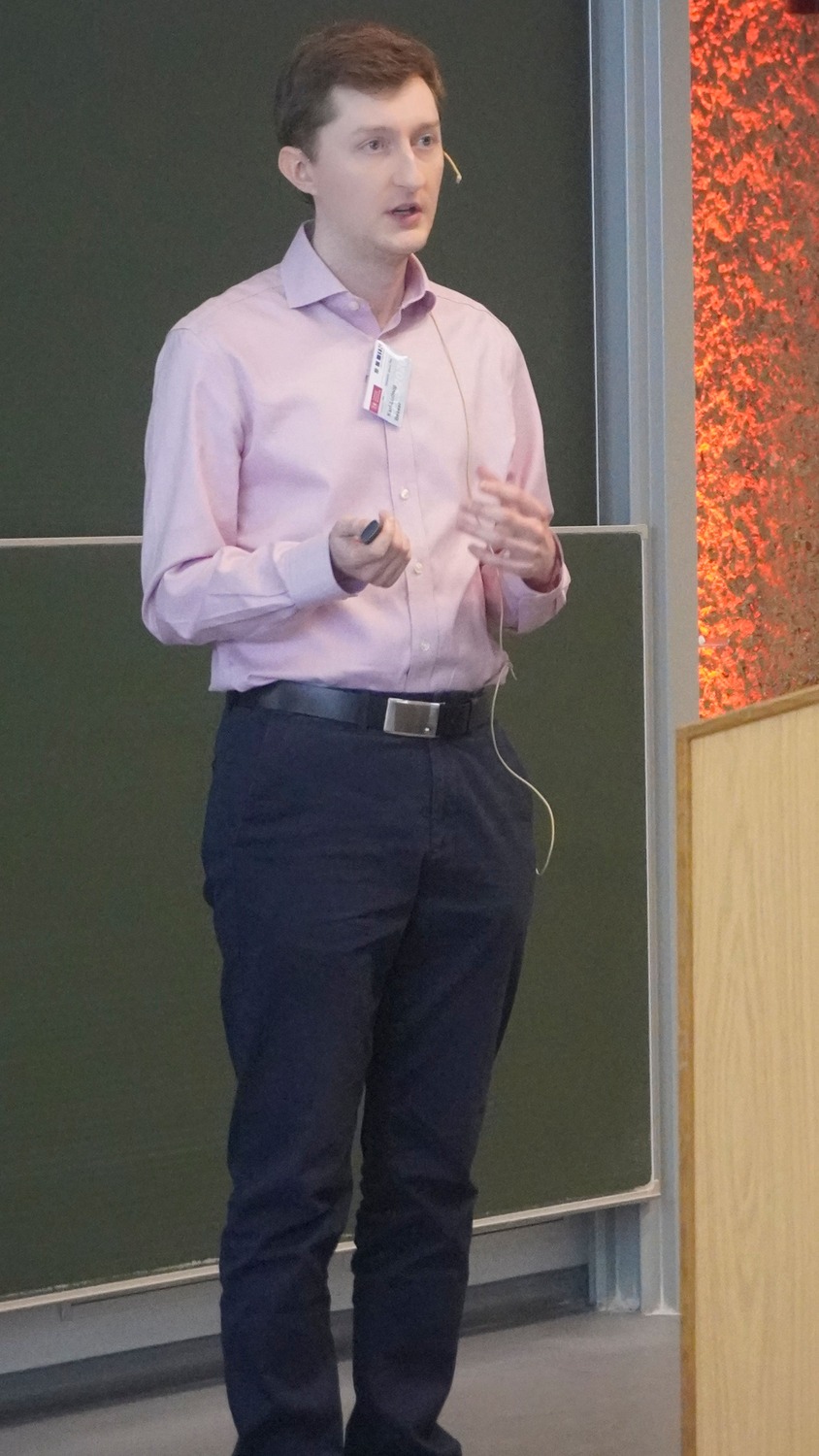}}
\caption{Karl-Ludwig Besser} \label{karl}
\end{floatingfigure}

Karl-Ludwig Besser (Fig.~\ref{karl}) presented recent advances on quantifying and optimizing resilience in wireless communication systems, with a particular focus on resource-aware system design. The talk emphasized resilience as a key performance metric and introduced a novel framework based on resource budgets.

To illustrate this concept, a system based on secret-key generation (SKG) was considered. In this model, secret key bits are generated via wireless SKG techniques and stored in a key budget. When secure communication is required, these key bits are consumed as a one-time pad for encryption and removed from the budget \cite{Besser2024}. This dynamic perspective enables modeling the system’s ability to sustain secure communication over time.

Building on this framework, the presentation introduced new physical-layer resilience metrics, including the *alert outage probability* and the *resilience outage probability*. These metrics quantify how well the system is prepared to handle critical or emergency situations, capturing the notion of survivability under resource constraints. Using tools from mathematical finance, theoretical bounds for these metrics were derived, providing analytical insights into system performance.

Furthermore, the talk investigated power allocation strategies and their trade-offs between resilience and energy consumption. It was shown how transmit power can be optimized to guarantee a desired level of resilience while minimizing energy usage \cite{Besser2025}.

Overall, the results provide a systematic approach for designing wireless systems that balance reliability, security, and resilience, offering practical guidelines for configuring system parameters under resource constraints.

\newpage
\subsection{Shun Watanabe - When is structured coding useful in goal-oriented communication? A case study of distributed function computation}
\begin{floatingfigure}[r]{6cm}
\mbox{\includegraphics[width=5.5cm, height=9.78cm]{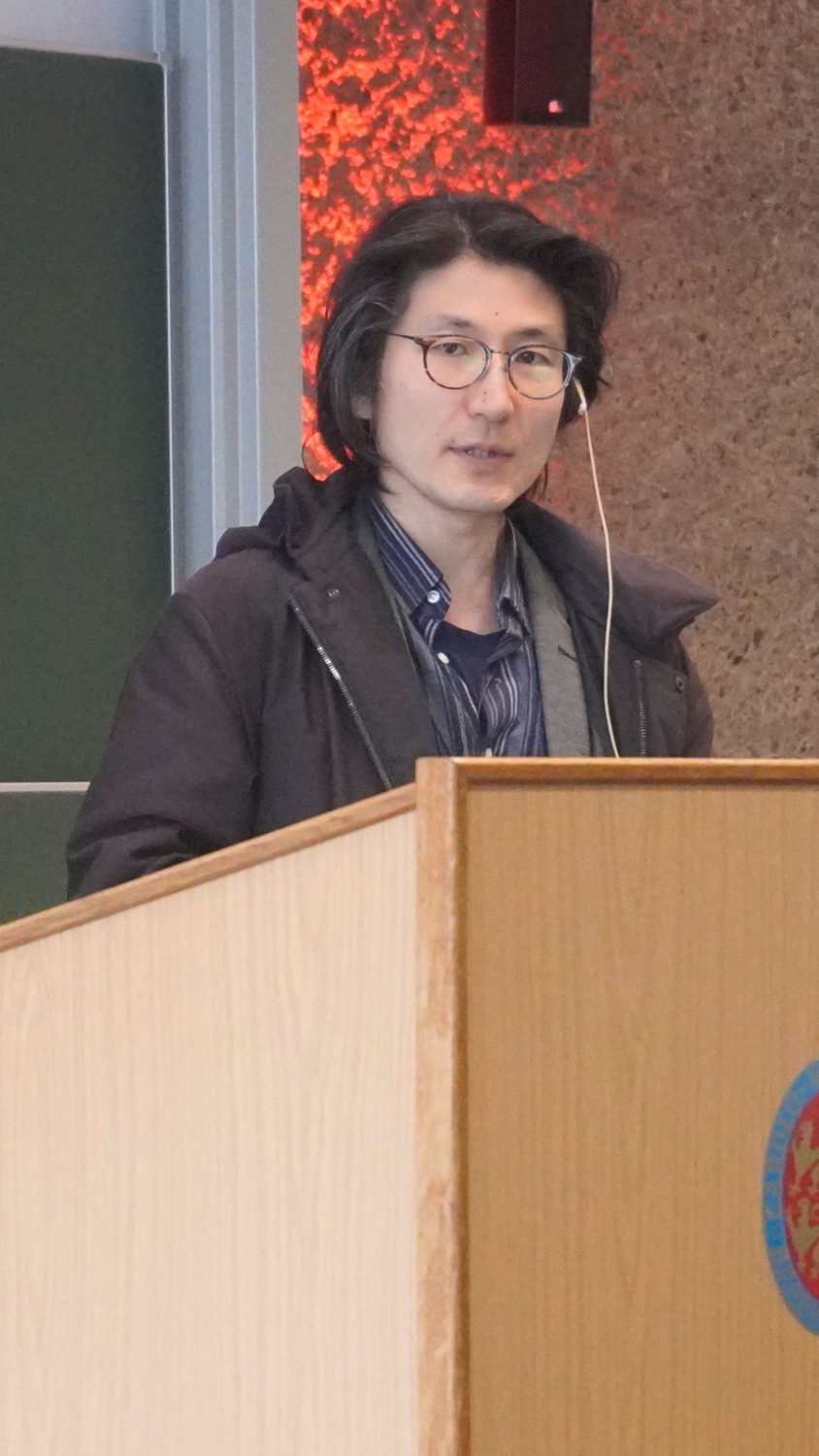}}
\caption{Shun Watanabe} \label{shun}
\end{floatingfigure}
Shun Watanabe (Fig.~\ref{shun}) discussed when structured coding can improve the efficiency of goal-oriented communication. The talk examined this question through the example of distributed function computation, where multiple terminals compress their observations so that a destination can compute a desired function.

In the classical setting, if the destination reconstructs the entire data, it can also compute the desired function. Therefore, the Slepian–Wolf rate region for distributed compression provides a natural achievable region for this problem. However, Körner and Marton showed that for certain functions, such as the modulo-sum, structured coding can achieve strictly larger rate regions than Slepian–Wolf coding~\cite{KM79}. Later, Han and Kobayashi derived necessary and sufficient conditions on functions for which the Slepian–Wolf region cannot be improved in the two-encoder case~\cite{HK87}. For more than two encoders, only partial characterizations are known, with improved sufficient conditions developed in later work~\cite{W20}.

The talk also considered the complementary question of identifying source conditions under which Slepian–Wolf coding remains optimal for a given function. For the modulo-sum problem, sufficient conditions for optimality are known~\cite{CK11,NW20}. In addition, Ahlswede and Han proposed a hybrid coding scheme combining random and structured coding that can enlarge the achievable region beyond both the Slepian–Wolf and Körner–Marton regions~\cite{AH}. However, this improvement does not necessarily increase the achievable sum rate~\cite{SGR15}. More recent numerical studies suggest that multi-letter versions of the Ahlswede–Han region can strictly improve upon the Slepian–Wolf region when certain conditions are violated~\cite{KW22}.

Overall, the presentation reviewed recent developments in distributed function computation and highlighted the potential importance of structured and multi-letter coding schemes for goal-oriented communication.

\newpage
\subsection{Boulat Bash - Emulation of entanglement distribution networks using quantum computer}
\begin{floatingfigure}[r]{6cm}
\mbox{\includegraphics[width=5.5cm, height=9.78cm]{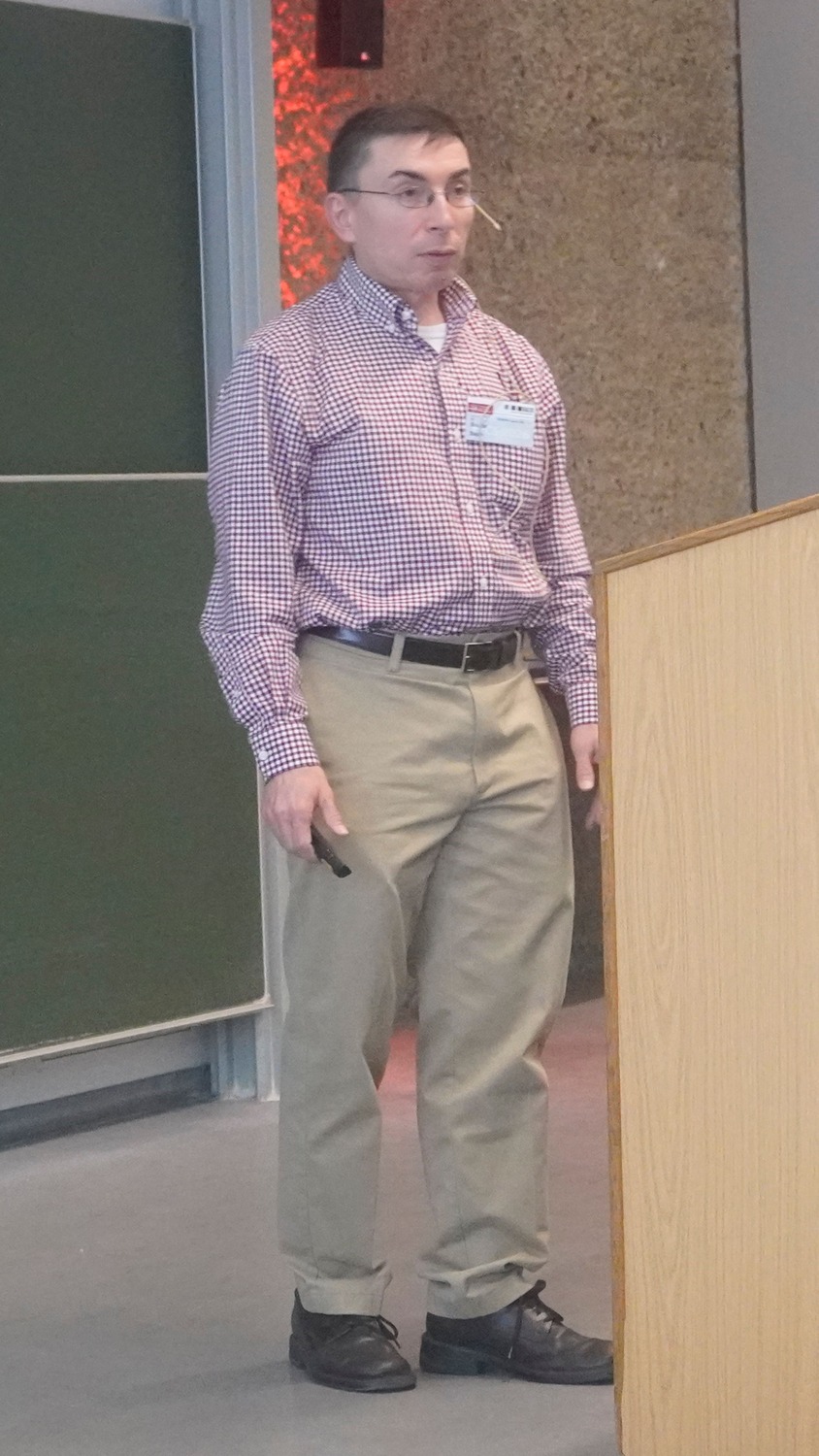}}
\caption{Boulat Bash} \label{bulat}
\end{floatingfigure}
Boulat Bash (Fig.~\ref{bulat}) presented work on the emulation of entanglement distribution networks using quantum computers, with the goal of supporting distributed quantum computing across multiple quantum processing units (QPUs). Current quantum computers are largely limited to local operations with classical communication (LOCC), which restricts the direct implementation of large-scale distributed quantum algorithms. The presented approach investigates how virtual entanglement can be used to emulate distributed quantum networks and study their performance under realistic imperfections.

The framework is based on circuit cutting applied to Bell pairs. In this approach, a distributed quantum circuit is decomposed into smaller separable circuits that can be executed independently on local quantum hardware. The results of these circuits are then classically combined to reproduce the outcome of the original circuit that would require entanglement across distant QPUs. This technique allows the emulation of entanglement-assisted distributed quantum computation using currently available quantum hardware.
The work further incorporates practical impairments into this framework. In particular, imperfect Bell-pair sources and classical communication delays are modeled using depolarizing noise channels. Several equivalent modeling approaches were studied, including Stinespring dilations, randomly applied Pauli errors, and quasi-probability decompositions. The performance of the resulting virtual entanglement links was evaluated using stabilizer measurements and graph-state fidelities as indicators of entanglement quality.
Experiments and simulations were conducted using IBM quantum hardware. The results show that an initial Bell-pair fidelity of roughly 90\% is required to reliably generate entanglement across the emulated network. Furthermore, classical communication delays introduce additional decoherence that limits the achievable network size. The study provides a first experimental framework for analyzing distributed quantum computing architectures and highlights the importance of both high-fidelity entanglement sources and low-latency classical communication for scalable quantum networks.
\newpage

\section{Lessons Learned and Open Problems}

The contributions presented at FFCS 2026 collectively point to a fundamental shift in how communication systems are understood and designed. While classical Shannon theory continues to provide the conceptual backbone, many of the discussed works demonstrate that future communication systems can no longer be adequately described by rate-centric transmission models alone. Instead, communication increasingly appears as part of a broader framework that integrates identification, inference, sensing, computation, and security under realistic physical and architectural constraints.

A central insight emerging across multiple contributions is the transition from reliable message transmission to identification- and inference-based communication. In many modern applications -- such as large-scale IoT systems, distributed autonomous agents, or nanoscale biomedical networks -- the primary goal is not to reconstruct a message in full detail, but rather to decide whether a specific hypothesis, event, or command applies. This shift leads to fundamentally different performance criteria and scaling laws. In particular, identification via channels enables a double-exponential growth in the number of distinguishable messages, and recent results indicate that this advantage can be maintained even in the presence of feedback, secrecy constraints, and sensing requirements. These findings suggest that identification-based paradigms will play a central role in the design of future communication systems, especially in scenarios characterized by massive device populations and stringent resource limitations.

Closely related to this development is the increasing integration of sensing and communication. Several contributions highlight that communication systems are no longer isolated from their environment but must actively acquire and process information about it. Integrated sensing and communication (ISAC) and related frameworks demonstrate that state information, feedback, and distributed observations can be leveraged not only to improve reliability but also to enable new functionalities such as environment-aware adaptation and common randomness generation. At the same time, this integration introduces new trade-offs between communication rate, estimation accuracy, latency, and resource consumption, which are not yet fully understood, particularly in multi-user and network settings.

Another important theme is the growing relevance of physical-layer models that go beyond classical electromagnetic communication. Quantum communication, molecular communication, and other physically grounded paradigms illustrate that the underlying physics of the system lead to significant new challenges, requiring innovative solutions. In quantum systems, entanglement and measurement-induced correlations enable fundamentally new capabilities but also impose strict constraints related to noise, decoherence, and implementability. In molecular and nanoscale communication, the medium itself -- characterized by diffusion, chemical reactions, and stochastic transport --becomes an integral part of the communication process. These developments indicate that future communication theory must be tightly coupled to physical modeling, requiring new abstractions that remain analytically tractable while faithfully capturing the relevant physical phenomena.

Security and trust also emerge as intrinsic design dimensions rather than secondary considerations. Many contributions study communication systems under adversarial conditions, including eavesdroppers, colluding users, and Byzantine behavior. A particularly notable insight is that, under certain conditions, security constraints do not necessarily reduce system performance. For instance, secure identification with feedback can achieve the same fundamental limits as its non-secure counterpart when the underlying secrecy capacity is positive. At the same time, the presence of adversaries introduces new challenges that require robust, adaptive, and often fundamentally different coding strategies.

Beyond these conceptual advances, several works highlight fundamental algorithmic and computational limitations. Results demonstrating the non-computability of certain capacity expressions or optimization procedures reveal that even when theoretical limits are well-defined, they may not be accessible through algorithmic means. This distinction between information-theoretic achievability and computational realizability points to a critical gap that must be addressed in order to translate theoretical insights into practical system designs.

Taken together, the contributions presented at FFCS 2026 suggest a number of important open problems. A deeper understanding of identification under finite blocklength and complexity constraints is required to enable practical implementations. The development of a unified theory for integrated sensing and communication, particularly in distributed and multi-user settings, remains an open challenge. In the quantum domain, extending existing results to scalable and heterogeneous network architectures is still largely unresolved. Similarly, communication models that incorporate physical and biological constraints, as in molecular systems, require new analytical tools and design principles. Finally, bridging the gap between theoretical limits and computationally efficient algorithms remains a central challenge across many areas.

Overall, FFCS 2026 highlights the emergence of a more holistic view of communication systems, in which communication, sensing, computation, and control are deeply intertwined. This evolution calls for new theoretical frameworks, closer integration with physical modeling, and increased interdisciplinary collaboration. It also suggests that the future of communication theory lies not only in extending classical results, but in rethinking its foundations to accommodate the complexity and diversity of next-generation systems.

\section*{Conclusions}

The mix of participants at the workshop was very successful, bringing together a diverse group of individuals. Researchers from information theory, quantum communication, molecular and nanoscale communication, semantic and goal-oriented communication, and resilient networked systems engaged in an unusually open and technically deep exchange. This breadth was not only visible in the program structure—spanning classical and quantum foundations, security, and new physical communication modalities, but also in the discussions that connected these areas and repeatedly challenged community boundaries.

Across the week, several recurring themes emerged. First, many contributions revisited fundamental limits under realistic constraints: limited resources (entanglement, memory, bandwidth, energy), imperfect hardware, uncertain environments, and adversarial behavior. Rather than treating constraints as secondary implementation details, the workshop highlighted them as central modeling ingredients that can reshape what is possible—sometimes even enabling new primitives such as secrecy, authentication, or identification. Second, multiple sessions emphasized that future communication systems will be evaluated increasingly by utility rather than by raw throughput alone. This perspective was reflected in work on identification, semantic communication, integrated sensing and communication, learning over unreliable networks, and multi-objective optimization, where “what must be achieved” matters as much as “how many bits can be transmitted.”

The workshop also made clear that the field is moving toward a closer integration of theory and system design. On the quantum side, presentations connected abstract rate regions, distillation methods, and device-independent security perspectives with architectural and technological roadmaps for quantum networks. On the classical side, topics such as covert communication, physical-layer security, post-quantum cryptographic code design, and reconfigurable arrays demonstrated how information-theoretic reasoning can directly inform robust system engineering. In parallel, molecular and biological signaling talks reinforced the importance of physically grounded channel models and of interdisciplinary dialogue, especially when communication mechanisms depart fundamentally from electromagnetic wave-based paradigms.

The poster session and lightning talks played an important role in broadening participation and capturing emerging directions. The posters complemented the invited program with early-stage ideas, new proof techniques, refined finite-blocklength perspectives, and novel applications—ranging from quantum multi-user models and optical security analyses to identification with feedback and integrated sensing concepts. The format encouraged high interaction density and helped connect junior researchers to senior experts, often leading to concrete follow-up discussions.

Overall, FFCS 2026 succeeded in its central objective: to create a forum where foundational questions about future communication systems can be addressed jointly across communities that often meet separately. The workshop demonstrated strong momentum toward (i) richer system models that include resources, constraints, and adversaries explicitly; (ii) new performance criteria aligned with tasks, semantics, and trust; and (iii) tighter loops between theory, simulation, and implementable architectures. The organizers anticipate that the collaborations initiated and strengthened during the week will translate into joint research projects, shared benchmarks and models, and follow-up workshops that continue building a coherent foundation for future communication systems.

\section*{Acknowledgments}
The authors would like to thank all contributors to the Foundations of Future Communication Systems (FFCS) 2026 workshop for their valuable input and inspiring presentations. This report is based on these contributions and on notes compiled during the workshop, and we particularly acknowledge the student contributors for their support in collecting, structuring, and preparing the material.
In particular, we gratefully acknowledge the contributions of all speakers, poster presenters, and participants, including Hadi Aghaee, Hossein Ahmadi, Houman Asgari, Liubov Bakhchova, Konrad Banaszek, Jessica Bariffi, Boulat Bash, Riccardo Bassoli, Karl-Ludwig Besser, Rawad Bitar, Igor Bernard, Igor Bjelakovic, Holger Boche, Stephan ten Brink, Stefano Buzzi, Minglai Cai, Yanling Chen, Ghislaine Coulter-de Wit, Matteo Nerini, Armin Dekorsy, Falko Dressler, Sandor Fekete, Frank Fitzek, Rami Ezzine, Marc Geitz, Ilja Gerhardt, Vida Gholamian, Fatma Gouiaa, Saikat Guha, Christoph Hirche, Pol Julià Farré, Daniel Kilper, Rodrigo Kloster Albarracín, Martin Korte, Thomas Kürner, Gohar Kyureghyan, Svenja Lage, Erik G. Larsson, Pin-Hsun Lin, Laura Luzzi, Karol Lukanowski, Maurizio Magarini, Bho Matthiesen, Kristin Michaelsen-Preusse, Marcel Mross, Mahtab Mirmohseni, Husein Natić, Kumar Nilesh, Prakash Narayan, Janis Nötzel, Nikhitha Nunavath, Arun Padakandla, Massimiliano Pierobon, Neha Sangwan, Paolo Santini, Maximilian Schäfer, Rafael F. Schaefer, Stefan Schmid, René Schwonnek, Aydin Sezgin, Mohammad Soleymani, Slawomir Stanczak, Nam Tran, Anton Trushechkin, Ugo Vaccaro, Tadashi Wadayama, Shun Watanabe, Stefan Wegele, Kläre Wienecke, Moritz Wiese, Andreas Winter, Gerhard Wunder, Jyun-Sian Wu, Lorenzo Zaniboni, Alessio Zappone, and Yaning Zhao.
Finally, we thank all participants for creating a stimulating and collaborative environment.
FFCS 2026 was supported by the German Research Foundation (DFG) through acquired funding. In addition, the workshop received generous sponsorship from the 6G-life project (financial support, invited presentation, and exhibition booth), Siemens Mobility (financial support and invited presentation), MDPI (Best Poster Award), and Springer (conference proceedings), whose contributions are gratefully acknowledged.
Christian Deppe and Vida Gholamian further acknowledge financial support from the Federal Ministry of Research, Technology and Space of Germany (BMFTR) within the program “Souverän. Digital. Vernetzt.” under the joint project 6G-life (project identification number: 16KIS2415). Christian Deppe gratefully acknowledges additional financial support from the BMFTR Quantum Program through the projects QD-CamNetz (Grant 16KISQ169), QUIET (Grant 16KISQ0170), QSTARS (Grant 16KIS2602), QR.N (Grant 16KIS2196), and Q-TREX (Grant 16KISR038). Further support was provided through QF2, funded by the DFG under Germany’s Excellence Strategy – EXC 2123/2 Quantum Frontiers (390837967), as well as the FOCAL project, funded by the European Union within the framework of the Marie Skłodowska-Curie Actions under Horizon Europe. Christian Deppe and Vida Gholamian also gratefully acknowledge support from the DFG under project DE1915/2-1.

\bibliographystyle{splncs04}
\bibliography{references}
\end{document}